\documentstyle[epsf]{mn}			

\title[Peculiar motions of early-type galaxies -- VII] 
{The peculiar motions of early-type galaxies in two distant regions -- VII. 
Peculiar velocities and bulk motions}

\author[Colless \etal]{Matthew Colless$^1$, R.P.\ Saglia$^2$,
David Burstein$^3$, Roger L.\ Davies$^4$, 
\newauthor  Robert K.\ McMahan Jr.$^5$ and Gary Wegner$^6$\\
$^1$ Research School of Astronomy \& Astrophysics, The Australian National
University, Weston Creek, Canberra, ACT 2611, Australia \\
$^2$ Institut f\"ur Astronomie und Astrophysik, Scheinerstra{\ss}e 1,
D-81679 Munich, Germany \\ 
$^3$ Dept of Physics and Astronomy, Arizona State University, Tempe, AZ
85287-1504, USA \\ 
$^4$ Dept of Physics, University of Durham, South Road, Durham, DH1 3LE, UK\\
$^5$ Dept of Physics and Astronomy, University of North Carolina,
CB\#3255 Phillips Hall, Chapel Hill, NC 27599-3255, USA \\
$^6$ Dept of Physics and Astronomy, Dartmouth College, Wilder Lab,
Hanover, NH 03755, USA}

\date{Accepted ---. Received ---; in original form ---.}

\newlength{\plotwidth}
\newlength{\fullwidth}
\newcommand{\etal}{\mbox{et~al.}}
\newcommand{\Mpc}{\mbox{\,h$_{50}^{-1}$\,Mpc}}

\newcommand{\degree}{\mbox{$^\circ$}}
\newcommand{\kpc}{\mbox{\,kpc}}
\newcommand{\kms}{\mbox{\,km\,s$^{-1}$}}
\newcommand{\mgb}{\mbox{Mg$b$}}

\newcommand{\mgtwo}{\mbox{Mg$_2$}}

\newcommand{\logsig}{\mbox{$\log\sigma$}}
\newcommand{\logRe}{\mbox{$\log R_e$}}
\newcommand{\SBe}{\mbox{$\langle S\!B_e \rangle$}}
\newcommand{\FP}{\mbox{$F\!P$}}
\newcommand{\dFP}{\mbox{$\delta F\!P$}}
\newcommand{\dZP}{\mbox{$\delta Z\!P$}}
\newcommand{\meanr}{\mbox{$\overline{r}$}}
\newcommand{\means}{\mbox{$\overline{s}$}}
\newcommand{\meanu}{\mbox{$\overline{u}$}}
\newcommand{\meanRe}{\mbox{$\overline{\log R_e}$}}
\newcommand{\meansig}{\mbox{$\overline{\log\sigma}$}}
\newcommand{\meanSBe}{\mbox{$\overline{\langle S\!B_e \rangle}$}}
\newcommand{\dr}{\mbox{$\delta r$}}
\newcommand{\ds}{\mbox{$\delta s$}}
\newcommand{\du}{\mbox{$\delta u$}}
\newcommand{\xx}{\mbox{$\vec{x}_i$}}
\newcommand{\xcut}{\mbox{$\vec{x}_{\rm cut}$}}
\newcommand{\DWi}{\mbox{$D_{Wi}$}}
\newcommand{\DoWj}{\mbox{$D^0_{Wj}$}}
\newcommand{\dWj}{\mbox{$\delta_{Wj}$}}
\newcommand{\Dni}{\mbox{$D_{ni}$}}
\newcommand{\dj}{\mbox{$\delta_j$}}
\newcommand{\DWcut}{\mbox{$D_{W{\rm cut}}$}}
\newcommand{\FPcut}{\mbox{$F\!P_{\rm cut}$}}
\newcommand{\scut}{\mbox{$s_{\rm cut}$}}
\newcommand{\Vpec}{\mbox{$V_{\rm pec}$}}
\newcommand{\dex}{\mbox{${\rm dex}$}}
\newcommand{\erf}{\mbox{${\rm erf}$}}
\newcommand{\head}[1]{\multicolumn{1}{c}{#1}}
\newcommand{\n}{\phantom{0}}
\newcommand{\noerror}{\phantom{$\pm$~0.00}}

\newcommand{\plotone}[1]
           {\centering \leavevmode \epsfxsize=\plotwidth \epsfbox{#1}}

\newcommand{\plotfull}[2]
           {\centering \leavevmode \epsfxsize=#1\fullwidth \epsfbox{#2}}

\begin{document}

\maketitle

\begin{abstract}

We present peculiar velocities for 84 clusters of galaxies in two large
volumes at distances between 6000 and 15000\kms\ in the directions of
Hercules-Corona Borealis and Perseus-Pisces-Cetus. These velocities are
based on Fundamental Plane (FP) distance estimates for early-type
galaxies in each cluster. We fit the FP using a maximum likelihood
algorithm which accounts for both selection effects and measurement
errors, and yields FP parameters with smaller bias and variance than
other fitting procedures. We obtain a best-fit FP with coefficients
consistent with the best existing determinations. We measure the bulk
motions of the sample volumes using the 50 clusters with the
best-determined peculiar velocities. We find the bulk motions in both
regions are small, and consistent with zero at about the 5\% level. The
EFAR results are in agreement with the small bulk motions found by Dale
\etal\ (1999) on similar scales, but are inconsistent with pure dipole
motions having the large amplitudes found by Lauer \& Postman (1994) and
Hudson \etal\ (1999). The alignment of the EFAR sample with the Lauer \&
Postman dipole produces a strong rejection of a large-amplitude bulk
motion in that direction, but the rejection of the Hudson \etal\ result
is less certain because their dipole lies at a large angle to the main
axis of the EFAR sample. We employ a window function covariance analysis
to make a detailed comparison of the EFAR peculiar velocities with the
predictions of standard cosmological models. We find the bulk motion of
our sample is consistent with most cosmological models that
approximately reproduce the shape and normalisation of the observed
galaxy power spectrum. We conclude that existing measurements of
large-scale bulk motions provide no significant evidence against
standard models for the formation of structure.

\end{abstract}

\begin{keywords}
galaxies: clustering --- galaxies: distances and redshifts --- galaxies:
elliptical and lenticular, cD --- galaxies: fundamental parameters ---
cosmology: large scale structure of universe
\end{keywords}

\section{INTRODUCTION}
\label{introduction}

This paper reports the main results of the EFAR project, which has
measured the peculiar motions of clusters of galaxies in two large
volumes at distances between 6000 and 15000\kms. The project was
initiated in the wake of early studies of peculiar motions which found
large-scale coherent flows over significant volumes of the local
universe (Dressler \etal\ 1987, Lynden-Bell \etal\ 1988). The primary
goal of the EFAR project was to test whether such large coherent motions
were to be found outside the local volume within 6000\kms. In the
following years, the velocity field within 6000\kms\ has been mapped by
several methods and in increasing detail so that today there is fair
agreement on the main features of the motions (recent results are given
in Giovanelli \etal\ 1998a,b, Dekel \etal\ 1999, Courteau \etal\ 2000,
Riess 2000, da Costa \etal\ 2000, Wegner \etal\ 2000 and Tonry \etal\
2000; see also the review by Dekel 2000). The bulk velocity within this
volume, and its convergence towards the frame of reference defined by
the Cosmic Microwave Background (CMB), appear to be consistent with the
broad range of currently-acceptable cosmological models (Dekel 2000,
Hudson \etal\ 2000).

However on larger scales there have been measurements of bulk motions
which, at face value, appear much greater than any acceptable model
would predict. The first of these was the measurement by Lauer \&
Postman (1994), using brightest cluster galaxies, of a bulk motion of
$\sim$700\kms\ towards $(l,b)$$\approx$(340\degree,$+50$\degree) for a
complete sample of Abell clusters out to 15000\kms. More recently, large
motions have been also be obtained for two smaller samples of clusters
at similar distances, for which peculiar velocities have been measured
by the more precise Fundamental Plane and Tully-Fisher estimators:
Hudson \etal\ (1999) find a motion of 630$\pm$200\kms\ towards
$(l,b)$=(260\degree,$-$1\degree) for the SMAC sample of 56 clusters at a
mean distance of $\sim$8000; Willick (1999) finds a motion of
720$\pm$280\kms\ towards $(l,b)$=(272\degree,$+$10\degree) for the LP10K
sample of 15 clusters at very similar distances. These two motions are
in good agreement with each other, but are nearly orthogonal to the
Lauer \& Postman motion (though similar in amplitude). In contrast, the
other extant study of peculiar motions on scales greater than 6000\kms,
the SCII Tully-Fisher survey of Dale \etal\ (1999a), finds a bulk flow
of less than 200\kms\ for a sample 52 Abell clusters with a mean
distance of $\sim$11000\kms. 

At these scales the robust prediction of most cosmological models is
that the bulk motion should be less than 300\kms\ with about 95\%
confidence. It is therefore of great interest to determine whether there
really are large coherent motions on scales of $\sim$10000\kms. The EFAR
peculiar motion survey probes the velocity field in the Hercules-Corona
Borealis and Perseus-Pisces-Cetus regions, which are almost
diametrically opposed on the sky and lie close to the axis of the bulk
motion found by Lauer \& Postman. With 84 clusters in these two regions
extending out to $\sim$15000\kms, the EFAR sample is well-suited to
testing for this particular bulk motion. Conversely, however, it is not
well-suited to testing for a bulk motion in the direction found for the
SMAC and LP10K samples, which is almost orthogonal to the major axis of
the EFAR sample. The main goal of this paper is to determine the
peculiar motions of the EFAR clusters and the consistency of the bulk
motion of the sample with both theory and other bulk motion measurements
on similar scales.

The structure of this paper is as follows: In \S2 we summarise the main
features of the data presented in Papers~I--IV of this series. In \S3 we
describe the maximum likelihood gaussian algorithm developed in
Paper~IV, which is used to determine the parameters of the Fundamental
Plane and obtain the distances and peculiar velocities for the clusters.
In \S4 we derive the best-fitting Fundamental Plane and critically
examine the random and systematic uncertainties in the fitted
parameters. In \S5 we derive the distances and peculiar velocities for
the clusters, testing them for possible systematic biases and comparing
them to the peculiar velocities obtained by other authors for the same
clusters. In \S6 we determine the bulk motion of the sample and compare
it, using a variety of methods, to the results of other studies and to
theoretical expectations. Our conclusions are given in \S7.

We use $H_0$=50\,km\,s$^{-1}$\,Mpc and $q_0$=0.5 unless otherwise
specified. All redshifts and peculiar velocities are given in the CMB
frame of reference.

\section{THE EFAR SAMPLE AND DATA}
\label{data}

Earlier papers in this series have described in detail the selection of
the clusters and galaxies in the EFAR sample (Wegner \etal\ 1996,
Paper~I), the spectroscopic data (Wegner \etal\ 1999, Paper~II; Colless
\etal\ 1999, Paper~V), the photoelectric and CCD photometry (Saglia
\etal\ 1997a, Paper~III; Colless \etal\ 1993) and the photometric
fitting procedures (Saglia \etal\ 1997b, Paper~IV; Saglia \etal\ 1993).
In this section we briefly summarise the main properties of the EFAR
database.

The clusters of galaxies in the EFAR sample are selected in two large,
distant (i.e.\ non-local) volumes: Hercules-Corona Borealis (HCB, 40
clusters, including Coma) and Perseus-Pisces-Cetus (PPC, 45 clusters).
These regions were chosen because they contain two of the richest
supercluster complexes (excluding the Great Attractor/Shapley
supercluster region) within 20000\kms. The clusters come from the ACO
catalogue (Abell \etal\ 1989), the list of Jackson (1982) and from scans
of Sky Survey prints by the authors. The nominal redshift range spanned
by the clusters is 6000\kms$<$$cz$$<$15000\kms. The distribution of the
EFAR clusters on the sky is shown in Figure~2 of Paper~I; their
distribution with respect to the major supercluster complexes is shown
in Figure~3 of Paper~I.

Galaxies were selected in each cluster for their apparently elliptical
morphology on Sky Survey prints, and for large apparent diameter. The
total sample includes 736 early-type galaxies in the 85 clusters.
Apparent diameters were measured visually for all early-type galaxies in
the cluster fields. The range in apparent visual diameter ($D_W$) is
from about 10~arcsec to over 60~arcsec. The sample selection function is
defined in terms of these visual diameters; in total, $D_W$ was measured
for 2185 early-type galaxies in the cluster fields. Selection functions
are determined separately for each cluster, and are approximated by
error functions in $\log D_W$. The mean value of the visual diameter is
$\langle \log D_W \rangle$=1.3 (i.e.\ 20~arcsec), and the dispersion in
$\log D_W$ is 0.3~dex (see Paper~I).

We obtained 1319 spectra for 714 of the galaxies in our sample,
measuring redshifts, velocity dispersions and the \mgb\ and \mgtwo\ Lick
linestrength indices (Paper~II). There are one or more repeat
observations for 45\% of the sample. The measurements from different
observing runs are calibrated to a common zeropoint or scale before
being combined, yielding a total of 706 redshifts, 676 velocity
dispersions, 676 \mgb\ linestrengths and 582 \mgtwo\ linestrengths. The
median estimated errors in the combined measurements are $\Delta
cz$=20\kms, $\Delta\sigma/\sigma$=9.1\%, $\Delta\mgb/\mgb$=7.2\% and
$\Delta\mgtwo$=0.015~mag. Comparison of our measurements with published
datasets shows no systematic errors in the redshifts or velocity
dispersions and only small zeropoint corrections to bring our
linestrengths onto the standard Lick system.

We have assigned sample galaxies to our target clusters (or to
fore/background clusters) by examining both the line-of-sight velocity
distributions and the projected distributions on the sky (Paper~II). The
velocity distributions were based on EFAR and ZCAT (Huchra \etal\ 1992)
redshifts for galaxies within 3\Mpc\ of the cluster centres. These
samples were also used to derive mean redshifts and velocity dispersions
for the clusters. The original selection was effective in choosing
cluster members, with 88\% of the galaxies with redshifts being members
of sample clusters and only 12\% lying in fore/background clusters or
the field. The median number of galaxies per cluster is 6.

We obtained R-band CCD photometry for 776 galaxies (Paper~III) and B and
R photoelectric photometry for 352 galaxies (Colless \etal\ 1993).
Comparison of the CCD and photoelectric photometry shows that we have
achieved a common zero-point to better than 1\%, and a photometric
precision of better than 0.03~mag per measurement. Circularised galaxy
light profiles were fitted with seeing-convolved models having both an
$R^{1/4}$ bulge and an exponential disk (Paper~IV). We find that only
14\% of the galaxies in our sample are well fitted by pure $R^{1/4}$
bulges and only about 1\% by pure exponential disks, with most of the
sample requiring both components to achieve a good fit. From these fits
we derive total R-band magnitudes $m_T$, $D_n$ diameters (at
20.5~mag~arcsec$^{-2}$), half-luminosity radii $R_e$, and average
effective surface brightnesses \SBe, for 762 galaxies. The total R
magnitudes span the range $m_T$=10.6--16.0 ($\langle m_T
\rangle$=13.85), the diameters span $D_n$=4.8--90~arcsec ($\langle D_n
\rangle$=20~arcsec), and the effective radii $R_e$ span 1.6--71~arcsec
($\langle R_e \rangle$=6.9~arcsec). For 90\% of our sample the precision
of the total magnitudes and half-luminosity radii is better than
0.15~mag and 25\% respectively. The errors on the combined quantity
$\FP=\log R_e-0.3\SBe$ which enters the Fundamental Plane equation are
always smaller than 0.03~dex. The visual selection diameters $D_W$
correlate well with the $D_n$ diameters (or, equivalently, with the
Fundamental Plane quantity \FP).

The morphological type classifications of the galaxies, based on all the
information available to us, reveal that 31\% of the sample objects,
visually selected from photographic images to be of early type, are in
fact spiral or barred galaxies. The 69\% of galaxies classified as
early-type can be subdivided into 8\% cD galaxies, 12\% E galaxies
(best-fit by a pure $R^{1/4}$ profile), and 48\% E/S0 galaxies (best-fit
by a disk plus bulge model).

All the EFAR project data is available from NASA's Astrophysical Data
Centre (http://adc.gsfc.nasa.gov) and the Centre de Donn\'{e}es
astronomiques de Strasbourg (http//:cdsweb.u-strasbg.fr). A summary
table with all the main parameters for every galaxy in the EFAR sample
is available at these locations as J/MNRAS/\textit{vol}/\textit{page}.
The contents of the summary table are described here in
Table~\ref{tab:efardata}.

\begin{table*}
\centering
\caption{Description of EFAR summary data table.}
\label{tab:efardata}
\vspace{3mm}
\begin{tabular}{rll}
Column 	& Code 		& Description [units] \vspace{6pt} \\  
 1 	& GIN  		& Galaxy Identification Number \\
 2 	& CID  		& Cluster Identification (see Paper~I)\\
 3 	& CAN  		& Cluster Assignment Number (see Paper~II) \\
 4 	& Clus 		& Cluster Name (corresponds to CID) \\
 5 	& Gal  		& Galaxy Name \\
 6 	& RAh  		& Right Ascension (J2000) [hours] \\
 7 	& RAm  		& Right Ascension (J2000) [minutes] \\
 8 	& RAs  		& Right Ascension (J2000) [seconds] \\
 9 	& Decd 		& Declination (J2000) [degrees] \\
10 	& Decm 		& Declination (J2000) [minutes] \\
11 	& Decs 		& Declination (J2000) [seconds] \\
12 	& $l$  		& Galactic longitude [degrees] \\
13 	& $b$  		& Galactic latitude [degrees] \\
14 	& Type 		& Morphological type \\
15 	& $D_n$ 	& Diameter enclosing a mean R-band SB of 20.5 mag arcsec$^{-2}$ [arcsec] \\
16 	& $\delta D_n$ 	& Error in $D_n$ [arcsec] \\
17 	& $D_n(20)$	& Diameter enclosing a mean R-band SB of 20.0 mag arcsec$^{-2}$ [arcsec] \\
18 	& $D_n(19.25)$ 	& Diameter enclosing a mean R-band SB of 19.25 mag arcsec$^{-2}$ [arcsec] \\
19 	& $R_e$		& Half-luminosity radius in the R-band [arcsec] \\
20 	& $R_e$(kpc) 	& Half-luminosity radius in the R-band [kpc, $H_0$=50, $q_0$=0.5] \\
21 	& SB$_e$	& R-band surface brightness at $R_e$ [mag arcsec$^{-2}$] \\
22 	& $\delta$SB$_e$& Photometric zero-point error on SB$_e$ [mag arcsec$^{-2}$] \\
23 	& \SBe\		& Mean R-band surface brightness inside $R_e$ [mag arcsec$^{-2}$] \\
24 	& $\delta$\SBe\ & Photometric zero-point error on \SBe\ [mag arcsec$^{-2}$] \\
25 	& $m_T$		& Total apparent R magnitude [mag] \\
26 	& $\delta m_T$  & Photometric zero-point error on $m_T$ [mag] \\
27 	& $R_{eB}$      & Bulge half-luminosity radius in the R-band [arcsec] \\
28 	& SB$_{eB}$	& Bulge R-band surface brightness at $R_{eB}$ [mag arcsec$^{-2}$] \\
29 	& $h$		& Disk scale-length in the R-band [arcsec] \\
30 	& $\mu_0$	& Disk central surface brightness in the R-band [mag arcsec$^{-2}$] \\
31 	& $h/R_{eB}$    & Ratio of bulge half-luminosity radius to disk scale-length \\
32 	& $D/B$         & Disk-to-bulge ratio (ratio of luminosity in disk to luminosity in bulge) \\
33 	& Fit        	& Type of fit (B=bulge, D=disk, BD=bulge+disk; other, see Paper~III) \\
34 	& $P$		& Quality of the photometric zero-point (P=0 good, P=1 bad; see Paper~III) \\
35 	& $Q$		& Global quality of the photometric fit (1=best, 2=fair, 3=poor; see Paper~III)\\
36 	& $B$$-$$R$	& $B$$-$$R$ colour [mag] \\
37 	& $\delta(B$$-$$R)$ & Error in $B$$-$$R$ colour [mag] \\
38 	& $\langle\epsilon(R_e)\rangle$ & Mean ellipticity inside $R_e$ \\
39 	& $A_R$		& Reddening in the R-band \\
40 	& $cz_{cl}$	& Cluster mean redshift [\kms] \\
41 	& $\delta cz_{cl}$ & Error in $cz_{cl}$ [\kms] \\
42 	& $cz$		& Galaxy redshift [\kms] \\
43 	& $\delta cz$	& Error in $cz$ [\kms] \\
44 	& $\sigma$      & Central velocity dispersion of galaxy [\kms] \\
45 	& $\delta\sigma$& Error in $\sigma$ [\kms] \\
46 	& \mgb\		& \mgb\ Lick linestrength index [\AA] \\
47 	& $\delta$\mgb\	& Error in \mgb\ [\AA] \\
48 	& \mgtwo\	& \mgtwo\ Lick linestrength index [mag] \\
49 	& $\delta$\mgtwo& Error in \mgtwo\ [mag] \\
50 	& $Q_s$		& Spectral quality (A=best, ..., E=worst; see Paper~II) \\
51 	& a/e		& Absorption/emission flag \\
52 	& $\log D_W$	& Logarithm of the $D_W$ diameter [arcsec] \\
53 	& $S(D_W)$	& Selection probability computed using $D_W$ (see \S\ref{method}) \\
54 	& $\log D_W(D_n)$ & Logarithm of $D_W$ computed from $D_n$ (see \S\ref{method}) [arcsec] \\
55 	& $S(D_W(D_n))$ & Selection probability computed from $D_W(D_n)$ (see \S\ref{method}) \\
\end{tabular}
\\ \vspace{5mm}
\parbox{0.75\textwidth}{The summary table is available as
J/MNRAS/\textit{vol}/\textit{page} from NASA's Astrophysical Data Centre
(ADC, http://adc.gsfc.nasa.gov) and from the Centre de Donn\'{e}es
astronomiques de Strasbourg (CDS, http://cdsweb.u-strasbg.fr).}
\end{table*}

\section{MAXIMUM LIKELIHOOD GAUSSIAN METHOD}
\label{method}

We use a maximum likelihood gaussian algorithm for fitting the FP and
determining relative distances and peculiar velocities. This algorithm,
which is described in detail in Paper~VI, was developed in order to deal
with the general deficiencies of previous approaches and with some
specific problems posed by the selection effects and measurement errors
in the EFAR sample. Previous methods for fitting the FP using forms of
multi-linear regression have not fully dealt with the intrinsic
distribution of galaxies in size, velocity dispersion and surface
brightness, nor with the simultaneous presence of measurement errors
with a wide range of values in all of these quantities. The maximum
likelihood gaussian algorithm properly accounts for all these factors,
and also handles complex selection effects in a straightforward way. The
selection criteria for the EFAR sample are well-determined, and involve
both the original sample selection based on galaxy size and {\it a
posteriori} limits imposed on both galaxy size and velocity dispersion.
A specific problem with the data is that the velocity dispersion
measurements include a significant fraction of cases where the errors,
though themselves well-determined, are large relative to the actual
value. There is also the fact that the numbers of galaxies observed per
cluster are relatively small, so a method is required which is both
efficient and robust against outliers (either unusual galaxies or errors
in the data). The extensive simulations carried out in Paper~VI
demonstrate that the maximum likelihood gaussian method is superior to
any of the classical linear regression approaches, minimising both the
bias and the variance of the fitted parameters, and performing well in
recovering the FP parameters and peculiar velocities when presented with
simulations of the EFAR dataset.

The maximum likelihood gaussian method assumes that each galaxy $i$ is
drawn from an underlying gaussian distribution in the three-dimensional
FP-space $(r\equiv\logRe, s\equiv\logsig, u\equiv\SBe)$. We also assume
that this underlying distribution is the same for each cluster $j$,
apart from a shift $\dj$ in the distance-dependent quantity $r$
resulting from the cluster's peculiar motion. We want to determine the
mean values $(\meanr, \means, \meanu)$ and the variance matrix $V$ which
characterise the galaxy distribution, along with the shifts $\dj$ due to
the clusters' peculiar velocities. We do this by maximising the
likelihood of the observed galaxy data over these parameters, while
properly accounting for all the various selection effects.

The probability density for the $i$th galaxy, in terms of
$\xx=(r_i-\meanr+\dj,s_i-\means,u_i-\meanu)$, is
\begin{equation}
\label{eqn:prob}
P(\xx)=\frac{\exp\left[-\frac{1}{2}\xx^T(V+E_i)^{-1}\xx\right]}
            {(2\pi)^{3/2}|V+E_i|^{1/2} f_i} \Theta(A\xx-\xcut) ~,
\end{equation}
where $V$ is the variance matrix of the underlying distribution and
$E_i$ is the error matrix of the measured quantities. The errors are
convolved with the intrinsic dispersion of the galaxy distribution to
give the observed distribution of the data. The exclusion function
$\Theta(\vec{y})=\prod\theta(y)$, where $\theta(y)=1$ if $y\ge0$ and 0
otherwise, accounts for parts of FP-space that are inaccessible because
of selection effects. For simplicity, we assume that these selection
effects apply to linear combinations of the variables, described by the
matrix $A$. The normalisation factor $f_i$ is such that $\int
P(\vec{x})\,d^3x=1$, and accounts for the selection effects described
the exclusion function $\Theta$. The likelihood of the observed sample
is
\begin{equation}
\label{eqn:like}
{\cal L} = \prod_i P(\xx)^{1/S(\xx)} ~,
\end{equation}
where $S(\xx)$ is the selection function giving the probability of
selecting a galaxy with parameters \xx. In order to correct for the
selection function, each object in the sample is included in the
likelihood product as if it were $1/S(\xx)$ objects.

The error matrix can be computed from the estimated errors
$(\dr_i,\ds_i,\dFP_i,\dZP_i)$, where \dFP\ is the error in the combined
quantity $\FP=r-\alpha u$ (with $\alpha\approx0.3$) and \dZP\ is the
photometric zeropoint error. In terms of these quantities, the error
matrix for galaxy $i$ is
\begin{equation}
\label{eqn:error}
E_i = \left( 
\begin{array}{ccc}
\dr_i^2 & 0 & \frac{(1+\alpha^2)\dr_i^2-\dFP_i^2}{\alpha(1+\alpha^2)} \\
0 & \ds_i^2  & 0 \\
\frac{(1+\alpha^2)\dr_i^2-\dFP_i^2}{\alpha(1+\alpha^2)} & 0 & \du_i^2 \\
\end{array}
\right) ~.
\end{equation}
Note that $\ds_i$ combines the estimated random errors in the velocity
dispersion measurements and the correlated errors between galaxies
introduced by the uncertainties in calibrating dispersions obtained in
different observing runs to a common system (see Paper~II). Likewise,
$\du_i$ is given by the quadrature sum of the error on the effective
surface brightness (from the fit to the galaxy's surface brightness
distribution) and the photometric zeropoint error (see Paper~III).
\begin{equation}
\label{eqn:du2}
\du_i^2 = \frac{(\alpha^2-1)\dFP_i^2+(1+\alpha^2)\dr_i^2}
               {\alpha^2(1+\alpha^2)} + \dZP_i^2
\end{equation}

For the EFAR sample, the selection function depends on galaxy diameter
and varies from cluster to cluster (see Paper~I). For galaxy $i$, a
member of cluster $j$, the selection probability is
\begin{equation}
\label{eqn:selfn}
S_i = S(\log\DWi) = \frac{1}{2} \left( 1+\erf\left[
                    \frac{\log\DWi-\log\DoWj}{\dWj} \right] \right) ~.
\end{equation}
The selection function for cluster $j$ is characterised by \DoWj, the
size at which the selection probability is 0.5, and by \dWj, the width
of the cutoff in the selection function. For early-type galaxies, the
visually estimated diameter \DWi\ correlates with the measured diameter
\Dni\ according to the relation $\log\Dni=0.80\log\DWi+0.26$, with a
scatter of 0.09~dex in $\log\Dni$ (see Paper~III). Because the visual
diameters given in Paper~I are individually uncertain, in computing
selection probabilities we actually use an estimate of \DWi\ obtained by
inverting this relation and inserting the accurately measured value of
\Dni.

In order to avoid biasing the FP fits and the estimated peculiar
velocities, it would be desirable to sample the same part of the FP
galaxy distribution in all clusters. However, because the clusters are
at different redshifts, the approximately constant apparent diameter
selection limit corresponds to actual diameter selection limits \DoWj\
for the clusters that vary by about a factor of 2--3 (the approximate
range of cluster redshifts; see Paper~1). We can limit this
redshift-dependent sampling bias by excluding the smaller galaxies,
which are only sampled in the nearer clusters. Guided by the simulations
of Paper~VI, we choose a selection limit \DWcut=12.6\kpc. This choice
balances the reduced bias of a higher \DWcut\ against the larger sample
size of a lower \DWcut\ (95\% of galaxies in the EFAR sample have
\DWi$\ge$12.6\kpc). Because of the good correlation between \DWi\ and
the combined quantity $\FP=r-0.3u$ (see Paper~III), this cut in \DWi\
corresponds to an approximate selection limit $\FPcut \approx
0.78\log\DWcut-6.14 \approx -5.28$.

Another selection limit is due to the difficulty of measuring velocity
dispersions smaller than a spectrograph's instrumental resolution. For
the spectrograph setups we used, only velocity dispersions greater than
about 100\kms\ could be reliably measured (see Paper~II). We therefore
impose a limit \scut=2, excluding galaxies with $\sigma$$<$100\kms. The
overall exclusion function for the EFAR sample is thus
$\Theta=\theta(s-\scut)\theta(\FP-\FPcut)$.

The mean of the distribution, (\meanr, \means, \meanu), the variance
matrix $V$, and the shifts \dj, are all determined by minimising
$-\ln\cal L$, which for the EFAR sample is given by
\begin{equation}
\label{eqn:loglike}
-\ln {\cal L} = \!\!\!\!\!\!
\sum^{s>s_{\rm cut}}_{\mbox{\tiny \FP}>\mbox{\tiny \FP}_{\rm cut}}\!\!\!\!\!\!\!
S_i^{-1}\!\!\left[0.5\xx^T(V\!+\!E_i)\xx\!+\!0.5\ln|V\!+\!E_i|\!+\!\ln\!f_i\right] 
\end{equation}
(where the constant term $1.5\ln(2\pi)$ has been dropped).The
normalisation $f_i$ is obtained by integrating the gaussian distribution
over the accessible volume defined by $s$$>$\scut\ and \FP$>$\FPcut. The
minimisation is performed using the simplex algorithm (Press \etal,
1986). 

The FP is defined as the plane $r=as+bu+c$ that passes through (\meanr,
\means, \meanu) and whose normal is the eigenvector of $V$ with the
smallest eigenvalue. For convenience, we define the second axis of the
galaxy distribution to be the unit vector within the FP that has zero
coefficient for $s$ (in fact, this turns out to be a reasonable
approximation to one of the remaining eigenvectors of $V$). The three
unit vectors giving the axes of the galaxy distribution can then be
written in terms of the FP constants as
\begin{eqnarray}
\label{eqn:evecs}
\hat{v}_1 & = & \hat{r}-a\hat{s}-b\hat{u} \nonumber \\
\hat{v}_2 & = & \hat{r}+\hat{u}/b \\
\hat{v}_3 & = & -\hat{r}/b-(1+b^2)\hat{s}/(ab)+\hat{u} ~, \nonumber
\end{eqnarray}
where $\hat{r}$, $\hat{s}$, and $\hat{u}$ are the unit vectors in the
directions of the FP-space axes. The eigenvalues of $V$ give the
dispersions $\sigma_1$, $\sigma_2$ and $\sigma_3$ of the galaxy
distribution in the directions of the eigenvectors; the smallest
eigenvalue, $\sigma_1$, is the intrinsic dispersion of the galaxies
about the FP.

The final step of the process is to recover each cluster's distance and
peculiar velocity. The mean galaxy size, $\meanr\equiv\meanRe$, provides
a standard scale which we can use to determine relative distances and
peculiar velocities. The offset \dj\ between the true mean galaxy size,
\meanRe, and the mean galaxy size observed for cluster $j$,
\meanRe$-$\dj, is a measure of the ratio of the true angular diameter
distance of a cluster, $D_j$, to the angular diameter distance
corresponding to its redshift, $D(z_j)$:
\begin{equation}
\label{eqn:dist}
\frac{D_j}{D(z_j)} = \frac{\dex(\meanRe)}{\dex(\meanRe-\dj)} = 10^{\dj} ~.
\end{equation}
The relation between angular diameter distance and redshift (Weinberg
1972) is given by
\begin{equation}
\label{eqn:daz}
D(z)=\frac{cz}{H_0(1+z)^2}\frac{1+z+\sqrt{1+2q_0z}}{1+q_0z+\sqrt{1+2q_0z}} ~.
\end{equation}
We assume $H_0$=50\,km\,s$^{-1}$\,Mpc, $q_0$=0.5, and compute all
redshifts and peculiar velocities in the CMB frame of reference. The
peculiar velocity of the cluster, $V_j$, is then obtained as
\begin{equation}
\label{eqn:pec}
V_j = \frac{cz_j-cz(D_j)}{1+z(D_j)},
\end{equation}
where $z(D_j)$ is the redshift corresponding to the true distance $D_j$
through the inverse of equation~(\ref{eqn:daz}). Note that we are not
using the low-redshift approximation $V = cz - H_0 D = cz(1 -
10^{\delta})$, which leads to small but systematic errors in the
peculiar velocities (e.g., at $cz$=15000\kms, the approximation leads to
a systematic peculiar velocity error of about $-$4\%).

These distances and peculiar velocities are relative, because the
standard scale is determined by assuming that the distance (or,
equivalently, peculiar velocity) of some standard cluster (or set of
clusters) is known. Distances and peculiar velocities are therefore in
fact relative to the true distance and peculiar velocity of this
standard.

\section{THE FUNDAMENTAL PLANE}
\label{fp}

\subsection{Best-fit solution and random errors}

We determine the parameters of the Fundamental Plane and the cluster
peculiar velocities in a two-step process. We first fit the Fundamental
Plane using only those clusters with 6 or more suitable galaxies having
reliable dispersions, effective radii and mean surface brightnesses (the
criteria are given below). We exclude clusters with fewer members
because the simulations of Paper~VI show that including less
well-sampled clusters increases the variance on the FP parameters. We
then determine peculiar velocities for {\em all} the clusters in a
second step, where we fix the FP parameters at the values determined in
the first step. This procedure results in more accurate and precise
peculiar velocities than a simultaneous global solution for the FP
parameters and the peculiar velocities.

In order to be included in the fit a galaxy had to satisfy the following
criteria: (1)~good quality photometric fit (Q=1 or Q=2; see Paper~III);
(2)~$\sigma \ge 100$\kms\ and $\delta\log\sigma\le0.5$ (see Paper~II);
(3)~a selection diameter $D_W\ge12.6$\kpc\ and a selection probability
$\ge$0.1. The first criterion excludes galaxies with unreliable
structural and photometric parameters (see Paper~III); the second
excludes galaxies with dispersions less that the typical instrumental
resolution or which have very large uncertainties; the third ensures
that the clusters have uniform selection criteria and that no individual
galaxy enters with a very high weight. No galaxy is excluded on the
basis of its morphological type. There were 31 clusters in the sample
with 6 or more galaxies satisfying these criteria.

As well as these {\it a priori} criteria, we also rejected a further 8
galaxies on the basis that they lie outside {\em both} the 3-$\sigma$
ellipse of the galaxy distribution in the \FP--\logsig\ plane when the
FP fit is obtained using all the galaxies in these 31 clusters meeting
the selection criteria (including themselves), {\em and} outside the
5-$\sigma$ ellipse of the galaxy distribution when the FP fit is
obtained excluding them. These galaxies are listed in
Table~\ref{tab:badfp}, which gives their galaxy ID number (GIN), their
cluster assignment number (CAN), their EFAR name, their morphological
type and, where appropriate, their NGC/IC numbers. The reasons why these
8 galaxies are poorly fitted by the FP distribution that satisfactorily
represents the other 255 galaxies fulfilling the selection criteria are
not apparent. Although three are spirals, the other five include two
ellipticals, two E/S0s and a cD. Three are members of A2151, including
the cD NGC~6041. Two of these galaxies (GINs 45 and 370) are in clusters
with data for 6 members; these two clusters (A160 and A1983) therefore
drop out of the sample of clusters to which we fit the FP. Also listed
in Table~\ref{tab:badfp} are another 3 galaxies in clusters with fewer
than 6 members that are excluded from further analysis because they lie
outside the 5-$\sigma$ ellipse of the best-fitting galaxy distribution.

\begin{table}
\centering
\caption{Galaxies excluded from the Fundamental Plane fits.}
\label{tab:badfp}
\begin{tabular}{rrlll}
GIN & CAN & Name      & Type & NGC/IC \vspace{6pt} \\  
\multicolumn{5}{l}{(i) Galaxies in clusters with $\ge$6 members} \\
 45 &  7  & A160 C    & E/S0 &        \\  
167 & 21  & A400 H    & E/S0 &        \\  
370 & 43  & A1983 2   & S    &        \\  
396 & 46  & J16-W B   & S    &        \\  
456 & 53  & A2147 D   & E    &        \\  
495 & 58  & A2151 A   & cD   & N6041  \\  
500 & 58  & A2151 F   & S    & I1185  \\  
501 & 58  & A2151 G   & E    & I1193  \vspace{6pt} \\  
\multicolumn{5}{l}{(ii) Galaxies in clusters with $<$6 members} \\
355 & 42  & J14-1 D   & S    &        \\   
489 & 57  & J18 C     & E    &        \\   
552 & 63  & A2162-S G & E/S0 &        \\   
\end{tabular}
\end{table}

The final sample of 29 clusters used to fit the FP parameters is listed
in Table~\ref{tab:fpclus}, which gives the cluster assignment number
(CAN), the cluster name, the mean heliocentric redshift and the number
of galaxies that enter the FP fit. Of these 29 clusters, 12 are in HCB
and 17 in PPC. They span the redshift range 6942\kms\ (Coma) to
20400\kms\ (A419), though most are in the range 9000--15000\kms.
However, they have similar selection diameters $D^0_W$, with minimum
values of the $D_W$ diameter in the range $\log D_W$(kpc)=1.0--1.3. The
Coma cluster sample is supplemented with the data of M\"{u}ller (1997;
see also M\"{u}ller \etal\ 1998, 1999), which were obtained using
essentially the same methodology. M\"{u}ller's photometric data have
been adjusted by adding 0.04~mag in order to bring them into agreement
with the EFAR data for galaxies in common.

\begin{table}
\centering
\caption{The Fundamental Plane cluster sample.}
\label{tab:fpclus}
\begin{tabular}{rlrr}
CAN & Name    & cz (km\,s$^{-1}$) & $N_{\FP}$ \vspace{6pt} \\  
  1 & A76     & 11888 &   6  \\  
  3 & A119    & 13280 &   6  \\  
 10 & J30     & 15546 &   6  \\  
 13 & A260    & 10944 &   8  \\  
 16 & J8      &  9376 &   8  \\  
 17 & A376    & 14355 &   7  \\  
 20 & A397    &  9663 &   8  \\  
 21 & A400    &  7253 &   6  \\  
 23 & A419    & 20400 &   6  \\  
 24 & A496    &  9854 &   6  \\  
 25 & J34     & 11021 &   8  \\  
 34 & A533    & 14488 &   6  \\  
 35 & A548-1  & 11866 &  19  \\  
 36 & A548-2  & 12732 &   6  \\  
 39 & J13     &  8832 &   8  \\  
 46 & J16W    & 11321 &   7  \\  
 48 & A2040   & 13455 &   6  \\  
 50 & A2063   & 10548 &   9  \\  
 53 & A2147   & 10675 &  10  \\  
 58 & A2151   & 11106 &  10  \\  
 59 & J19     & 12693 &   7  \\  
 65 & A2197   &  9137 &   9  \\  
 66 & A2199   &  9014 &   9  \\  
 68 & A2247   & 11547 &   7  \\  
 70 & J22     & 10396 &  10  \\  
 80 & A2593-N & 12399 &  18  \\  
 82 & A2634   &  9573 &  12  \\  
 83 & A2657   & 12252 &   7  \\  
 90 & Coma    &  6942 &  20  \\  
\end{tabular}
\end{table}

In fitting the FP we assume $H_0$=50\,km\,s$^{-1}$\,Mpc$^{-1}$ and
$q_0$=0.5. We fix the zeropoint of the FP by forcing the mean of the FP
shifts of the 29 clusters to be zero---i.e.\ we fix \meanRe\ by
requiring $\sum\dj$=0. This results in a peculiar velocity for Coma of
only $-$29\kms, so our FP zeropoint is essentially identical to that
obtained by setting the peculiar velocity of Coma to be zero, as is
often done. The effective radii and mean surface brightnesses used were
the total $R_e$ and \SBe\ (rather than the bulge-only $R_{eB}$ and
$\langle S\!B_{eB} \rangle$) given in Paper~III. In applying absorption
corrections (taken to be $2.6E_{\rm B-V}/4.0$) we have adopted the mean
of the absorption corrections derived from Burstein \& Heiles (1982,
1984; BH) and Schlegel \etal\ (1998, SFD; with $E_{\rm B-V}$ offset by
$-$0.02~mag, the mean offset from BH given by SFD). The above
assumptions and cluster/galaxy selection criteria yield our best fit to
the FP. This best fit is given as case~1 in Table~\ref{tab:fpfits},
which lists the number of clusters and galaxies in the fit, the FP
coefficients $a$, $b$ and $c$, and the means and dispersions describing
the galaxy distribution: \meanRe, \meansig, \meanSBe, $\sigma_1$,
$\sigma_2$ and $\sigma_3$. The table also explores the effects of the
various assumptions and selection criteria, giving the FP fits obtained
for a wide range of alternative cases.

\begin{table*}
\centering
\caption{The parameters of the Fundamental Plane derived for various cases.}
\label{tab:fpfits}
\begin{tabular}{rrrcccccccccl}
Case & $N_{\rm cl}$ & $N_{\rm gal}$ & $a$ & $b$ & $c$ & \meanRe & \meansig &\meanSBe & $\sigma_1$ & $\sigma_2$ & $\sigma_3$ & Notes \vspace{6pt} \\   
 1 & 29 & 255 & 1.223 & 0.3358 & $-$8.664 & 0.7704 & 2.304 & 19.71 & 0.0638 & 1.995 & 0.6103 & standard fit  \\
 2 & 29 & 271 & 1.286 & 0.3439 & $-$8.975 & 0.7621 & 2.298 & 19.72 & 0.0688 & 1.958 & 0.6201 & includes $Q$=3 photometry \\
 3 & 29 & 261 & 1.201 & 0.3265 & $-$8.430 & 0.7840 & 2.306 & 19.74 & 0.0671 & 2.057 & 0.6202 & includes rejected galaxies \\
 4 & 29 & 255 & 1.232 & 0.3373 & $-$8.721 & 0.7686 & 2.300 & 19.73 & 0.0642 & 1.992 & 0.6138 & uses BH absorption corrections \\
 5 & 29 & 255 & 1.183 & 0.3292 & $-$8.422 & 0.7961 & 2.315 & 19.69 & 0.0632 & 2.019 & 0.5901 & uses SFD absorption corrections \\
 6 & 29 & 255 & 1.220 & 0.3349 & $-$8.639 & 0.7739 & 2.306 & 19.71 & 0.0638 & 1.996 & 0.6043 & uses 0.64SFD+0.36BH corrections \\
 7 & 29 & 235 & 1.235 & 0.3357 & $-$8.690 & 0.7750 & 2.300 & 19.74 & 0.0642 & 2.014 & 0.6161 & excludes $\delta\logsig$$>$0.1 \\
 8 & 29 & 255 & 1.082 & 0.3221 & $-$8.062 & 0.8159 & 2.329 & 19.74 & 0.0612 & 2.057 & 0.5394 & no $D_W$ cut is applied \\
 9 & 29 & 275 & 1.132 & 0.3224 & $-$8.184 & 0.7773 & 2.297 & 19.73 & 0.0675 & 2.122 & 0.6827 & \DWcut=6.3\kpc \\
10 & 29 & 244 & 1.300 & 0.3388 & $-$8.906 & 0.7446 & 2.292 & 19.69 & 0.0637 & 2.040 & 0.6220 & \DWcut=14.1\kpc \\
11 & 29 & 222 & 1.247 & 0.3303 & $-$8.607 & 0.7369 & 2.265 & 19.74 & 0.0696 & 2.001 & 0.7176 & \DWcut=15.9\kpc \\
12 & 29 & 255 & 1.077 & 0.3014 & $-$7.665 & 0.7511 & 2.310 & 19.66 & 0.0458 & 1.575 & 0.5286 & excludes galaxies with $\ln{\cal L}$$<$0 \\
13 & 29 & 256 & 1.207 & 0.3359 & $-$8.625 & 0.7725 & 2.299 & 19.72 & 0.0625 & 1.981 & 0.6326 & uses $q_0$=0 \\
14 & 29 & 258 & 1.204 & 0.3472 & $-$8.846 & 0.6745 & 2.206 & 19.77 & 0.0634 & 1.965 & 0.8923 & uses galaxies with $S_i$$>$0.01 \\
15 & 29 & 241 & 1.080 & 0.3239 & $-$8.081 & 0.8099 & 2.315 & 19.73 & 0.0575 & 2.135 & 0.6276 & uses galaxies with $S_i$$>$0.2 \\
16 & 29 & 255 & 1.221 & 0.3309 & $-$8.553 & 0.8302 & 2.331 & 19.75 & 0.0646 & 2.108 & 0.5981 & uses no selection weighting \\
17 & 29 & 255 & 1.223 & 0.3345 & $-$8.629 & 0.7895 & 2.305 & 19.73 & 0.0637 & 2.001 & 0.6091 & mean FP shift set to +0.01 \\
18 & 29 & 255 & 1.215 & 0.3342 & $-$8.628 & 0.7700 & 2.307 & 19.73 & 0.0636 & 1.999 & 0.6056 & mean FP shift set to $-$0.01 \\
19 & 29 & 255 & 1.227 & 0.3359 & $-$8.648 & 0.7990 & 2.302 & 19.71 & 0.0639 & 1.991 & 0.6136 & mean FP shift set to +0.03 \\
20 & 29 & 255 & 1.226 & 0.3359 & $-$8.704 & 0.7418 & 2.303 & 19.71 & 0.0639 & 1.992 & 0.6104 & mean FP shift set to $-$0.03 \\
21 & 29 & 255 & 1.227 & 0.3374 & $-$8.707 & 0.7735 & 2.304 & 19.72 & 0.0639 & 2.249 & 0.4334 & also fit third axis of FP \\
22 & 29 & 255 & 1.247 & 0.3341 & $-$8.694 & 0.7721 & 2.301 & 19.75 & 0.0564 & 2.192 & 0.6402 & uses uniform errors for all galaxies \\
23 & 66 & 397 & 1.206 & 0.3274 & $-$8.452 & 0.8021 & 2.307 & 19.77 & 0.0634 & 2.051 & 0.6619 & uses clusters with $N_{\rm gal}$$\ge$3 \\
24 & 52 & 355 & 1.208 & 0.3272 & $-$8.460 & 0.7969 & 2.306 & 19.78 & 0.0644 & 2.035 & 0.6564 & uses clusters with $N_{\rm gal}$$\ge$4 \\
25 & 39 & 304 & 1.244 & 0.3265 & $-$8.531 & 0.7927 & 2.306 & 19.77 & 0.0651 & 2.084 & 0.6139 & uses clusters with $N_{\rm gal}$$\ge$5 \\
26 & 31 & 265 & 1.228 & 0.3329 & $-$8.616 & 0.7839 & 2.305 & 19.74 & 0.0643 & 1.994 & 0.6060 & uses clusters with $N_{\rm gal}$$\ge$6 \\
27 & 16 & 173 & 1.109 & 0.3432 & $-$8.525 & 0.7487 & 2.299 & 19.59 & 0.0661 & 1.765 & 0.5890 & uses clusters with $N_{\rm gal}$$\ge$8 \\
28 &  7 &  99 & 0.992 & 0.3526 & $-$8.425 & 0.7652 & 2.326 & 19.52 & 0.0544 & 1.864 & 0.5564 & uses clusters with $N_{\rm gal}$$\ge$10 \\
29 & 29 & 222 & 1.330 & 0.3351 & $-$8.904 & 0.7776 & 2.320 & 19.68 & 0.0668 & 2.009 & 0.5470 & excludes spirals \\
30 & 66 & 348 & 1.284 & 0.3327 & $-$8.737 & 0.8186 & 2.330 & 19.73 & 0.0660 & 1.966 & 0.5488 & excludes spirals; $N_{\rm gal}$$\ge$3 \\
31 & 52 & 310 & 1.293 & 0.3323 & $-$8.756 & 0.8097 & 2.330 & 19.72 & 0.0675 & 1.947 & 0.5404 & excludes spirals; $N_{\rm gal}$$\ge$4 \\
32 & 39 & 267 & 1.352 & 0.3291 & $-$8.835 & 0.7966 & 2.323 & 19.72 & 0.0678 & 2.006 & 0.5452 & excludes spirals; $N_{\rm gal}$$\ge$5 \\
33 & 31 & 232 & 1.333 & 0.3300 & $-$8.804 & 0.8020 & 2.322 & 19.73 & 0.0673 & 2.027 & 0.5414 & excludes spirals; $N_{\rm gal}$$\ge$6 \\
34 & 29 & 223 & 1.147 & 0.3198 & $-$8.174 & 0.7558 & 2.300 & 19.68 & 0.0646 & 1.861 & 0.6102 & excludes cD galaxies \\
35 & 29 & 199 & 1.241 & 0.3125 & $-$8.250 & 0.7568 & 2.319 & 19.62 & 0.0672 & 1.831 & 0.5426 & excludes spirals and cDs \\
\end{tabular}
\end{table*}

\begin{figure}
\plotone{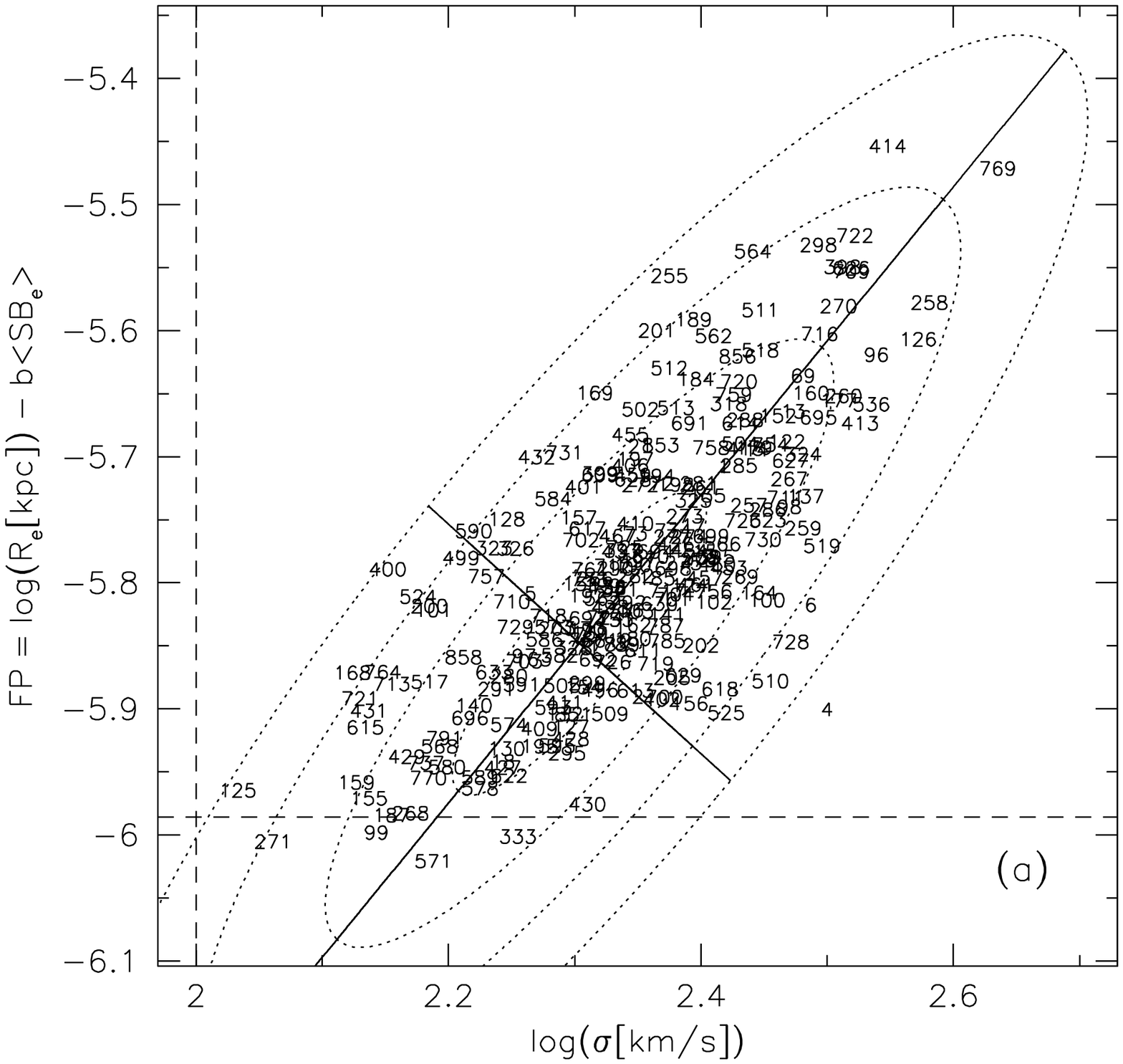} \\ ~ \\
\plotone{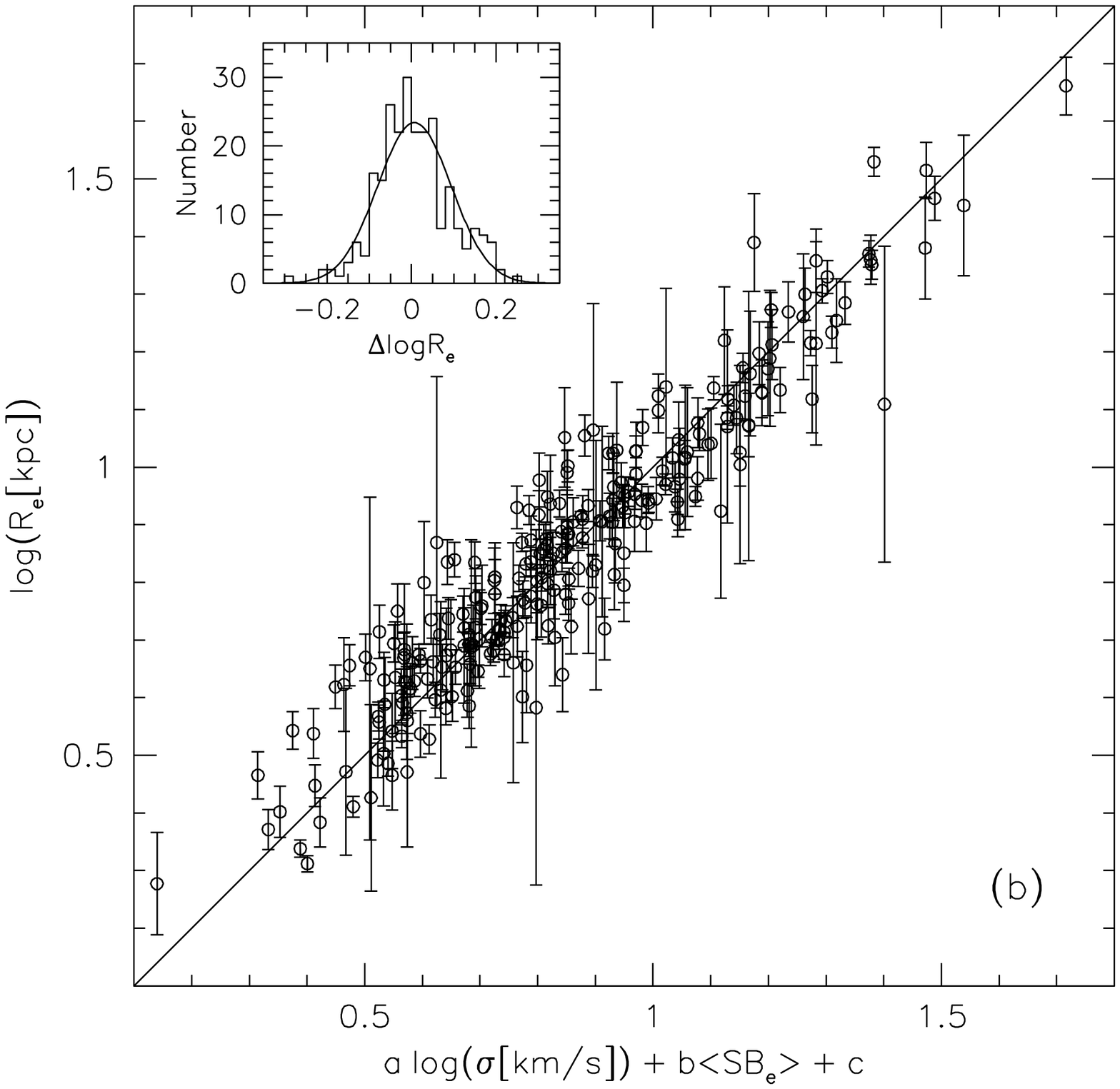}
\caption{The best-fitting FP solution (case~1) for 255 galaxies
belonging to the 29 clusters with 6 or more members. (a)~The projection
of the galaxies (marked by their GINs) in the \logsig--\FP\ plane (where
$\FP=r-bu$). The dashed lines are the cut in \logsig\ and the
approximate cut in \FP. The best-fitting gaussian distribution is shown
by the projections of its major and minor axes and its 1, 2, 3 and
4-$\sigma$ contours. (b)~The scatter of \logRe\ about the FP predictor
for \logRe, namely $a\logsig+b\SBe+c$. The rms scatter about the 1-to-1
line is 0.087~dex (an rms distance error of 20\% per galaxy). The inset
histogram of residuals $\Delta\logRe$ has a gaussian with an rms of
0.087~dex overlaid.}
\label{fig:fpall}
\end{figure}

Case~1 is our best-fit solution. The EFAR FP, based on 29 clusters and
255 galaxies, has $a$=1.223$\pm$0.087, $b$=0.336$\pm$0.013 and
$c$=$-$8.66$\pm$0.33. The intrinsic scatter about this FP is
$\sigma_1$=0.064$\pm$0.006, corresponding in to an intrinsic error in
estimating distances of 15\%\footnote{Logarithmic errors, $\epsilon$,
are converted to linear errors, $\varepsilon$, according to
$\varepsilon=(10^{+\epsilon}-10^{-\epsilon})/2$.}.
Figure~\ref{fig:fpall}a shows the projection of the galaxy distribution
in the \logsig--\FP\ plane (where $\FP=r-bu$). The hard cut in \logsig\
and the approximate cut in \FP\ are indicated by dashed lines. The shape
of the best-fitting galaxy distribution is shown by the projections of
its major and minor axes and its 1, 2, 3 and 4-$\sigma$ contours.
Figure~\ref{fig:fpall}b shows the scatter of \logRe\ about the FP
predictor for \logRe, namely $a\logsig+b\SBe+c$. The rms scatter about
the 1-to-1 relation (the solid line) is 0.087~dex, which is larger than
$\sigma_1$ because of the errors in the measurements. (Allowing for the
estimated measurement errors, the reduced $\chi^2$ is 1.01, which is a
consistency check on the fitted value of $\sigma_1$.) Thus although the
intrinsic rms precision of distance estimates from the FP is 0.064~dex
(15\%), the effective rms precision for the EFAR sample when the
intrinsic scatter and the measurement errors are combined is 0.087~dex
(20\%).

\begin{figure}
\plotone{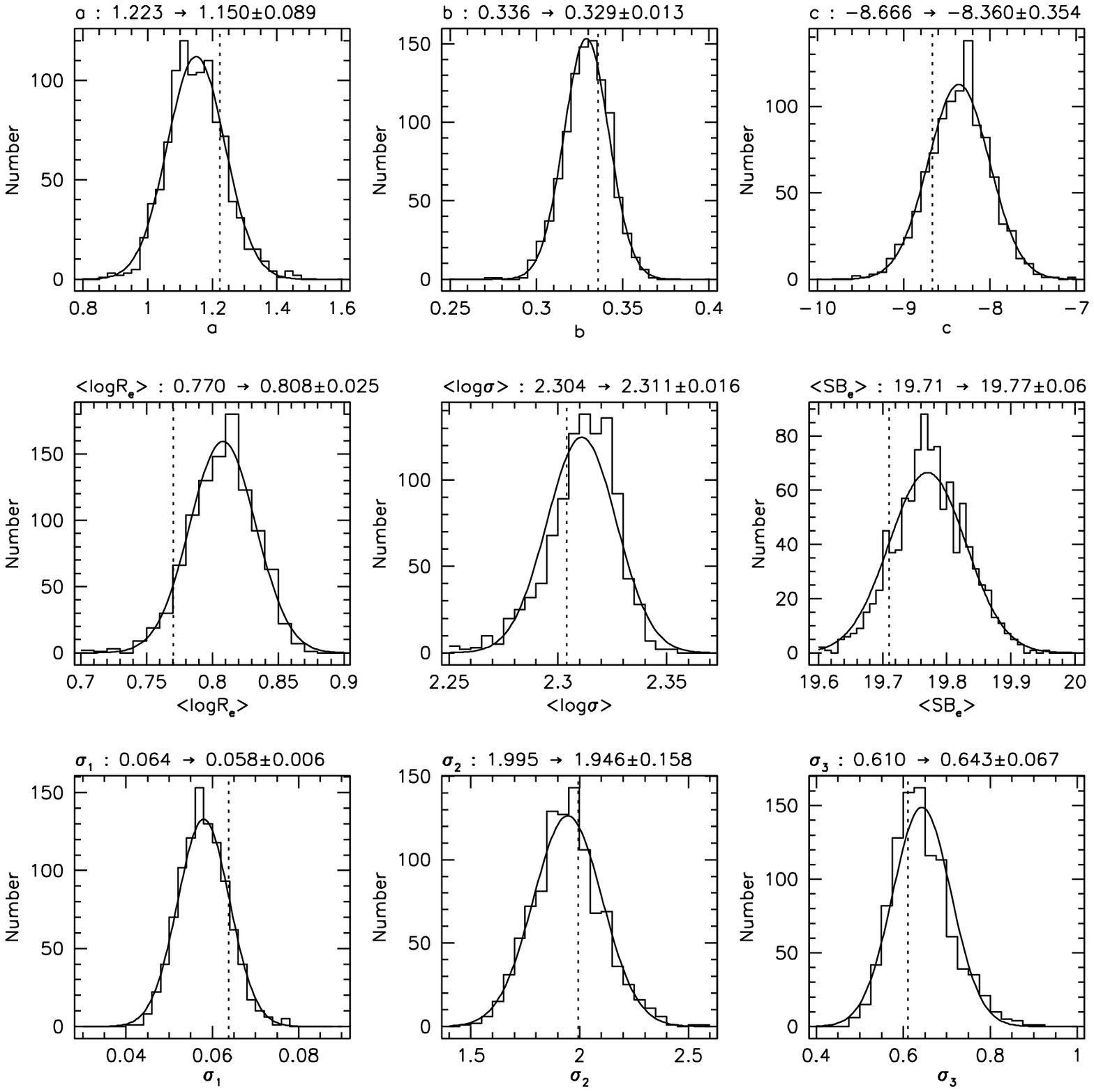}
\caption{The distributions of the FP parameters $a$, $b$, $c$, \meanRe,
\meansig, \meanSBe, $\sigma_1$, $\sigma_2$ and $\sigma_3$ resulting from
fitting 1000 simulations of the best-fit FP. The input parameters of the
simulations are given at the head of each panel (and indicated by the
vertical dotted line), followed by the mean and rms of the fits to the
simulations (the curve is the gaussian with this mean and rms). }
\label{fig:simresults}
\end{figure}

The random errors given above for the best-fit parameters are based on
1000 simulations of the recovery of the FP from the EFAR dataset
(assuming {\em no} peculiar velocities) using the maximum likelihood
gaussian algorithm, as described in Paper~VI.
Figure~\ref{fig:simresults} shows the distributions of the fitted
parameters from these 1000 simulations: the dotted vertical line is the
input value of the parameter and the smooth curve is the gaussian with
the same mean and rms as the fits. There are residual biases in the
fitted parameters, as shown by the offsets between the input parameters
and the mean of the fits: $a$ is biased low by 6\%, $b$ is biased low by
2\%, $c$ is biased high by 4\%; \meanRe, \meansig and \meanSBe are all
biased high, by 0.036~dex, 0.007~dex and 0.05~mag respectively; the
scatter about the FP is under-estimated by 0.006~dex, or 1.4\%. These
biases are all less than or comparable to the rms width of the
distribution, so that although they are statistically significant (i.e.\
much greater than the standard error in the mean, rms/$\sqrt{1000}$),
they do not dominate the random error in the fitted parameters. We do
not correct for these biases since they are small and have negligible
impact on the derived distances and peculiar velocities (see
\S\ref{distances} below).

\subsection{Variant cases and systematic errors}

All the other cases listed in Table~\ref{tab:fpfits} are variants of
this standard case, as briefly described in the Notes column of
Table~\ref{tab:fpfits}. Case~2 includes galaxies with poorer quality
($Q$=3) photometry and less reliable structural parameters, increasing
the scatter about the FP. Case~3 includes the outlier galaxies rejected
from the standard sample, and also has a larger FP scatter. Cases~4--6
show that applying alternative prescriptions for the absorption
corrections (BH corrections, SFD corrections without an offset, and
corrections based on a 36:64 weighting of BH and SFD) has no significant
effect on the FP fit. Case~7 shows that applying a stricter constraint
on the errors in the velocity dispersions, excluding galaxies for which
$\delta\logsig$$>$0.1, also has no effect. Cases~8--11 correspond to
different cuts in $D_W$ (no cut and \DWcut=6.3, 14.1 and 15.9\kpc\
respectively); there is a slight flattening of the FP slope $a$ for
lower cuts. Case~12 excludes not only the galaxies rejected from the
standard fit, but also galaxies with low likelihoods ($\ln{\cal
L}$$<$0); this results in a highly biased fit, with both $a$ and $b$
significantly lower than in the standard case, and with an artificially
lowered FP scatter. Case~13 shows that assuming a $q_0$=0 cosmology has
no impact on the FP fit. Cases~14 and~15 examine the effect of a lower
($S_i$$>$0.01) and a higher ($S_i$$>$0.2) limit on the allowed selection
probabilities. The former case has highly deviant values for \meanRe,
\meansig\ and \meanSBe\ due to over-weighting a few galaxies with low
selection probabilities; the latter case has biased values of $a$, $b$
and $c$ due to ignoring the tail of the selection function. Case~16
ignores the selection probabilities altogether and applies a uniform
weight to all galaxies, resulting in an effective over-weighting of the
larger galaxies and biasing the mean values of \meanRe\ and \meansig\ to
higher values. Cases~17--20 show that setting the mean FP shift to
+0.01, $-$0.01, +0.03 and $-$0.03~dex respectively (rather than to zero,
as in the standard case) has no effect on the fitted FP. 

Case~21 permits an extra degree of freedom by allowing the orientation
of the major axis of the galaxy distribution {\em within} the FP to be
fitted, rather than specified {\it a priori}. The unit vectors of the
galaxy distribution for the standard case, given by
equation~(\ref{eqn:evecs}), are \\ ~ \\
\begin{eqnarray}
\label{eqn:evecs1}
\hat{v}_1 & = & +1.000\hat{r}-1.223\hat{s}-0.336\hat{u} \nonumber \\
\hat{v}_2 & = & +1.000\hat{r}+0.000\hat{s}+2.978\hat{u} \\
\hat{v}_3 & = & -2.978\hat{r}-2.710\hat{s}+1.000\hat{u} ~, \nonumber
\end{eqnarray}
while the true eigenvectors, obtained by fitting with the extra degree
of freedom, are
\begin{eqnarray}
\label{eqn:evecs21}
\hat{v}_1 & = & +1.000\hat{r}-1.227\hat{s}-0.337\hat{u} \nonumber \\
\hat{v}_2 & = & +1.000\hat{r}-0.032\hat{s}+2.964\hat{u} \\
\hat{v}_3 & = & -3.176\hat{r}-2.863\hat{s}+1.000\hat{u} ~. \nonumber
\end{eqnarray}
The coefficient of $\hat{s}$ in the second eigenvector is small,
justifying the simplifying approximation of setting it to zero used in
equation~(\ref{eqn:evecs}). The FP itself is very close to the standard
fit, while the axes within the FP have coefficients differing from the
standard values by no more than a few percent; $\sigma_1$ stays the
same, while $\sigma_2$ is maximised and $\sigma_3$ is minimised.

Case~22 replaces the individual error estimates for all measured
quantities with uniform errors; this has little effect on the FP, but
under-estimates the intrinsic scatter about the plane. Cases~23--28
explore the effects of varying the minimum number of galaxies required
for a cluster to be included in the fit, from 3, 4, 5, 6 and 8 up to 10.
Note that this is the number of galaxies in the cluster before excluding
outliers; hence case~26 differs from case~1 in having 31 clusters rather
than 29. The simulations of Paper~VI suggested that a spuriously small
estimate for $\sigma_1$ could in principle result when clusters with few
galaxies are included in the fit, as offsetting the FP with a spurious
peculiar velocity could suppress the apparent scatter. However this
effect is not observed in fitting the actual data, and the FP fits are
consistent with the errors on the best fit for samples with a minimum
number of galaxies per cluster of between 3 and 8. A significantly
flatter FP slope is found only for the set of clusters with 10 or more
galaxies, where there are only 7 clusters and 99 galaxies in the fit and
correspondingly larger uncertainties. Case~29 is the same as the
standard case except that spirals are excluded, so that the FP is fitted
only to galaxies with E, E/S0 and cD morphological types. The FP slope
for these early-type galaxies is steeper, with $a$=1.33. Cases~30--33
are similar to case~29, but with the minimum number of galaxies required
for a cluster to be included in the fit varied from 3 to 6. Cases~34
and~35 are the same as the standard case except that the fit is
restricted, respectively, to exclude cD galaxies and both cD galaxies
and spirals. Removing cDs flattens $a$ and lowers $b$, in contrast to
case~29; removing both cDs and spirals restores the FP to the
intermediate values obtained by including both populations.

\begin{figure}
\plotone{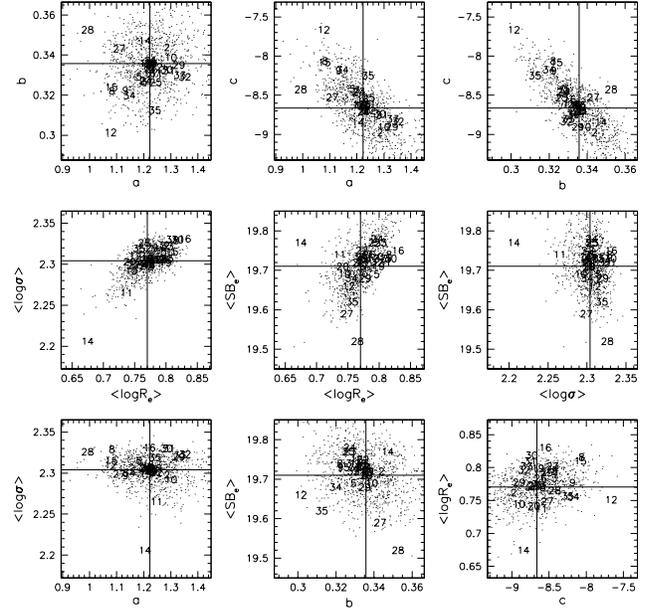}
\caption{The fitted FP parameters for each case in
Table~\ref{tab:fpfits}, showing the distributions and correlations for
various pairs of parameters. Each case is numbered as in the table. The
dots are the distribution of fits obtained for 1000 simulations of the
standard case after removing the effects of the residual biases.}
\label{fig:solpairs}
\end{figure}

\begin{table*}
\centering
\caption{Determinations of the Fundamental Plane.}
\label{tab:fplit}
\begin{tabular}{lccccl}
Source                     & Band  &        A        &           B        &$\Delta$& Fit method \vspace{6pt} \\
Dressler \etal\ (1987)     & $B$   & 1.33~$\pm$~0.05 & $-$0.83~$\pm$~0.03 &  20\%  & inverse \\
Djorgovski \& Davis (1987) & $r_G$ & 1.39~$\pm$~0.14 & $-$0.90~$\pm$~0.09 &  20\%  & 2-step inverse \\
Lucey \etal\ (1991)        & $B$   & 1.27~$\pm$~0.07 & $-$0.78~$\pm$~0.09 &  13\%  & inverse \\
Guzm\'{a}n \etal\ (1993)   & $V$   & 1.14~\noerror   & $-$0.79~\noerror   &  17\%  & forward \\
J{\o}rgensen \etal\ (1996) & $r$   & 1.24~$\pm$~0.07 & $-$0.82~$\pm$~0.02 &  17\%  & orthogonal \\
Hudson \etal\ (1997)       & $R$   & 1.38~$\pm$~0.04 & $-$0.82~$\pm$~0.03 &  20\%  & inverse \\
Scodeggio \etal\ (1997)    & $I$   & 1.25~$\pm$~0.02 & $-$0.79~$\pm$~0.03 &  20\%  & mean regression \\
Pahre \etal\ (1998)        & $K$   & 1.53~$\pm$~0.08 & $-$0.79~$\pm$~0.03 &  21\%  & orthogonal \\
M\"{u}ller \etal\ (1998)   & $R$   & 1.25~\noerror   & $-$0.87~\noerror   &  19\%  & orthogonal \\
Gibbons \etal\ (2000)      & $R$   & 1.39~$\pm$~0.04 & $-$0.84~$\pm$~0.01 &  19\%  & inverse \\
EFAR (this paper)          & $R$   & 1.22~$\pm$~0.09 & $-$0.84~$\pm$~0.03 &  20\%  & ML gaussian \\
\end{tabular}
\end{table*}

Figure~\ref{fig:solpairs} shows the fitted values in each case for
various pairs of the parameters, in order to show their distributions
and correlations. The cases are numbered following
Table~\ref{tab:fpfits}. The dots show the distribution of fits obtained
for 1000 simulations of the standard case (case~1) after removing the
effects of the residual biases. The main point to note is that, with
only a few exceptions (noted above), the systematic differences in the
fits derived for difference cases are comparable to the random errors in
the determination of the parameters for the standard case. We conclude
that the uncertainties in our best-fit FP parameters are dominated by
the random errors and not by systematic effects from the fitting method.
In particular we conclude that the following inputs have relatively
little effect on the fitted FP: the absorption correction, the
cosmological model, the assumed mean FP shift and the choice of the
second and third FP axes. Our standard case provides an optimum fit to
the FP because: (i)~it excludes the galaxies with poor structural
parameters and velocity dispersion measurements which artificially
inflate the scatter about the FP and the uncertainty in the FP
parameters; (ii)~it applies a selection function cutoff that balances
over-weighting a small number of galaxies against biasing the results by
ignoring galaxies with low selection probabilities; (iii)~it uses
clusters with 6 or more galaxies to avoid artificially reducing FP
scatter by confusing scatter with peculiar velocities while yet
retaining a sufficiently large overall number of galaxies to keep the
random errors in the FP parameters small.

\subsection{Comparison with previous work}

Table~\ref{tab:fplit} compares the best-fit EFAR FP with previous
determinations in the literature, noting both the passband to which the
relation applies and the method of the fit. To match the usage in most
of this literature, we present the FP in the form
$R_e\propto\sigma_0^A\langle\Sigma\rangle_e^B$, where $\sigma_0$ is the
central velocity dispersion and $\langle\Sigma\rangle_e$ is the mean
surface brightness (in linear units) within the effective radius $R_e$.
The exponents of this relation are related to the coefficients of our FP
relation, $\logRe = a\logsig + b\SBe + c$, by $A$=$a$ and $B$=$-$2.5$b$.
The table also quotes the fractional distance error, $\Delta$,
corresponding to the rms scatter about the FP in $R_e$. In most cases
the determination of the FP is limited to galaxies with
$\sigma$$>$100\kms. The forward and inverse fitting methods are linear
regressions with, respectively, \logRe\ and \logsig\ as the independent
variable; orthogonal fitting minimises the residuals orthogonal to the
FP, while mean regression averages the fits obtained by taking each of
\logRe, \logsig\ and \SBe\ as the independent variable.

The first point to note is that all the fitted values of $B$ are
consistent within the errors, regardless of passband and fitting method.
The second point to note is that this is not true for $A$, which has a
higher value in the $K$-band FP fit of Pahre \etal\ (1998) than in any
of the optical fits. The third point is that, within the optical FP
fits, the forward and inverse fits give, respectively, lower and higher
values of $A$ than the orthogonal and mean regressions and the maximum
likelihood gaussian method. This is consistent with the analysis and
simulations of the methods carried out in Paper~VI: for samples in which
the errors in $\sigma$ dominate and/or selection cuts are applied in
$R_e$ (as is the case for most of these datasets), the value of $A$ will
be under-estimated by a forward fit and over-estimated by an inverse
fit. Orthogonal and mean regressions reduce these biases, with the least
bias being produced with the maximum likelihood method. We conclude that
apparent differences between FP fits in optical passbands are due to
differences in the fitting methods that have been applied.

There is also consistency on the observed scatter about the FP as
represented by the fractional distance error, $\Delta$. With the
exception of Lucey \etal\ (1991), the observed errors are all in the
range 17\% to 21\%. This is consistent with (i.e.\ larger than) the
estimated intrinsic scatter about the FP of 15\% that we derive from the
EFAR sample, and the range corresponds to the range of measurement
errors in the various studies, which account for between 8\% and 15\% of
the observed scatter.

\section{DISTANCES AND PECULIAR VELOCITIES}
\label{distances}

In order to determine distances and peculiar velocities, we re-apply the
maximum likelihood gaussian algorithm to the whole cluster sample. This
time we fix the parameters of the intrinsic galaxy distribution at their
best-fit values (case~1 of Table~\ref{tab:fpfits}) and fit only for the
shift of the FP for each cluster.

\subsection{Sample}

We remove outliers (interlopers from the cluster foreground or
background, objects which genuinely do not lie on the FP, and objects
with bad data) by excluding the galaxies that deviate most from the
fitted FP until all clusters have FP fits with $\chi^2/\nu$$<$3. To
check that this procedure is conservative, we visually inspected each
cluster's distribution of $D_g-D_c$ (individual galaxy distances
relative to the overall cluster distance, from the residuals about the
best-fit FP) with respect to $cz_g-cz_c$ (individual galaxy redshifts
relative to the overall cluster redshift). The rejected galaxies were
invariably clear outliers in these distributions. In all, 36 galaxies
were rejected using this procedure, including all the galaxies rejected
from the FP fit (see Table~\ref{tab:badfp}). The list of galaxies
excluded from the peculiar velocity fits is given in
Table~\ref{tab:badpv}. There were three clusters with $\chi^2/\nu$$>$3
(CAN~2=A85 with 4 galaxies, CAN~55=P386-2 with 2 galaxies, CAN~79=A2589
with 5 galaxies) for which half or more of the galaxies had to be
rejected in order to obtain a good FP fit, so that it was difficult to
determine which galaxies were the outliers. Although we give distances
and peculiar velocities for these clusters below (using all the
available galaxies), we exclude them from further analysis.

\begin{table}
\centering
\caption{Galaxies excluded from the peculiar velocity fits.}
\label{tab:badpv}
\begin{tabular}{rrlcrrl}
GIN & CAN & Name     & & GIN & CAN & Name \vspace{6pt} \\    
 45 &   7 & A160 C   & & 489 &  57 & J18 C     \\
 52 &   7 & A160 J   & & 495 &  58 & A2151 A   \\
 78 &  11 & A193 A   & & 500 &  58 & A2151 F   \\
125 &  16 & J8 D     & & 501 &  58 & A2151 G   \\
128 &  16 & J8 G     & & 519 &  59 & A2152 I   \\
156 &  20 & A397 F   & & 525 &  59 & A2152 1   \\
167 &  21 & A400 H   & & 552 &  63 & A2162-S G \\
184 &  23 & A419 H   & & 562 &  65 & A2197 A   \\
187 &  23 & A419 1   & & 564 &  65 & A2197 C   \\
189 &  24 & A496 A   & & 584 &  66 & A2199 F   \\
200 &  25 & J34 E    & & 590 &  66 & A2199 L   \\
201 &  25 & J34 F    & & 711 &  80 & A2593-S C \\
271 &  35 & A548-1 F & & 713 &  80 & A2593-S E \\
355 &  42 & J14-1 D  & & 721 &  82 & A2634 F   \\
370 &  43 & A1983 2  & & 728 &  82 & A2634 2   \\
396 &  46 & J16-W B  & & 730 &  83 & A2657 B   \\
432 &  50 & A2063 G  & & 731 &  83 & A2657 C   \\
456 &  53 & A2147 D  & & 756 &  90 & COMA 133  \\
\end{tabular}
\end{table}

\subsection{Bias corrections}

To the extent that its assumptions are justified, the maximum likelihood
gaussian algorithm accounts for the effects of biases on the estimated
distances which are due to the selection function of the galaxies within
each cluster. (We refer to this bias as `selection bias' rather than
`Malmquist bias' because, following the usage of Strauss \& Willick
(1995), the effect is due to the selection criteria and not the
line-of-sight density distribution.) As discussed in Paper~VI, however,
the sample selection function parameter \DoWj\ varies with cluster
redshift, introducing a redshift-dependent bias in the peculiar velocity
estimates. Although this bias is reduced by the selection limit
\DWi$>$\DWcut\ imposed on galaxy sizes (see \S\ref{method}), clusters
with \DoWj$>$\DWcut\ are nonetheless sampled differently to clusters
with \DoWj$\le$\DWcut. This difference in the way the FP galaxy
distribution is sampled in different clusters leads to a residual bias
in the clusters' fitted FP offsets and peculiar velocities as a function
of \DoWj\ (or redshift, with which \DoWj\ is closely correlated). 

\begin{figure}
\plotone{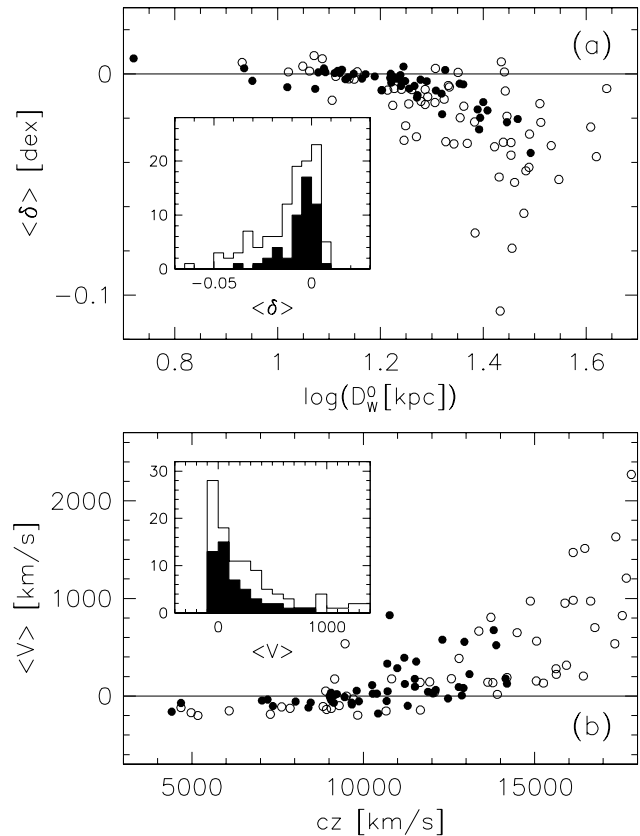}
\caption{The residual selection bias determined from 1000 simulations of
the EFAR dataset. (a)~The bias in the FP offsets $\langle\dj\rangle$ for
each cluster as a function of the cluster's selection function parameter
\DoWj. The inset histogram shows the distribution of bias corrections
$\langle\delta\rangle$. (b)~The corresponding bias in the cluster
peculiar velocities $\langle V_{\rm pec} \rangle$ as a function of cluster
redshift $cz$. The inset histogram shows the distribution of bias
corrections $\langle V_{\rm pec} \rangle$. The filled symbols and the shaded
histogram show the subsample of clusters used in the peculiar velocity
analysis.}
\label{fig:pvbias}
\end{figure}

This effect is investigated in detail through simulations in Paper~VI.
Figure~\ref{fig:pvbias} shows the residual selection bias determined
from 1000 simulations of the EFAR dataset. For clusters with redshifts
below the sample mean the bias in the peculiar velocities is small and
negative, while for clusters at redshifts above the sample mean it is
positive and increases rapidly with redshift. We correct this systematic
bias individually for each cluster by subtracting the mean error in the
FP offset determined from 1000 simulations of the EFAR dataset before
computing the cluster distances and peculiar velocities. The size of the
corrections are shown in the inset histograms of
Figure~\ref{fig:pvbias}. For the subsample of clusters included in
subsequent analyses of the peculiar velocities (whose selection is
discussed below), the amplitude of the bias correction is less than
250\kms\ for 40 of the 50 clusters. The random errors in the peculiar
velocities are typically of order 1000\kms, while the uncertainties in
the peculiar velocity bias corrections for these clusters are typically
less than 50\kms. To the extent that the simulated datasets match the
real distribution of galaxies in the FP, therefore, the bias corrections
should not significantly increase the random errors in the peculiar
velocities.

\subsection{Results}

The individual FP fits are shown in Figure~\ref{fig:fpplots}, where the
fixed parameters of the galaxy distribution used to fit the FP shift are
given at the top of the plot. Each panel corresponds to a cluster,
labelled by its CAN; the 29 clusters used to derive the parameters of
the galaxy distribution are indicated by bold labels. The area of each
point is proportional to the selection weight of the galaxy; the
corresponding GINs are given at left. The solid line is the major axis
of the global fit to the FP, and the cross on this line the centre of
the global galaxy distribution, (\meansig, \meanRe$-$$b$\meanSBe). The
dotted lines and ellipse are the major and minor axes and the 3$\sigma$
contour of the cluster's FP, vertically offset from the global FP by the
cluster's FP shift. The cluster's mean redshift $cz$, distance $D$, and
peculiar velocity \Vpec, each with its estimated error, are given at the
bottom of the panel. The distances and peculiar velocities are corrected
for the residual selection bias discussed above.

\begin{figure*}
\plotfull{1.0}{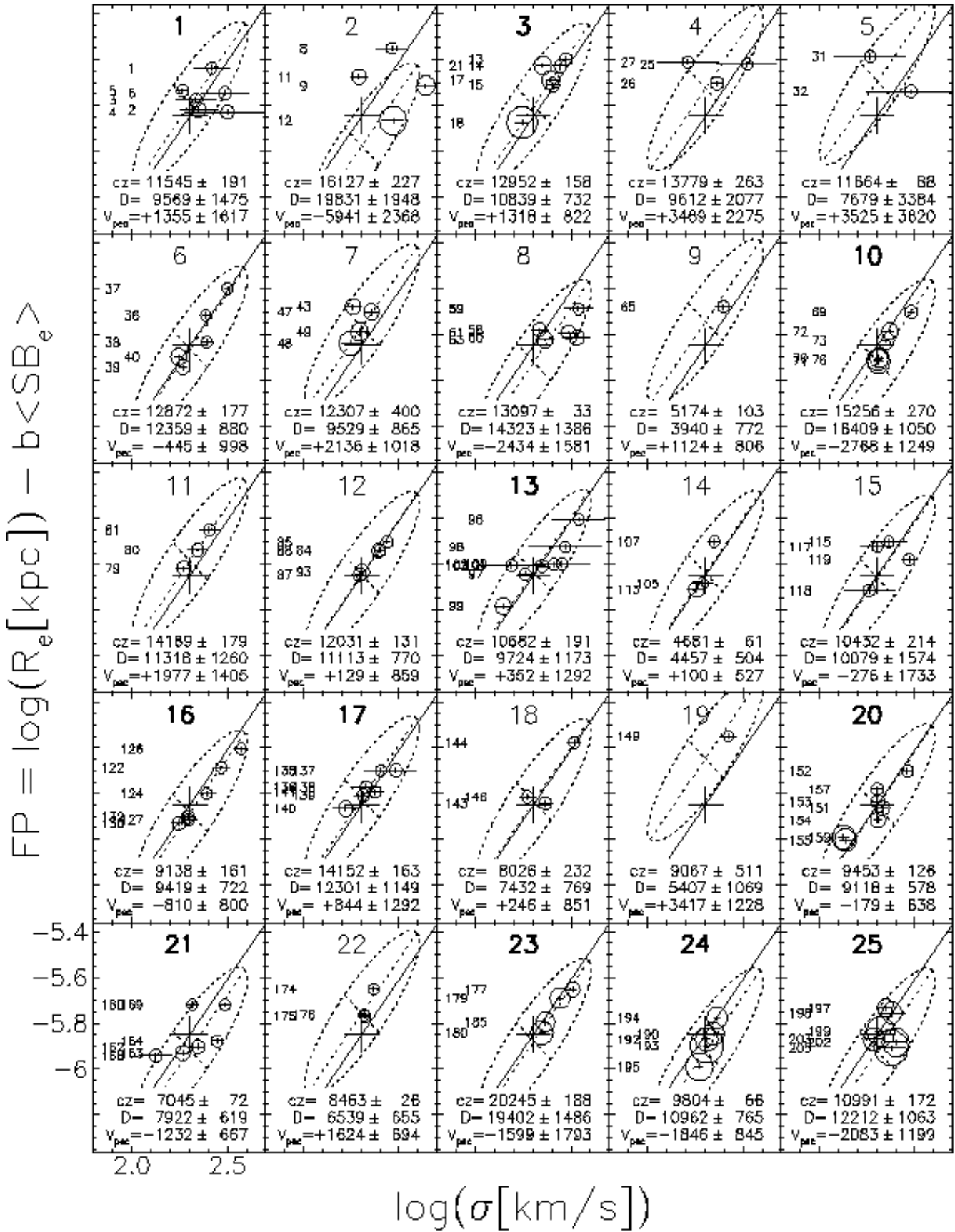}
\caption{The FP distributions for each individual cluster. See text for
description.}
\label{fig:fpplots}
\end{figure*}
\addtocounter{figure}{-1}
\begin{figure*}
\plotfull{1.0}{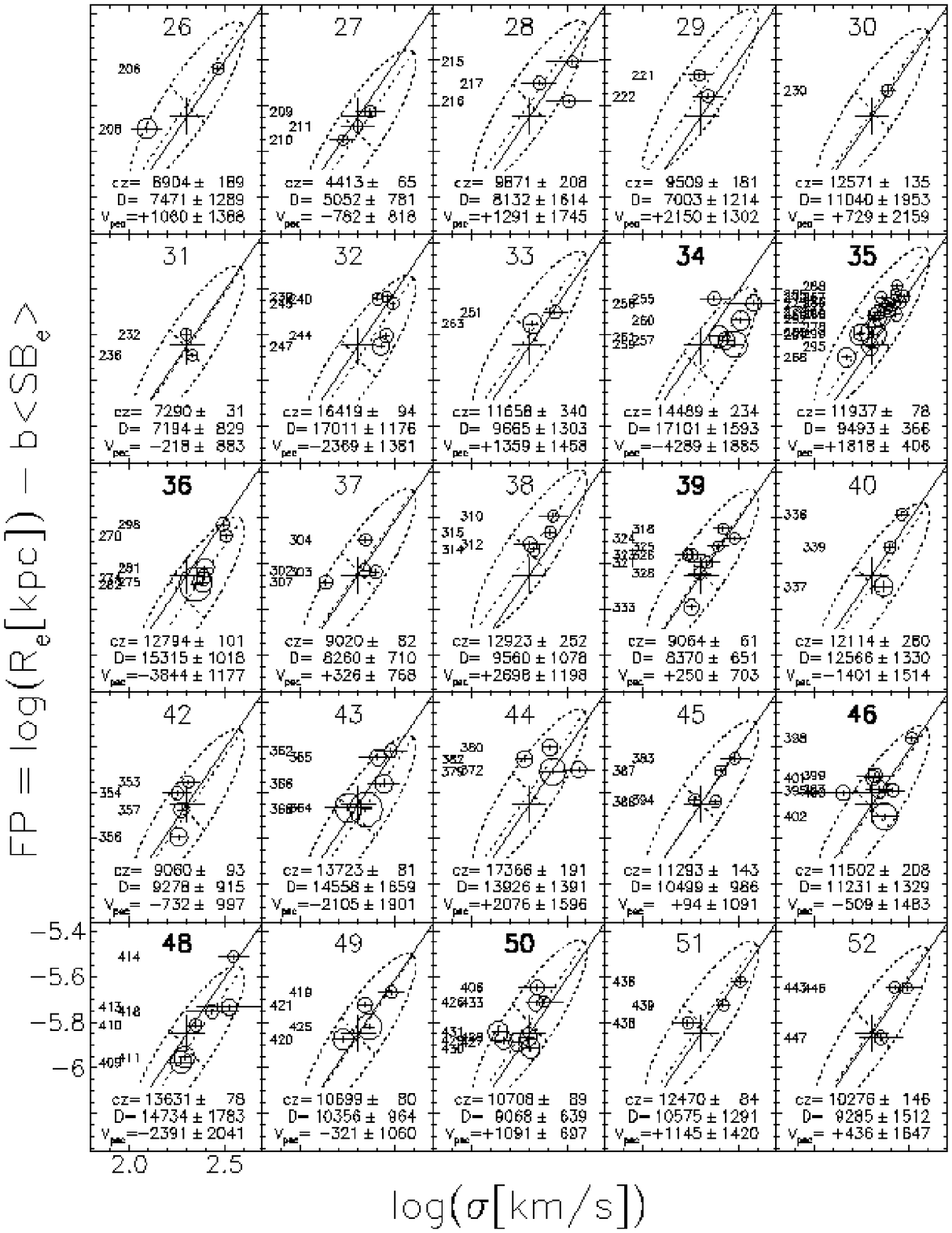}
\caption{Continued.}
\end{figure*}
\addtocounter{figure}{-1}
\begin{figure*}
\plotfull{1.0}{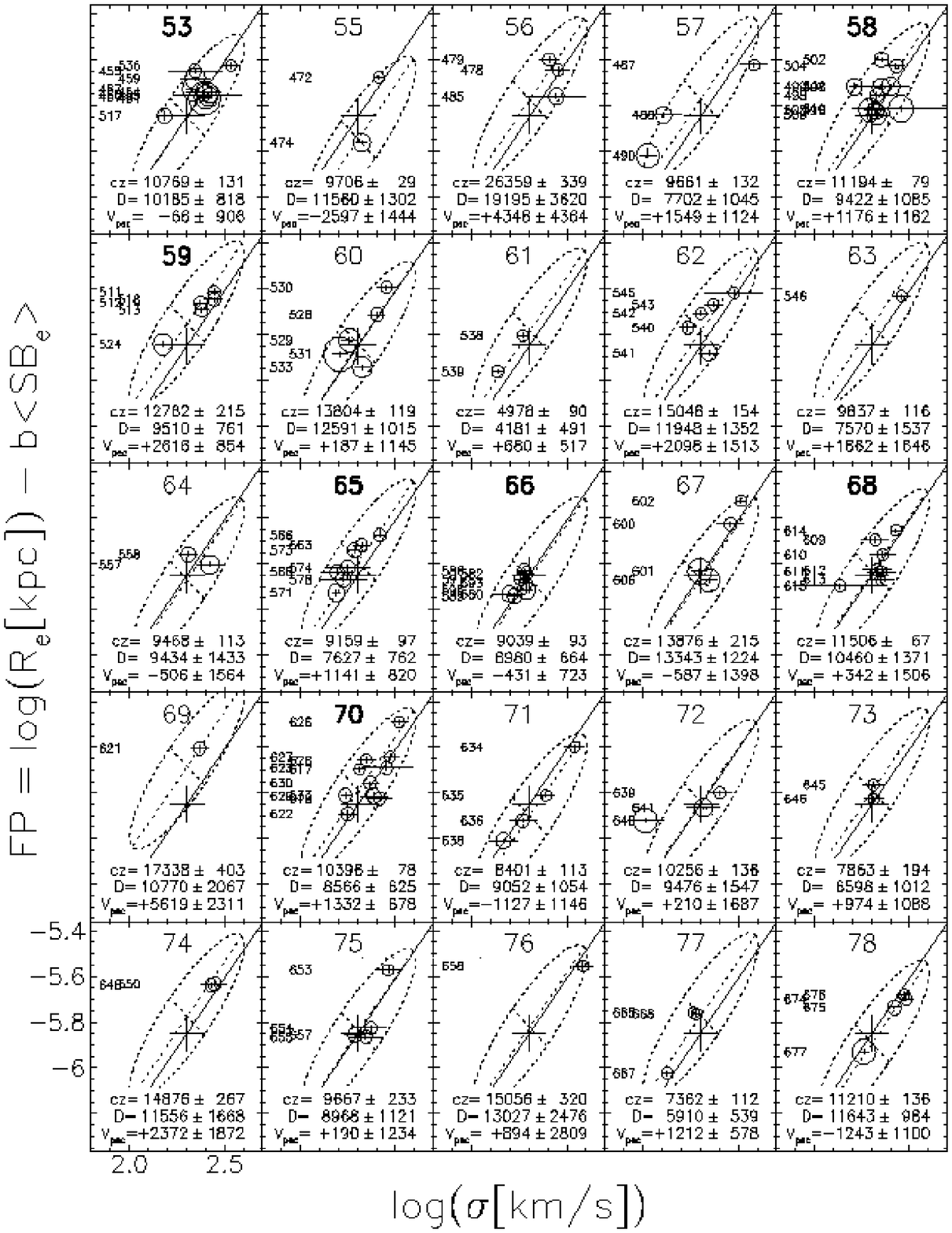}
\caption{Continued.}
\end{figure*}
\addtocounter{figure}{-1}
\begin{figure*}
\plotfull{1.0}{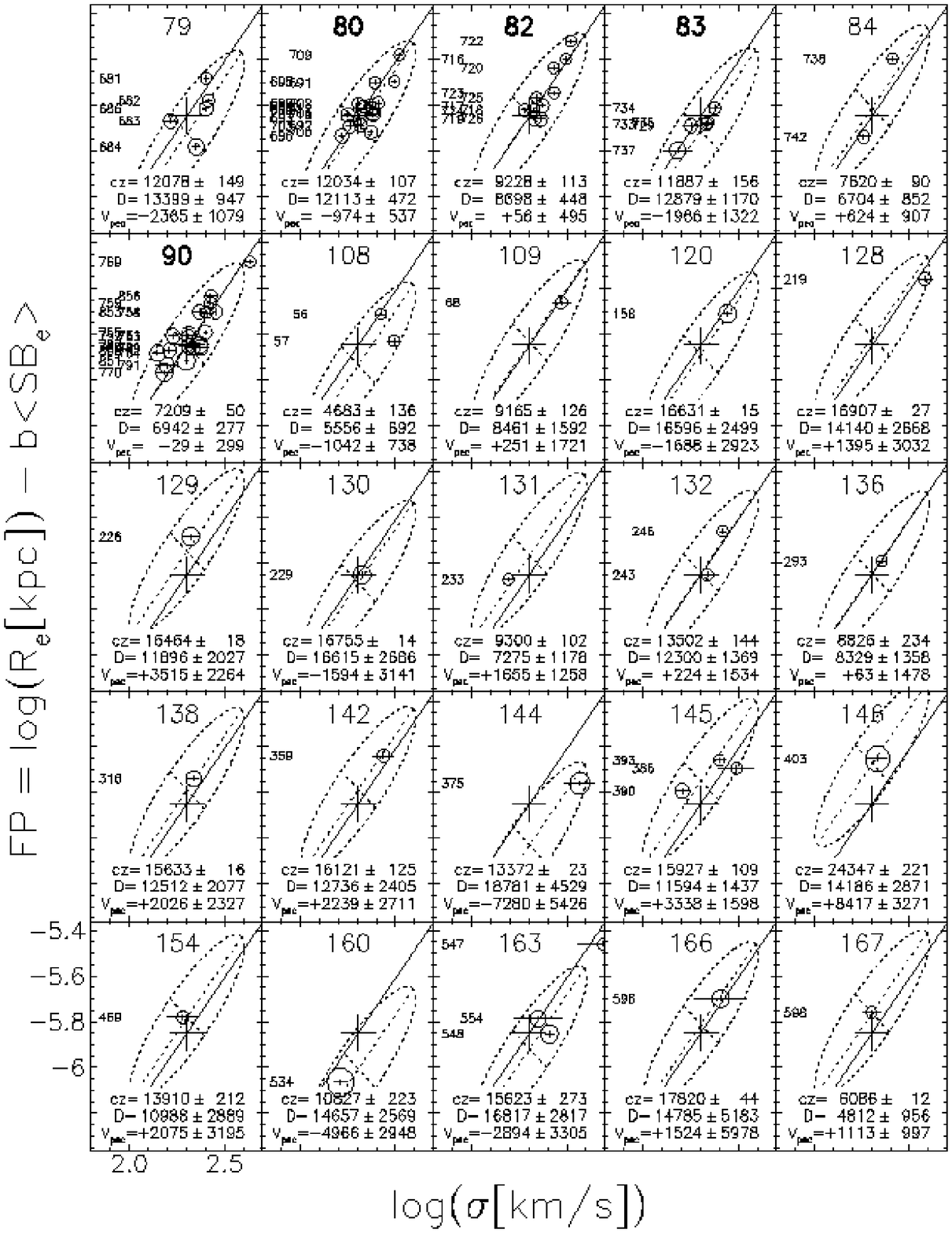}
\caption{Continued.}
\end{figure*}
\addtocounter{figure}{-1}
\begin{figure*}
\plotfull{1.0}{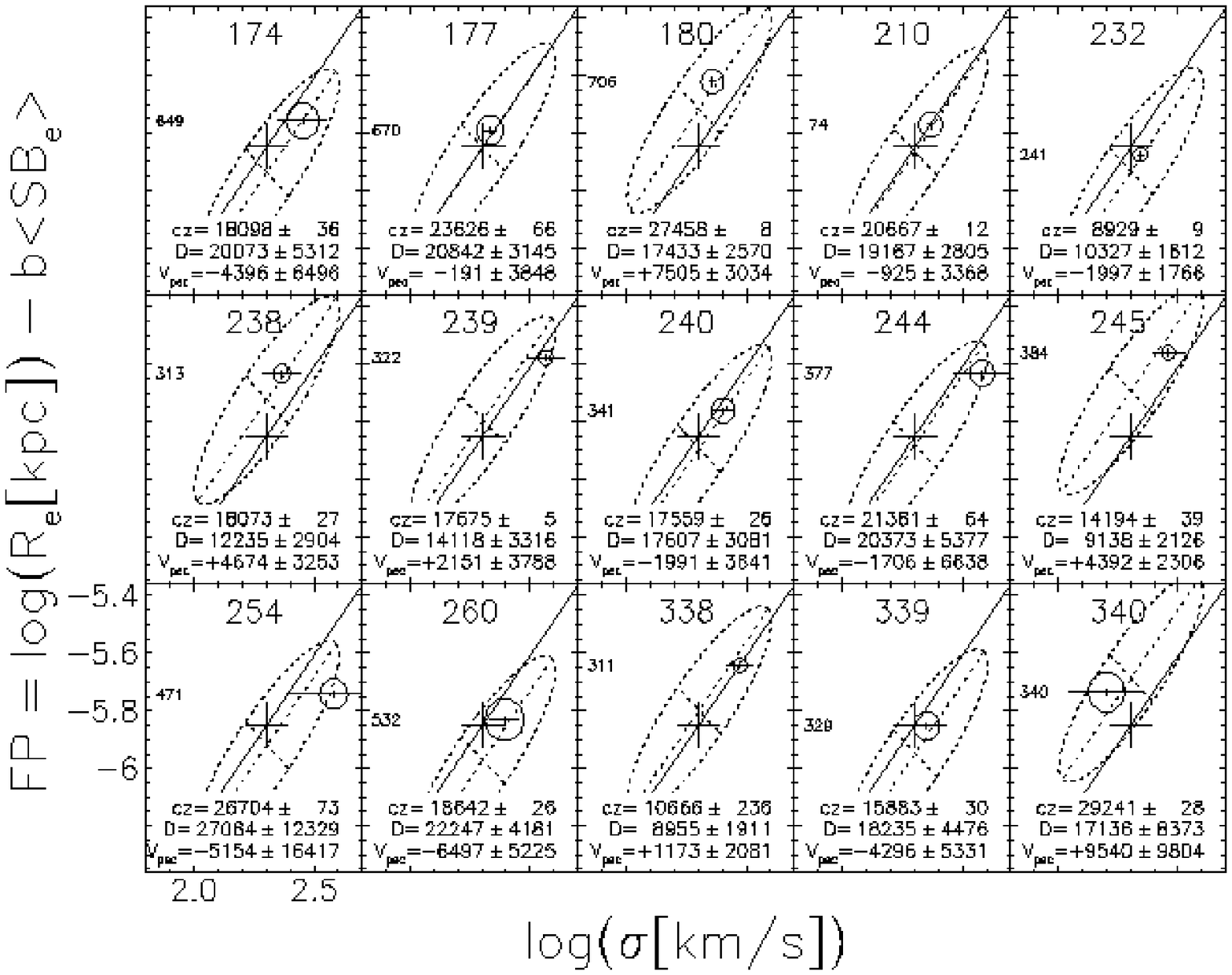}
\caption{Continued.}
\end{figure*}

The results are summarised in Table~\ref{tab:efarpv}, which for each
cluster gives: CAN, cluster name (in parentheses for fore- and
background groups), number of galaxies used in the distance
determination, Galactic longitude and latitude, the bias-corrected FP
shift $\delta$ and its uncertainty, the bias correction
$\epsilon_\delta$ that was subtracted from the raw value of $\delta$,
the cluster redshift $cz$ and its uncertainty $\Delta cz$, and the
bias-corrected values of the cluster distance $D$ and its uncertainty
$\Delta D$, the redshift $cz_D$ corresponding to $D$, and the peculiar
velocity $V$ and its uncertainty $\Delta V$. Note that some clusters are
missing from this list: CAN~81 because it has been combined with CAN~80
(see Paper~II); CANs~41, 47, 54 and a number of the fore- and background
groups (CANs$\,>\,$100) because no cluster members meet the selection
criteria.

\begin{table*}
\centering
\caption{Redshifts, distances and peculiar velocities for the EFAR clusters in the CMB frame}
\label{tab:efarpv}
\begin{tabular}{rlrrrrrrrrrrrrrrr}
\head{CAN} & \head{Name} & \head{$N_{\rm g}$} & \head{$l$} & \head{$b$} & \head{$\delta$} & \head{$\Delta\delta$}
& \head{$\epsilon_\delta$} & \head{$cz$} & \head{$\Delta cz$} & \head{$D$} & \head{$\Delta D$} & \head{$cz_{D}$} 
& \head{$\Delta cz_{D}$} & \head{$V$} & \head{$\Delta V$}  \vspace{6pt} \\
{\bf  1}*  & A76       &  6 & 117.57 & $-$56.02 & $-$0.0528 & 0.0697 & $-$0.0182 & 11545 & 191 &  9569 &  1475 & 10143 &  1658 & $+$1355 &  1617 \\
{\rm  2}\n & A85       &  4 & 115.23 & $-$72.04 & $+$0.1297 & 0.0466 & $-$0.0310 & 16127 & 227 & 19831 &  1948 & 22514 &  2522 & $-$5941 &  2368 \\
{\bf  3}*  & A119      &  6 & 125.80 & $-$64.07 & $-$0.0452 & 0.0309 & $-$0.0204 & 12952 & 158 & 10839 &   732 & 11582 &   837 & $+$1318 &   822 \\
{\rm  4}\n & J3        &  3 & 125.87 & $-$35.86 & $-$0.1222 & 0.0953 & $-$0.0103 & 13779 & 263 &  9612 &  2077 & 10192 &  2336 & $+$3469 &  2275 \\
{\rm  5}\n & J4        &  2 & 125.72 & $-$49.85 & $-$0.1525 & 0.1902 & $-$0.0142 & 11664 &  88 &  7679 &  3384 &  8043 &  3713 & $+$3525 &  3620 \\
{\rm  6}*  & A147      &  5 & 132.02 & $-$60.34 & $+$0.0143 & 0.0311 & $-$0.0021 & 12872 & 177 & 12359 &   880 & 13337 &  1027 &  $-$445 &   998 \\
{\rm  7}*  & A160      &  4 & 130.33 & $-$46.82 & $-$0.0805 & 0.0416 & $-$0.0218 & 12307 & 400 &  9529 &   865 & 10098 &   972 & $+$2136 &  1018 \\
{\rm  8}*  & A168      &  5 & 134.36 & $-$61.61 & $+$0.0714 & 0.0432 & $-$0.0107 & 13097 &  33 & 14323 &  1386 & 15658 &  1661 & $-$2434 &  1581 \\
{\rm  9}\n & A189      &  1 & 140.13 & $-$59.99 & $-$0.1053 & 0.0832 & $+$0.0069 &  5174 & 103 &  3940 &   772 &  4034 &   810 & $+$1124 &   806 \\
{\bf 10}\n & J30       &  6 & 151.84 & $-$75.04 & $+$0.0694 & 0.0282 & $-$0.0052 & 15256 & 270 & 16409 &  1050 & 18192 &  1295 & $-$2768 &  1249 \\
{\rm 11}*  & A193      &  3 & 136.94 & $-$53.26 & $-$0.0630 & 0.0492 & $-$0.0073 & 14189 & 179 & 11318 &  1260 & 12131 &  1450 & $+$1977 &  1405 \\
{\rm 12}*  & J32       &  5 & 156.21 & $-$69.05 & $-$0.0045 & 0.0302 & $-$0.0024 & 12031 & 131 & 11113 &   770 & 11896 &   884 &  $+$129 &   859 \\
{\bf 13}*  & A260      &  8 & 137.00 & $-$28.17 & $-$0.0142 & 0.0524 & $-$0.0011 & 10682 & 191 &  9724 &  1173 & 10317 &  1322 &  $+$352 &  1292 \\
{\rm 14}*  & A262      &  3 & 136.59 & $-$25.09 & $-$0.0095 & 0.0486 & $+$0.0033 &  4681 &  61 &  4457 &   504 &  4578 &   532 &  $+$100 &   527 \\
{\rm 15}*  & J7        &  4 & 143.10 & $-$22.18 & $+$0.0111 & 0.0673 & $+$0.0007 & 10432 & 214 & 10079 &  1574 & 10718 &  1782 &  $-$276 &  1733 \\
{\bf 16}*  & J8        &  6 & 150.69 & $-$34.33 & $+$0.0360 & 0.0331 & $+$0.0019 &  9138 & 161 &  9419 &   722 &  9975 &   811 &  $-$810 &   800 \\
{\bf 17}*  & A376      &  7 & 147.11 & $-$20.52 & $-$0.0258 & 0.0413 & $-$0.0074 & 14152 & 163 & 12301 &  1149 & 13270 &  1339 &  $+$844 &  1292 \\
{\rm 18}*  & J9        &  3 & 143.01 & $-$11.22 & $-$0.0133 & 0.0449 & $-$0.0001 &  8026 & 232 &  7432 &   769 &  7773 &   842 &  $+$246 &   851 \\
{\rm 19}\n & J33       &  1 & 195.20 & $-$58.30 & $-$0.2018 & 0.0851 & $+$0.0009 &  9067 & 511 &  5407 &  1069 &  5586 &  1141 & $+$3417 &  1228 \\
{\bf 20}*  & A397      &  7 & 161.84 & $-$37.33 & $+$0.0080 & 0.0276 & $-$0.0005 &  9453 & 126 &  9118 &   578 &  9638 &   647 &  $-$179 &   638 \\
{\bf 21}*  & A400      &  6 & 170.28 & $-$45.00 & $+$0.0687 & 0.0338 & $+$0.0014 &  7045 &  72 &  7922 &   619 &  8311 &   682 & $-$1232 &   667 \\
{\rm 22}*  & J28       &  3 & 183.86 & $-$50.08 & $-$0.0908 & 0.0433 & $+$0.0008 &  8463 &  26 &  6539 &   655 &  6802 &   709 & $+$1624 &   694 \\
{\bf 23}\n & A419      &  4 & 214.31 & $-$59.00 & $+$0.0313 & 0.0340 & $-$0.0077 & 20245 & 188 & 19402 &  1486 & 21962 &  1912 & $-$1599 &  1793 \\
{\bf 24}*  & A496      &  5 & 209.59 & $-$36.49 & $+$0.0730 & 0.0306 & $-$0.0044 &  9804 &  66 & 10962 &   765 & 11722 &   876 & $-$1846 &   845 \\
{\bf 25}*  & J34       &  6 & 213.90 & $-$34.95 & $+$0.0732 & 0.0390 & $-$0.0127 & 10991 & 172 & 12212 &  1063 & 13166 &  1238 & $-$2083 &  1199 \\
{\rm 26}\n & J10       &  2 & 197.18 & $-$25.49 & $-$0.0539 & 0.0758 & $-$0.0073 &  8904 & 189 &  7471 &  1289 &  7815 &  1411 & $+$1060 &  1388 \\
{\rm 27}*  & P597-1    &  3 & 198.62 & $-$24.50 & $+$0.0699 & 0.0657 & $+$0.0070 &  4413 &  65 &  5052 &   781 &  5208 &   830 &  $-$782 &   818 \\
{\rm 28}*  & J35       &  3 & 217.47 & $-$33.61 & $-$0.0595 & 0.0869 & $-$0.0066 &  9871 & 208 &  8132 &  1614 &  8543 &  1782 & $+$1291 &  1745 \\
{\rm 29}\n & J34/35    &  2 & 216.40 & $-$34.19 & $-$0.1091 & 0.0761 & $-$0.0070 &  9509 & 181 &  7003 &  1214 &  7305 &  1321 & $+$2150 &  1302 \\
{\rm 30}\n & P777-1    &  1 & 218.49 & $-$32.70 & $-$0.0251 & 0.0789 & $-$0.0135 & 12571 & 135 & 11040 &  1953 & 11812 &  2238 &  $+$729 &  2159 \\
{\rm 31}\n & P777-2    &  2 & 220.77 & $-$32.62 & $+$0.0125 & 0.0493 & $+$0.0052 &  7290 &  31 &  7194 &   829 &  7513 &   905 &  $-$218 &   883 \\
{\rm 32}\n & P777-3    &  5 & 219.72 & $-$31.71 & $+$0.0560 & 0.0306 & $-$0.0071 & 16419 &  94 & 17011 &  1176 & 18938 &  1463 & $-$2369 &  1381 \\
{\rm 33}\n & A533-1    &  2 & 224.95 & $-$33.54 & $-$0.0524 & 0.0600 & $-$0.0114 & 11658 & 340 &  9665 &  1303 & 10251 &  1467 & $+$1359 &  1458 \\
{\bf 34}\n & A533      &  6 & 223.18 & $-$33.65 & $+$0.1079 & 0.0430 & $-$0.0219 & 14489 & 234 & 17101 &  1593 & 19050 &  1984 & $-$4289 &  1885 \\
{\bf 35}\n & A548-1    & 18 & 230.28 & $-$24.43 & $-$0.0698 & 0.0170 & $-$0.0059 & 11937 &  78 &  9493 &   366 & 10057 &   412 & $+$1818 &   406 \\
{\bf 36}\n & A548-2    &  6 & 230.40 & $-$25.97 & $+$0.1099 & 0.0300 & $-$0.0146 & 12794 & 101 & 15315 &  1018 & 16854 &  1237 & $-$3844 &  1177 \\
{\rm 37}*  & J11       &  4 & 118.21 & $+$63.43 & $-$0.0157 & 0.0375 & $-$0.0018 &  9020 &  82 &  8260 &   710 &  8683 &   785 &  $+$326 &   768 \\
{\rm 38}*  & J12       &  4 &  50.52 & $+$78.23 & $-$0.0988 & 0.0495 & $-$0.0054 & 12923 & 252 &  9560 &  1078 & 10132 &  1212 & $+$2698 &  1198 \\
{\bf 39}*  & J13       &  8 &  28.27 & $+$75.54 & $-$0.0119 & 0.0337 & $+$0.0004 &  9064 &  61 &  8370 &   651 &  8805 &   721 &  $+$250 &   703 \\
{\rm 40}*  & J36       &  3 & 332.77 & $+$49.31 & $+$0.0461 & 0.0465 & $-$0.0054 & 12114 & 280 & 12566 &  1330 & 13578 &  1556 & $-$1401 &  1514 \\
{\rm 42}*  & J14-1     &  4 &   8.80 & $+$58.73 & $+$0.0330 & 0.0432 & $-$0.0039 &  9060 &  93 &  9278 &   915 &  9816 &  1025 &  $-$732 &   997 \\
{\rm 43}\n & A1983     &  5 &  18.59 & $+$59.60 & $+$0.0597 & 0.0531 & $-$0.0272 & 13723 &  81 & 14558 &  1659 & 15940 &  1994 & $-$2105 &  1901 \\
{\rm 44}\n & A1991     &  4 &  22.74 & $+$60.52 & $-$0.0530 & 0.0487 & $-$0.0438 & 17366 & 191 & 13926 &  1391 & 15184 &  1658 & $+$2076 &  1596 \\
{\rm 45}*  & J16       &  4 &   6.81 & $+$48.20 & $-$0.0035 & 0.0407 & $+$0.0004 & 11293 & 143 & 10499 &   986 & 11195 &  1122 &   $+$94 &  1091 \\
{\bf 46}*  & J16W      &  7 &   5.08 & $+$49.63 & $+$0.0183 & 0.0524 & $-$0.0089 & 11502 & 208 & 11231 &  1329 & 12032 &  1527 &  $-$509 &  1483 \\
{\bf 48}\n & A2040     &  6 &   9.08 & $+$51.15 & $+$0.0676 & 0.0534 & $-$0.0070 & 13631 &  78 & 14734 &  1783 & 16151 &  2148 & $-$2391 &  2041 \\
{\rm 49}*  & A2052     &  4 &   9.42 & $+$50.11 & $+$0.0125 & 0.0420 & $-$0.0160 & 10699 &  80 & 10356 &   964 & 11032 &  1095 &  $-$321 &  1060 \\
{\bf 50}*  & A2063     &  8 &  12.80 & $+$49.70 & $-$0.0455 & 0.0308 & $-$0.0033 & 10708 &  89 &  9068 &   639 &  9581 &   714 & $+$1091 &   697 \\
{\rm 51}*  & A2107     &  3 &  34.41 & $+$51.51 & $-$0.0406 & 0.0532 & $-$0.0031 & 12470 &  84 & 10575 &  1291 & 11281 &  1471 & $+$1145 &  1420 \\
{\rm 52}*  & J17       &  3 &  66.25 & $+$49.99 & $-$0.0184 & 0.0715 & $-$0.0067 & 10276 & 146 &  9285 &  1512 &  9824 &  1694 &  $+$436 &  1647 \\
{\bf 53}*  & A2147     & 10 &  28.91 & $+$44.53 & $+$0.0026 & 0.0382 & $-$0.0357 & 10769 & 131 & 10185 &   818 & 10838 &   927 &   $-$66 &   906 \\
{\rm 55}\n & P386-2    &  2 &  40.53 & $+$45.09 & $+$0.1002 & 0.0491 & $-$0.0024 &  9706 &  29 & 11560 &  1302 & 12410 &  1503 & $-$2597 &  1444 \\
{\rm 56}\n & A2148     &  3 &  41.97 & $+$47.23 & $-$0.0736 & 0.0887 & $-$0.0316 & 26359 & 339 & 19195 &  3620 & 21696 &  4643 & $+$4348 &  4364 \\
{\rm 57}*  & J18       &  3 &  39.95 & $+$46.50 & $-$0.0743 & 0.0591 & $-$0.0032 &  9661 & 132 &  7702 &  1045 &  8069 &  1147 & $+$1549 &  1124 \\
{\bf 58}*  & A2151     & 10 &  31.47 & $+$44.64 & $-$0.0469 & 0.0520 & $-$0.0165 & 11194 &  79 &  9422 &  1085 &  9978 &  1218 & $+$1176 &  1182 \\
{\bf 59}*  & J19       &  5 &  29.06 & $+$43.50 & $-$0.0966 & 0.0351 & $-$0.0047 & 12782 & 215 &  9510 &   761 & 10077 &   855 & $+$2616 &   854 \\
{\rm 60}*  & P445-1    &  5 &  31.19 & $+$46.17 & $-$0.0057 & 0.0374 & $-$0.0251 & 13804 & 119 & 12591 &  1015 & 13607 &  1188 &  $+$187 &  1145 \\
{\rm 61}\n & P445-2    &  2 &  28.77 & $+$45.63 & $-$0.0632 & 0.0499 & $+$0.0083 &  4978 &  90 &  4181 &   491 &  4288 &   517 &  $+$680 &   517 \\
\end{tabular}
\end{table*}
\addtocounter{table}{-1}
\begin{table*}
\centering
\caption{\it Continued.}
\begin{tabular}{rlrrrrrrrrrrrrrr}
\head{CAN} & \head{Name} & \head{$N_{\rm g}$} & \head{$l$} & \head{$b$} & \head{$\delta$} & \head{$\Delta\delta$}
& \head{$\epsilon_\delta$} & \head{$cz$} & \head{$\Delta cz$} & \head{$D$} & \head{$\Delta D$} & \head{$cz_{D}$} 
& \head{$\Delta cz_{D}$} & \head{$V$} & \head{$\Delta V$}  \vspace{6pt} \\
{\rm 62}\n & A2162-N   &  5 &  50.36 & $+$46.10 & $-$0.0629 & 0.0499 & $-$0.0066 & 15048 & 154 & 11948 &  1352 & 12859 &  1569 & $+$2098 &  1513 \\
{\rm 63}\n & A2162-S   &  1 &  48.36 & $+$46.03 & $-$0.0892 & 0.0872 & $+$0.0010 &  9837 & 116 &  7570 &  1537 &  7925 &  1685 & $+$1862 &  1646 \\
{\rm 64}\n & J20       &  2 &  56.54 & $+$45.58 & $+$0.0221 & 0.0710 & $-$0.0309 &  9468 & 113 &  9434 &  1433 &  9991 &  1608 &  $-$506 &  1564 \\
{\bf 65}*  & A2197     &  7 &  64.68 & $+$43.50 & $-$0.0566 & 0.0435 & $-$0.0015 &  9159 &  97 &  7627 &   762 &  7987 &   836 & $+$1141 &   820 \\
{\bf 66}*  & A2199     &  7 &  62.92 & $+$43.70 & $+$0.0198 & 0.0320 & $+$0.0001 &  9039 &  93 &  8980 &   664 &  9483 &   741 &  $-$431 &   723 \\
{\rm 67}*  & J21       &  4 &  77.51 & $+$41.64 & $+$0.0174 & 0.0419 & $-$0.0198 & 13876 & 215 & 13343 &  1224 & 14492 &  1447 &  $-$587 &  1398 \\
{\bf 68}*  & A2247     &  7 & 114.45 & $+$31.01 & $-$0.0127 & 0.0576 & $-$0.0060 & 11506 &  67 & 10460 &  1371 & 11150 &  1560 &  $+$342 &  1506 \\
{\rm 69}\n & P332-1    &  1 &  49.95 & $+$35.22 & $-$0.1640 & 0.0874 & $-$0.0217 & 17338 & 403 & 10770 &  2067 & 11503 &  2361 & $+$5619 &  2311 \\
{\bf 70}*  & J22       & 10 &  49.02 & $+$35.93 & $-$0.0581 & 0.0319 & $-$0.0026 & 10396 &  78 &  8566 &   625 &  9023 &   694 & $+$1332 &   678 \\
{\rm 71}*  & J23       &  4 &  85.81 & $+$35.40 & $+$0.0535 & 0.0501 & $+$0.0025 &  8401 & 113 &  9052 &  1054 &  9564 &  1177 & $-$1127 &  1146 \\
{\rm 72}*  & J24       &  3 &  69.51 & $+$32.08 & $-$0.0088 & 0.0722 & $-$0.0096 & 10256 & 136 &  9476 &  1547 & 10038 &  1737 &  $+$210 &  1687 \\
{\rm 73}\n & J25       &  2 &  91.82 & $+$30.22 & $-$0.0565 & 0.0661 & $+$0.0013 &  7863 & 194 &  6598 &  1012 &  6865 &  1096 &  $+$974 &  1088 \\
{\rm 74}\n & J26       &  2 &  69.59 & $+$26.60 & $-$0.0728 & 0.0680 & $-$0.0329 & 14876 & 267 & 11556 &  1668 & 12405 &  1925 & $+$2372 &  1872 \\
{\rm 75}*  & J27       &  4 &  80.41 & $+$23.15 & $-$0.0084 & 0.0541 & $+$0.0001 &  9667 & 233 &  8968 &  1121 &  9470 &  1251 &  $+$190 &  1234 \\
{\rm 76}\n & J38       &  1 &  36.09 & $-$44.90 & $-$0.0256 & 0.0887 & $-$0.0306 & 15056 & 320 & 13027 &  2476 & 14119 &  2914 &  $+$894 &  2809 \\
{\rm 77}*  & P522-1    &  3 &  81.75 & $-$41.26 & $-$0.0770 & 0.0393 & $+$0.0026 &  7362 & 112 &  5910 &   539 &  6124 &   579 & $+$1212 &   578 \\
{\rm 78}*  & A2572     &  4 &  94.28 & $-$38.95 & $+$0.0444 & 0.0374 & $-$0.0075 & 11210 & 136 & 11643 &   984 & 12505 &  1137 & $-$1243 &  1100 \\
{\rm 79}\n & A2589     &  5 &  94.64 & $-$41.23 & $+$0.0751 & 0.0309 & $-$0.0022 & 12078 & 149 & 13399 &   947 & 14558 &  1121 & $-$2365 &  1079 \\
{\bf 80}*  & A2593-N   & 16 &  93.44 & $-$43.19 & $+$0.0328 & 0.0170 & $-$0.0016 & 12034 & 107 & 12113 &   472 & 13051 &   549 &  $-$974 &   537 \\
{\bf 82}*  & A2634     & 10 & 103.50 & $-$33.08 & $-$0.0026 & 0.0225 & $-$0.0021 &  9228 & 113 &  8698 &   448 &  9169 &   498 &   $+$56 &   495 \\
{\bf 83}*  & A2657     &  5 &  96.73 & $-$50.25 & $+$0.0644 & 0.0396 & $-$0.0024 & 11887 & 156 & 12879 &  1170 & 13945 &  1375 & $-$1966 &  1322 \\
{\rm 84}\n & A2666     &  2 & 106.71 & $-$33.80 & $-$0.0365 & 0.0545 & $+$0.0035 &  7620 &  90 &  6704 &   852 &  6981 &   924 &  $+$624 &   907 \\
{\bf 90}*  & COMA      & 19 &  58.00 & $+$88.00 & $+$0.0017 & 0.0173 & $+$0.0018 &  7209 &  50 &  6942 &   277 &  7238 &   302 &   $-$29 &   299 \\
{\rm108}\n & (A168)    &  2 & 134.36 & $-$61.61 & $+$0.0861 & 0.0532 & $+$0.0055 &  4683 & 136 &  5556 &   692 &  5745 &   740 & $-$1042 &   738 \\
{\rm109}\n & (A189)    &  1 & 140.13 & $-$59.99 & $-$0.0118 & 0.0851 & $-$0.0191 &  9165 & 126 &  8461 &  1592 &  8906 &  1765 &  $+$251 &  1721 \\
{\rm120}\n & (A397)    &  1 & 161.84 & $-$37.33 & $+$0.0402 & 0.0709 & $-$0.0314 & 16631 &  15 & 16596 &  2499 & 18423 &  3089 & $-$1688 &  2923 \\
{\rm128}\n & (J35)     &  1 & 217.47 & $-$33.61 & $-$0.0359 & 0.0812 & $+$0.0000 & 16907 &  27 & 14140 &  2668 & 15439 &  3188 & $+$1395 &  3032 \\
{\rm129}\n & (J34/35)  &  1 & 216.40 & $-$34.19 & $-$0.1005 & 0.0830 & $-$0.0466 & 16464 &  18 & 11896 &  2027 & 12798 &  2350 & $+$3515 &  2264 \\
{\rm130}\n & (P777-1)  &  1 & 218.49 & $-$32.70 & $+$0.0377 & 0.0746 & $-$0.0240 & 16755 &  14 & 16615 &  2686 & 18447 &  3322 & $-$1594 &  3141 \\
{\rm131}\n & (P777-2)  &  1 & 220.77 & $-$32.62 & $-$0.0834 & 0.0698 & $+$0.0007 &  9300 & 102 &  7275 &  1178 &  7602 &  1286 & $+$1655 &  1258 \\
{\rm132}\n & (P777-3)  &  2 & 219.72 & $-$31.71 & $-$0.0070 & 0.0482 & $+$0.0000 & 13502 & 144 & 12300 &  1369 & 13268 &  1596 &  $+$224 &  1534 \\
{\rm136}\n & (A548-2)  &  1 & 230.40 & $-$25.97 & $-$0.0031 & 0.0706 & $-$0.0011 &  8826 & 234 &  8329 &  1358 &  8760 &  1503 &   $+$63 &  1478 \\
{\rm138}\n & (J12)     &  1 &  50.52 & $+$78.23 & $-$0.0580 & 0.0740 & $-$0.0124 & 15633 &  16 & 12512 &  2077 & 13515 &  2428 & $+$2026 &  2327 \\
{\rm142}\n & (J14-1)   &  1 &   8.80 & $+$58.73 & $-$0.0625 & 0.0936 & $-$0.0490 & 16121 & 125 & 12736 &  2405 & 13778 &  2819 & $+$2239 &  2711 \\
{\rm144}\n & (A1991)   &  1 &  22.74 & $+$60.52 & $+$0.1807 & 0.1143 & $-$0.0374 & 13372 &  23 & 18781 &  4529 & 21166 &  5774 & $-$7280 &  5426 \\
{\rm145}\n & (J16)     &  3 &   6.81 & $+$48.20 & $-$0.0985 & 0.0553 & $-$0.0119 & 15927 & 109 & 11594 &  1437 & 12449 &  1659 & $+$3338 &  1598 \\
{\rm146}\n & (J16W)    &  1 &   5.08 & $+$49.63 & $-$0.1752 & 0.0870 & $+$0.0000 & 24347 & 221 & 14186 &  2871 & 15494 &  3433 & $+$8417 &  3271 \\
{\rm154}\n & (P386-1)  &  1 &  37.09 & $+$47.81 & $-$0.0679 & 0.1159 & $-$0.0129 & 13910 & 212 & 10988 &  2889 & 11752 &  3309 & $+$2075 &  3195 \\
{\rm160}\n & (P445-1)  &  1 &  31.19 & $+$46.17 & $+$0.1585 & 0.0783 & $-$0.0134 & 10827 & 223 & 14657 &  2569 & 16059 &  3092 & $-$4966 &  2948 \\
{\rm163}\n & (A2162-S) &  3 &  48.36 & $+$46.03 & $+$0.0706 & 0.0750 & $-$0.0138 & 15623 & 273 & 16817 &  2817 & 18697 &  3493 & $-$2894 &  3305 \\
{\rm166}\n & (A2199)   &  1 &  62.92 & $+$43.70 & $-$0.0371 & 0.1774 & $-$0.0719 & 17820 &  44 & 14785 &  5183 & 16213 &  6248 & $+$1524 &  5978 \\
{\rm167}\n & (J21)     &  1 &  77.51 & $+$41.64 & $-$0.0867 & 0.0852 & $+$0.0025 &  6086 &  12 &  4812 &   956 &  4953 &  1013 & $+$1113 &   997 \\
{\rm174}\n & (J26)     &  1 &  69.59 & $+$26.60 & $+$0.0896 & 0.1401 & $-$0.0788 & 18098 &  36 & 20073 &  5312 & 22828 &  6899 & $-$4396 &  6496 \\
{\rm177}\n & (P522-1)  &  1 &  81.75 & $-$41.26 & $+$0.0033 & 0.0710 & $-$0.0300 & 23626 &  66 & 20842 &  3145 & 23832 &  4130 &  $-$191 &  3848 \\
{\rm180}\n & (A2593-N) &  1 &  93.44 & $-$43.19 & $-$0.1306 & 0.0696 & $-$0.0324 & 27458 &   8 & 17433 &  2570 & 19464 &  3215 & $+$7505 &  3034 \\
{\rm210}\n & (J30)     &  1 & 151.84 & $-$75.04 & $+$0.0180 & 0.0676 & $-$0.0234 & 20667 &  12 & 19167 &  2805 & 21659 &  3596 &  $-$925 &  3368 \\
{\rm232}\n & (P777-3)  &  1 & 219.72 & $-$31.71 & $+$0.0855 & 0.0673 & $+$0.0009 &  8929 &   9 & 10327 &  1612 & 10999 &  1831 & $-$1997 &  1766 \\
{\rm238}\n & (J12)     &  1 &  50.52 & $+$78.23 & $-$0.1249 & 0.1134 & $-$0.0422 & 18073 &  27 & 12235 &  2904 & 13192 &  3382 & $+$4674 &  3253 \\
{\rm239}\n & (J13)     &  1 &  28.27 & $+$75.54 & $-$0.0540 & 0.1141 & $-$0.0477 & 17675 &   5 & 14118 &  3316 & 15412 &  3961 & $+$2151 &  3788 \\
{\rm240}\n & (J36)     &  1 & 332.77 & $+$49.31 & $+$0.0445 & 0.0816 & $-$0.0284 & 17559 &  26 & 17607 &  3081 & 19681 &  3863 & $-$1991 &  3641 \\
{\rm244}\n & (A1991)   &  1 &  22.74 & $+$60.52 & $+$0.0315 & 0.1564 & $-$0.1206 & 21381 &  64 & 20373 &  5377 & 23219 &  7014 & $-$1706 &  6638 \\
{\rm245}\n & (J16)     &  1 &   6.81 & $+$48.20 & $-$0.1560 & 0.1049 & $-$0.0198 & 14194 &  39 &  9138 &  2126 &  9660 &  2377 & $+$4392 &  2306 \\
{\rm254}\n & (P386-1)  &  1 &  37.09 & $+$47.81 & $+$0.0711 & 0.2459 & $-$0.1072 & 26704 &  73 & 27084 & 12329 & 32415 & 17745 & $-$5154 & 16417 \\
{\rm260}\n & (P445-1)  &  1 &  31.19 & $+$46.17 & $+$0.1227 & 0.0970 & $-$0.0630 & 18642 &  26 & 22247 &  4181 & 25696 &  5604 & $-$6497 &  5225 \\
{\rm338}\n & (J12)     &  1 &  50.52 & $+$78.23 & $-$0.0493 & 0.0933 & $-$0.0066 & 10666 & 236 &  8955 &  1911 &  9455 &  2132 & $+$1173 &  2081 \\
{\rm339}\n & (J13)     &  1 &  28.27 & $+$75.54 & $+$0.0993 & 0.1160 & $-$0.0367 & 15883 &  30 & 18235 &  4476 & 20472 &  5662 & $-$4296 &  5331 \\
{\rm340}\n & (J36)     &  1 & 332.77 & $+$49.31 & $-$0.1612 & 0.1989 & $+$0.0000 & 29241 &  28 & 17136 &  8373 & 19093 & 10428 & $+$9540 &  9804 \\
\end{tabular}
\vspace{12pt} \\
\parbox{0.95\textwidth}{Note: clusters in the Fundamental Plane sample
(Table~\ref{tab:fpclus}) have their CANs in bold; clusters in the
peculiar velocity sample are marked with an asterisk.}
\\ \vspace{5mm}
\parbox{0.95\textwidth}{This table is also available as
J/MNRAS/\textit{vol}/\textit{page} from NASA's Astrophysical Data Centre
(ADC, http://adc.gsfc.nasa.gov) and from the Centre de Donn\'{e}es
astronomiques de Strasbourg (CDS, http://cdsweb.u-strasbg.fr).}
\end{table*}

\subsection{Tests and comparisons}

Gibbons \etal\ (2000) have suggested that the large peculiar velocities
found for some clusters are due to poor FP fits. For a heterogeneous
sample of 20 clusters drawn from their own observations and the
literature, they find that nearly half are poorly-fit by a FP and have
twice the rms scatter of the well-fit clusters. The half of their
clusters that have good FP fits all have peculiar velocities that are
consistent with them being at rest in the CMB frame; the poorly-fit
clusters show a much larger range of peculiar velocities. Gibbons \etal\
suggest that the large peculiar velocities detected for some clusters
may result from those clusters being poorly fit (for whatever reason) by
the global FP. The origin of the poor fits is not known, but the
possibilities include intrinsic FP variations between clusters, failure
to identify and remove interlopers, observational errors, the
heterogeneity of the data, and combinations of these effects.

\begin{figure}
\plotone{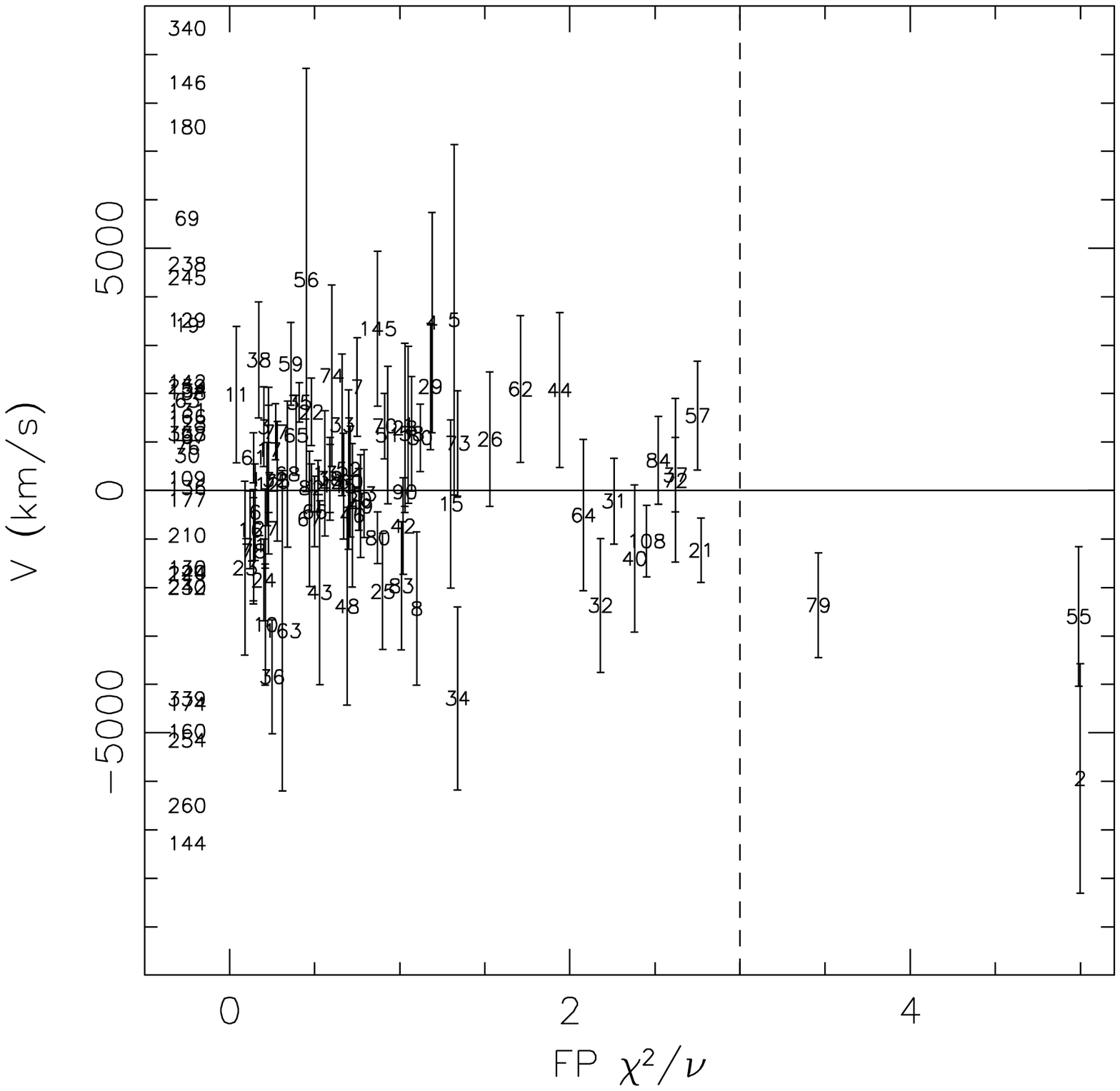}
\caption{Cluster peculiar velocities as a function of the
goodness-of-fit of their best-fit FP. Clusters indicated by their CANs;
those with only single members have no $\chi^2/\nu$ and are plotted at
the left of the figure to show their peculiar velocities.}
\label{fig:mergechi2}
\end{figure}

We therefore need to test whether some of the peculiar velocities we
derive from the EFAR dataset are due to poor fits to the FP rather than
genuine peculiar velocities. Figure~\ref{fig:mergechi2} shows the
peculiar velocities of the EFAR clusters as a function of the
goodness-of-fit of their best-fit FP (as measured by the reduced
$\chi^2$ statistic). As noted above, even after removing outliers, there
are still three clusters with very poor FP fits ($\chi^2/\nu$$>$3; in
fact CAN~2=A85 actually has $\chi^2/\nu$=11, but is plotted at
$\chi^2/\nu$=5 for convenience). All three of these clusters have large
negative peculiar velocities, detected at nominal significance levels of
1.8--2.5$\sigma$. The poor quality of the FP fits raises considerable
doubts about the reality of the peculiar velocity estimates, however,
and we therefore omit these clusters from all subsequent analysis. The
remaining clusters generally have acceptable fits
($\chi^2/\nu$$\approx$1). There are 10 clusters with $\chi^2/\nu$=2--3,
but none of these have significant peculiar velocities (the strongest
detection is at the 1.8$\sigma$ level). Apart from the three clusters
with $\chi^2/\nu$$>$3, the clusters are all adequately fitted by the
global FP, and there is no evidence for any increased scatter in the
peculiar velocities for poorer FP fits.

Another possible source of systematic errors are the small biases in the
recovered parameters of the best-fitting FP (see \S\ref{fp} above). If
we apply the corrections for these biases derived from our simulations
(Figure~\ref{fig:simresults}) and re-derive the peculiar velocities with
this bias-corrected FP, we find that the peculiar velocities of the
clusters are not significantly altered: the peculiar velocity of Coma
changes by +14\kms, and the rms difference in peculiar velocity between
our standard solution and the bias-corrected solution is only 67\kms.

\begin{figure}
\plotone{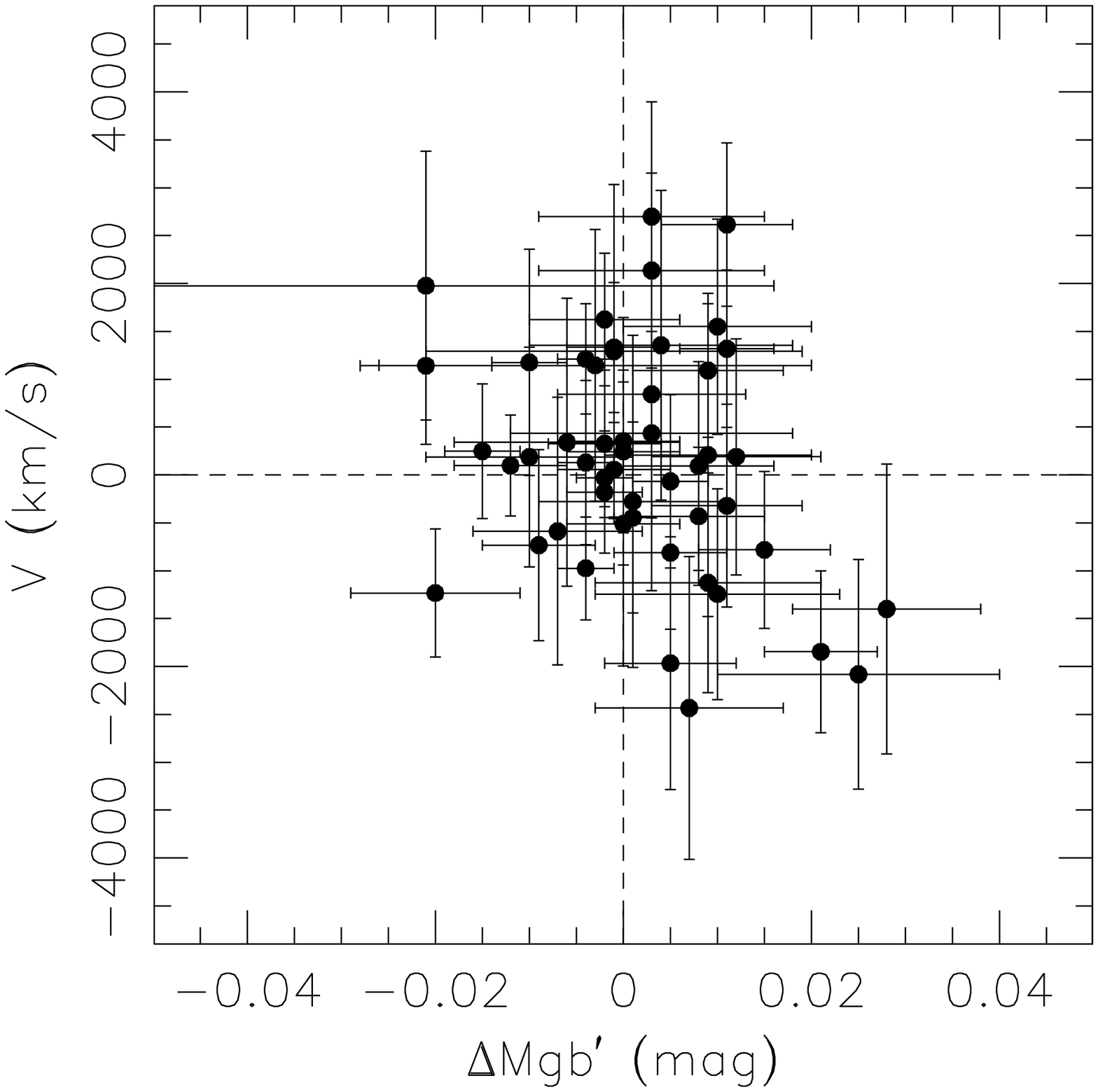}
\caption{The distribution of cluster peculiar velocities with respect to
the cluster offsets from the global Mg--$\sigma$ relation of Paper~V.}
\label{fig:dmgdfp}
\end{figure}

We can also attempt to test whether differences in the mean stellar
populations between clusters produce spurious peculiar velocities, by
looking for a correlation between the peculiar velocities and the offset
of each cluster from the global Mg--$\sigma$ relation derived in
Paper~V. The correlation coefficient for the distribution (shown in
Figure~\ref{fig:dmgdfp}) is $-$0.30, but 1000 simulations of the
observed distribution show that, allowing for the estimated errors, this
value does not indicate a correlation significant at the 95\% level.
However, while there is no positive evidence that stellar population
differences are leading to spurious peculiar velocities, this test
cannot rule out this possibility. Figure~10 of Paper~V shows that the
joint distribution of residuals about the FP and Mg--$\sigma$ relations
is consistent with simple stellar population models if one invokes
sufficiently large (and possibly correlated) scatter in the ages and
metallicities of the galaxies. Against this possibility we can set the
generally good agreement between the distance estimates obtained from
the FP and other methods (such as the Tully-Fisher relation and surface
brightness fluctuations) which have different dependences on the stellar
populations.

\begin{figure}
\plotone{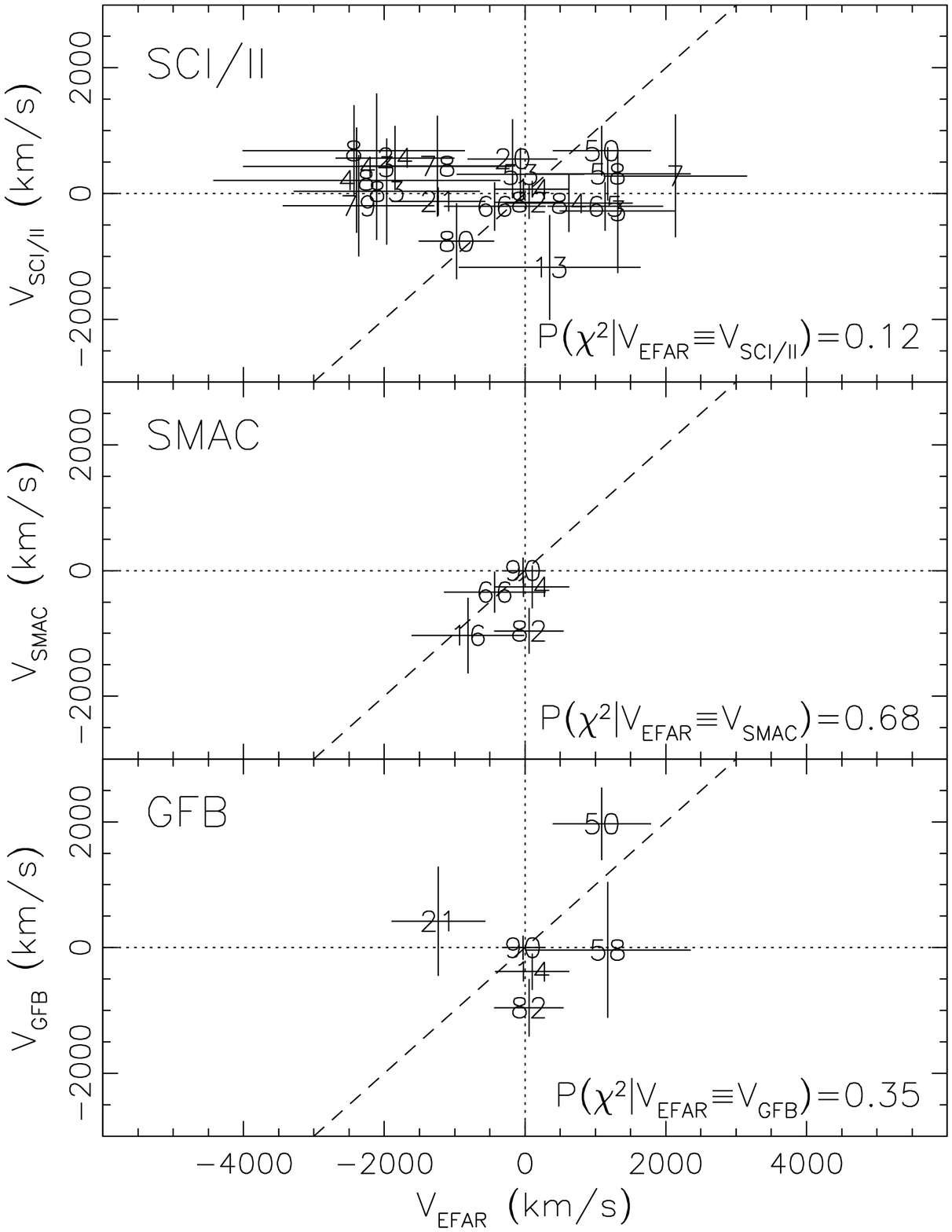}
\caption{Comparisons of EFAR peculiar velocities of clusters in common
with SCI/II (Giovanelli \etal\ 1998b, Dale \etal\ 1999b), SMAC (Hudson
\etal\ 1997) and GFB (Gibbons \etal\ 2000).}
\label{fig:vcomp}
\end{figure}

Finally, we can perform a direct comparison between the peculiar
velocities we measure and those obtained by other groups for the same
clusters. Figure~\ref{fig:vcomp} shows comparisons with the Tully-Fisher
estimates of Giovanelli \etal\ (1998b; SCI) and Dale \etal\ (1999b; SCII),
and the FP estimates of Hudson \etal\ (1997; SMAC) and Gibbons \etal\
(2000; GFB). The flattening in the $V_{\rm EFAR}$--$V_{\rm other}$
distributions is due to the fact that the uncertainties in the EFAR
peculiar velocities are generally larger than those of the other
measurements---although the error per galaxy is similar in all cases,
the EFAR sample typically has a smaller number of galaxies per cluster.
A $\chi^2$-test shows that the peculiar velocity measurements are
consistent within the errors in all three comparisons.

\section{BULK MOTIONS}
\label{motions}

\subsection{Cluster sample}

\begin{figure}
\plotone{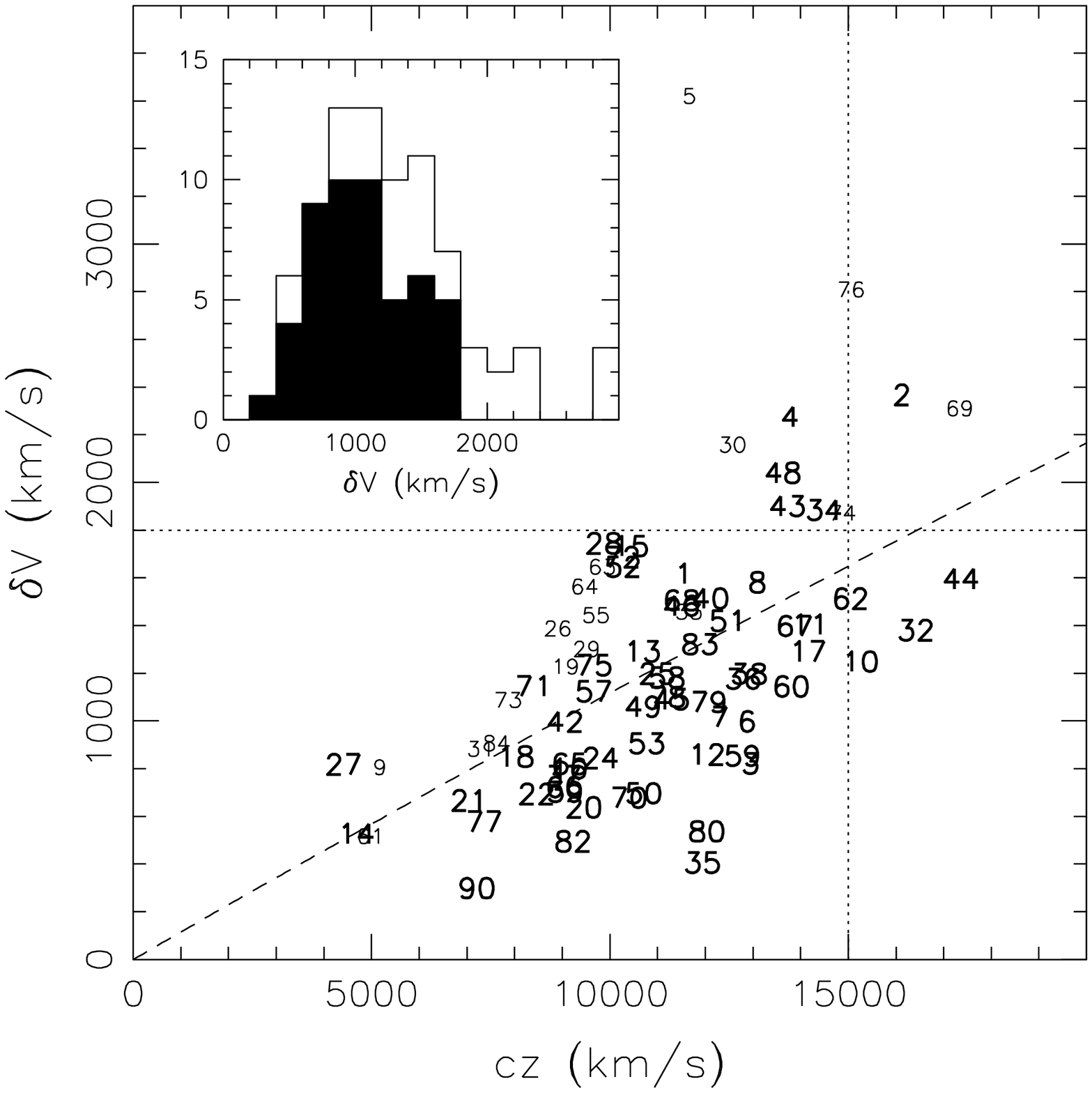}
\caption{The selection of clusters for the peculiar velocity analysis.
The cluster's peculiar velocity errors are plotted as a function of
their redshifts. Each cluster is marked by its CAN, with clusters having
3 or more galaxies in a larger font; fore- and background groups
(CAN$>$100) are not shown. The selection limits in $cz$ and $\delta V$
are indicated by the dotted lines. The distribution of peculiar velocity
errors is shown in the inset: the open histogram is for all clusters,
the filled histogram for the selected clusters.}
\label{fig:pvsel}
\end{figure}

In analysing the peculiar motions of the clusters in the EFAR sample we
confine ourselves to the subsample of clusters with 3 or more galaxies
($N_g$$\ge$3), $cz$$\le$15000\kms\ and $\delta V$$\le$1800\kms. These
criteria are illustrated in Figure~\ref{fig:pvsel}, and are chosen
because: (i)~they eliminate all the fore- and background clusters, for
which the selection functions have not been directly measured and are
only poorly approximated by the selection function of the main cluster
onto which they are projected; (ii)~they eliminate the clusters with
only 1 or 2 galaxies in the FP fit, where it is not possible to check if
galaxies are cluster interlopers or FP outliers; (iii)~they eliminate
the higher-redshift clusters, which have proportionally higher
uncertainties in their peculiar velocities (and in any case sample the
volume beyond $cz$=15000\kms\ too sparsely to be useful); (iv)~they
eliminate clusters with large uncertainties in their peculiar
velocities, resulting from large measurement errors for individual
galaxies exacerbated by a small number of galaxies in the
cluster---restricting the subsample to $\delta V$$\le$1800\kms\ (the
peculiar velocity error for a cluster with a FP distance from 3 galaxies
with a distance error per galaxy of 20\%) represents a compromise
between using clusters with better-determined peculiar velocities and
keeping the largest possible cluster sample.

We also eliminate from the sample the three clusters which were
identified in the previous section as having unacceptably poor FP fits
(CAN~2=A85, CAN~55=P386-2 and CAN~79=A2589); two of these would be
eliminated in any case: A85 because it has $cz$$>$15000\kms, and P386-2
because it has only two galaxies. We also eliminate the two components
of A548 (CAN~35=A548-1 and CAN~36=A548-2), since the substructure in
this region (Zabludoff \etal\ 1993, Davis \etal\ 1995) makes cluster
membership problematic and since the high relative velocity of the two
main subclusters is not relevant to the large-scale motions we are
investigating (Watkins 1997).

\begin{figure}
\plotone{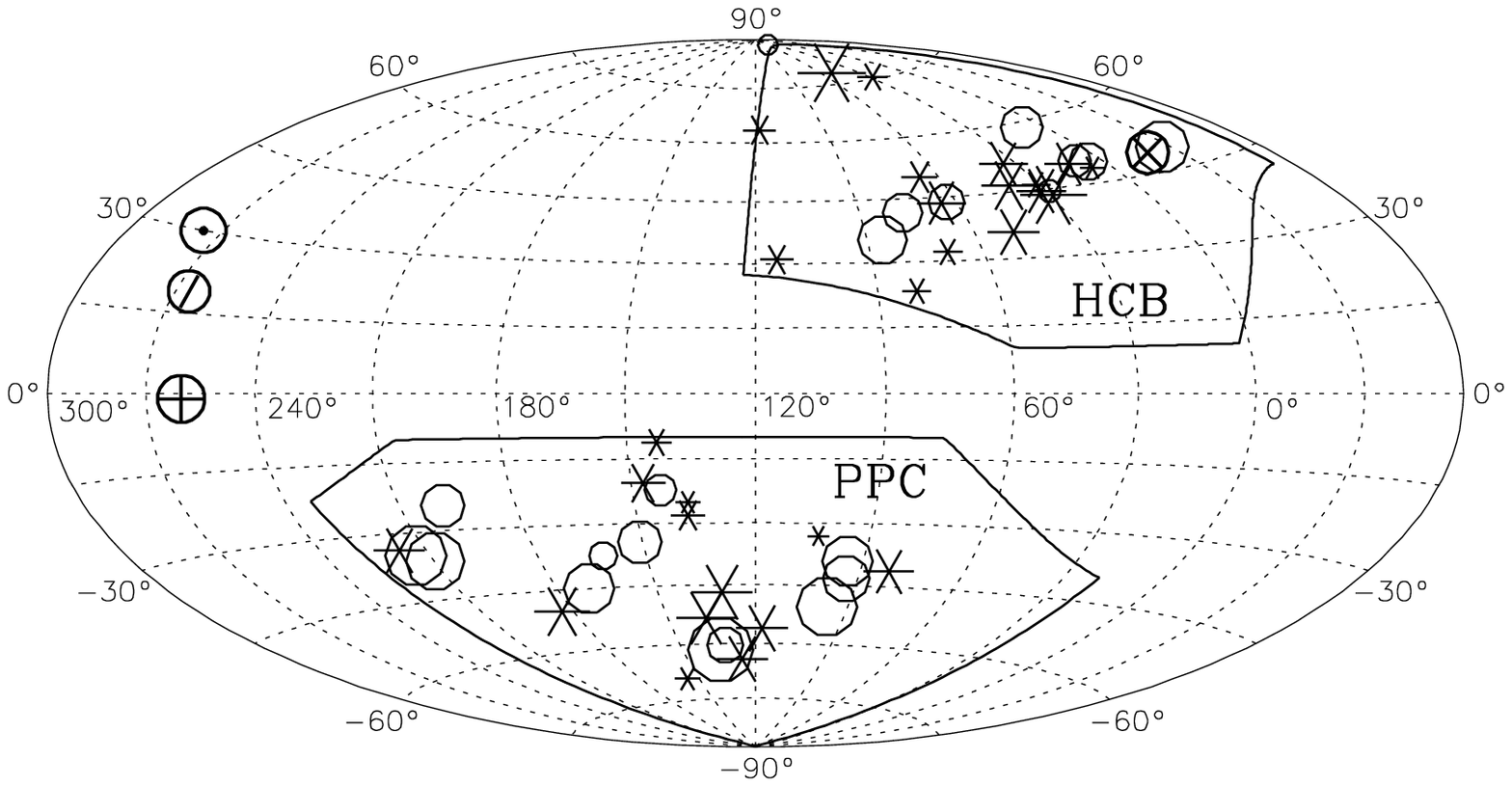}
\caption{The projection on the sky in Galactic coordinates of the EFAR
peculiar velocities in the CMB frame. Clusters with positive (negative)
peculiar velocities are indicated by asterisks (circles); marker sizes
are related to the amplitude of the peculiar velocity. Other markers
show the directions with respect to the CMB frame of the Local Group
dipole ($\odot$), the Lauer \& Postman (1994) dipole ($\otimes$), the
SMAC (Hudson \etal\ 1999) dipole ($\oplus$), and the LP10K (Willick
1999) dipole ($\oslash$).}
\label{fig:pvsky}
\end{figure}

The subsample selected in this way for the analysis of the peculiar
motions comprises 50 clusters (25 in HCB, 25 in PPC); they are indicated
by an asterisk in Table~\ref{tab:efarpv}. The distribution of the
peculiar velocity uncertainties for this subsample is shown in the inset
to Figure~\ref{fig:pvsel}); the median peculiar velocity error is
1060\kms. Figure~\ref{fig:pvsky} shows the projection of the sample on
the sky in Galactic coordinates, with the amplitude of the clusters'
peculiar velocities in the CMB frame indicated by the size of the
symbols. Inflowing clusters (circles) and outflowing clusters
(asterisks) are fairly evenly distributed over the survey regions. The
median direction of the clusters belonging to the peculiar velocity
sample in the HCB region is ($l$,$b$)=(42\degree,48\degree), and in the
PPC region is ($l$,$b$)=(152\degree,-36\degree); the angle between these
two directions is 128\degree.

\subsection{Bulk motions}

\begin{figure}
\plotone{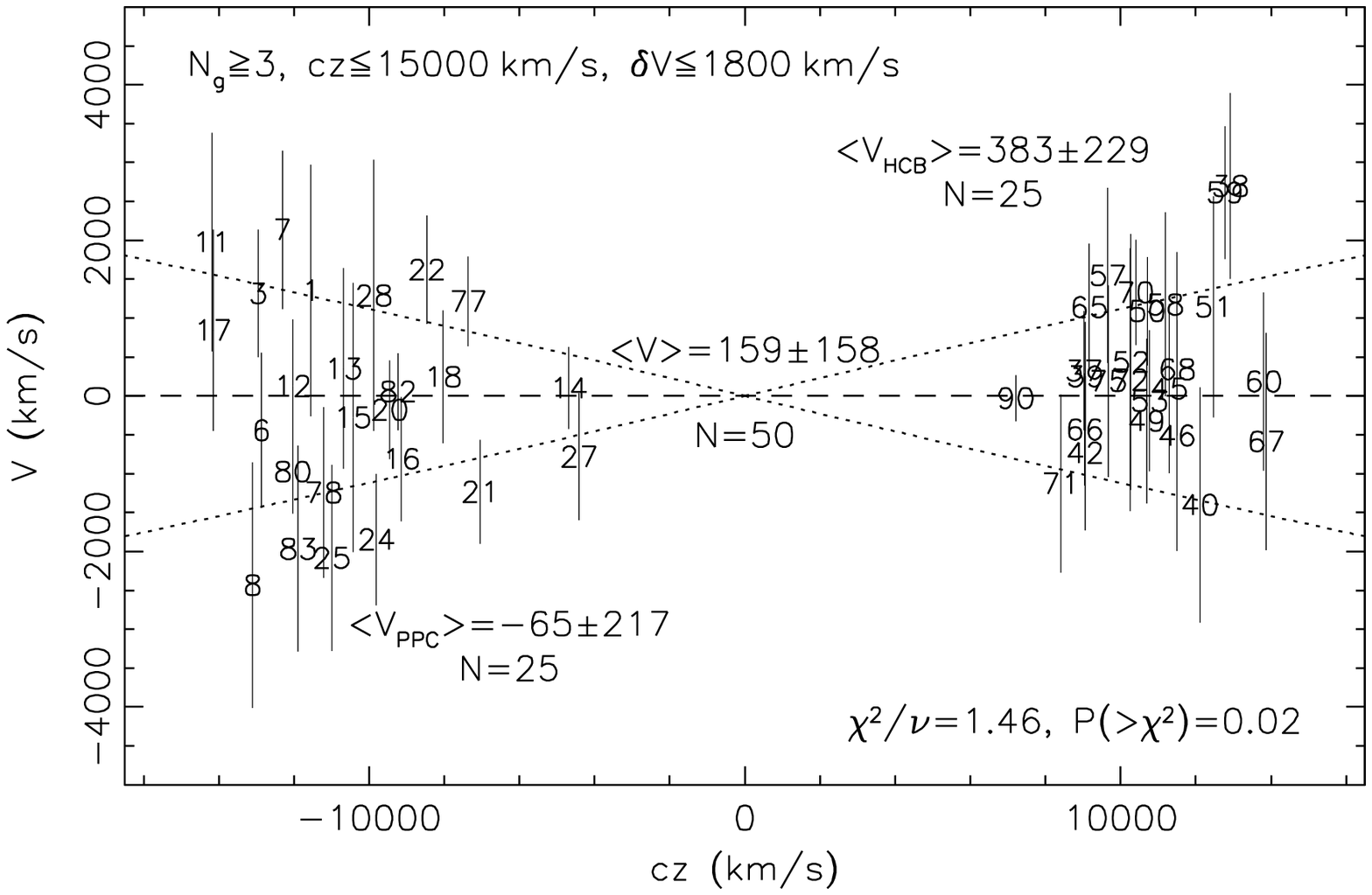}
\caption{Peculiar velocities of the EFAR clusters as a function of
redshift. The clusters in PPC are given negative redshifts, but in all
cases positive peculiar velocities indicate outflow and negative
peculiar velocities inflow. Both redshifts and peculiar velocities are
in the CMB frame. Clusters are indicated by the CANs. Peculiar velocity
errors are shown, but redshift errors (which are small) are omitted for
clarity. The dotted curves correspond to the typical $\pm$1$\sigma$
peculiar velocity errors for clusters with peculiar velocities based on
3 galaxies. The unweighted mean peculiar velocity, and the number of
sample clusters, are shown for the HCB and PPC regions separately and
for the sample as a whole. The $\chi^2$ probability that the observed
peculiar velocities are consistent with strictly zero motions is also
given.}
\label{fig:pvcz}
\end{figure}

The peculiar velocities of the sample clusters as a function of redshift
are shown in Figure~\ref{fig:pvcz}. The mean peculiar velocity of the
whole sample ($\langle V \rangle$=159$\pm$158\kms) is consistent, within
the errors, with no net inflow or outflow. This need not have been the
case, as the FP zeropoint is based on the 29 clusters listed in
Table~\ref{tab:fpclus}, which make up only 26 of the 50 clusters in the
peculiar velocity sample. The mean peculiar velocities of each of the
two sample regions separately are also consistent with zero inflow or
outflow: $\langle V_{\rm HCB} \rangle$=$+$383$\pm$229\kms; $\langle
V_{\rm PPC} \rangle$=$-$65$\pm$217\kms. A $\chi^2$ test shows that the
observed peculiar velocities are consistent with strictly zero motions
(i.e.\ no bulk or random motions at all) at the 2\% level. If the one
cluster with a 3$\sigma$ peculiar velocity detection (J19, CAN=59) is
omitted, this rises to 8\%. If the peculiar velocity errors were
under-estimated by 5\% (10\%), then the fit is consistent at the 6\%
(15\%) level. If random thermal motions with an rms of 250\kms\
(500\kms) are assumed, then the fit is consistent at the 5\% (30\%)
level. There is, therefore, no evidence in the EFAR sample for
significant bulk motions in the HCB or PPC volumes.

\begin{figure}
\plotone{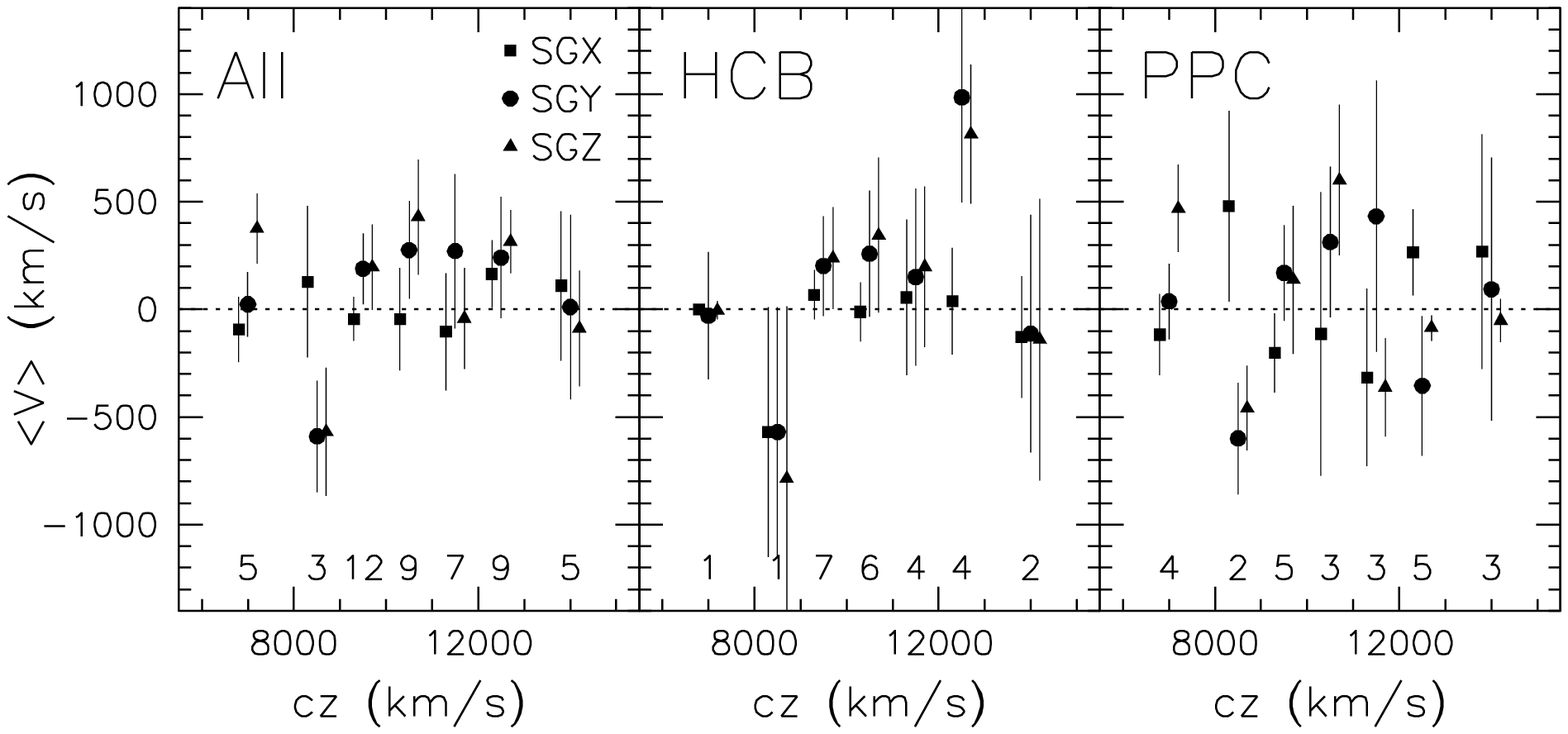}
\caption{The mean peculiar velocity in radial shells. The clusters are
grouped into 7 redshift ranges: the first is 4000--8000\kms, the next
five cover 8000\kms\ to 13000\kms\ in 1000\kms\ steps, and the last is
13000--15000\kms. The left panel shows the whole sample of 50 clusters,
the middle panel shows the 25 HCB clusters, and the right panel shows
the 25 PPC clusters. The Supergalactic X, Y and Z components are shown
as filled squares, circles and triangles respectively (with small
offsets in redshift for clarity). The number of clusters in each
redshift range is indicated at the bottom of each panel.}
\label{fig:pvbin}
\end{figure}

The components in Supergalactic coordinates of the mean peculiar
velocity in redshift shells are shown in Figure~\ref{fig:pvbin}. There
is no sign of any trend with redshift in the mean peculiar velocity,
either for the whole sample or for the two regions separately. None of
the components of the mean peculiar velocity are significant in any
redshift bin apart from the 12000--13000\kms\ bin in HCB, which is due
to J19 (CAN=59)---cf.\ Figure~\ref{fig:pvcz}.

\begin{figure}
\plotone{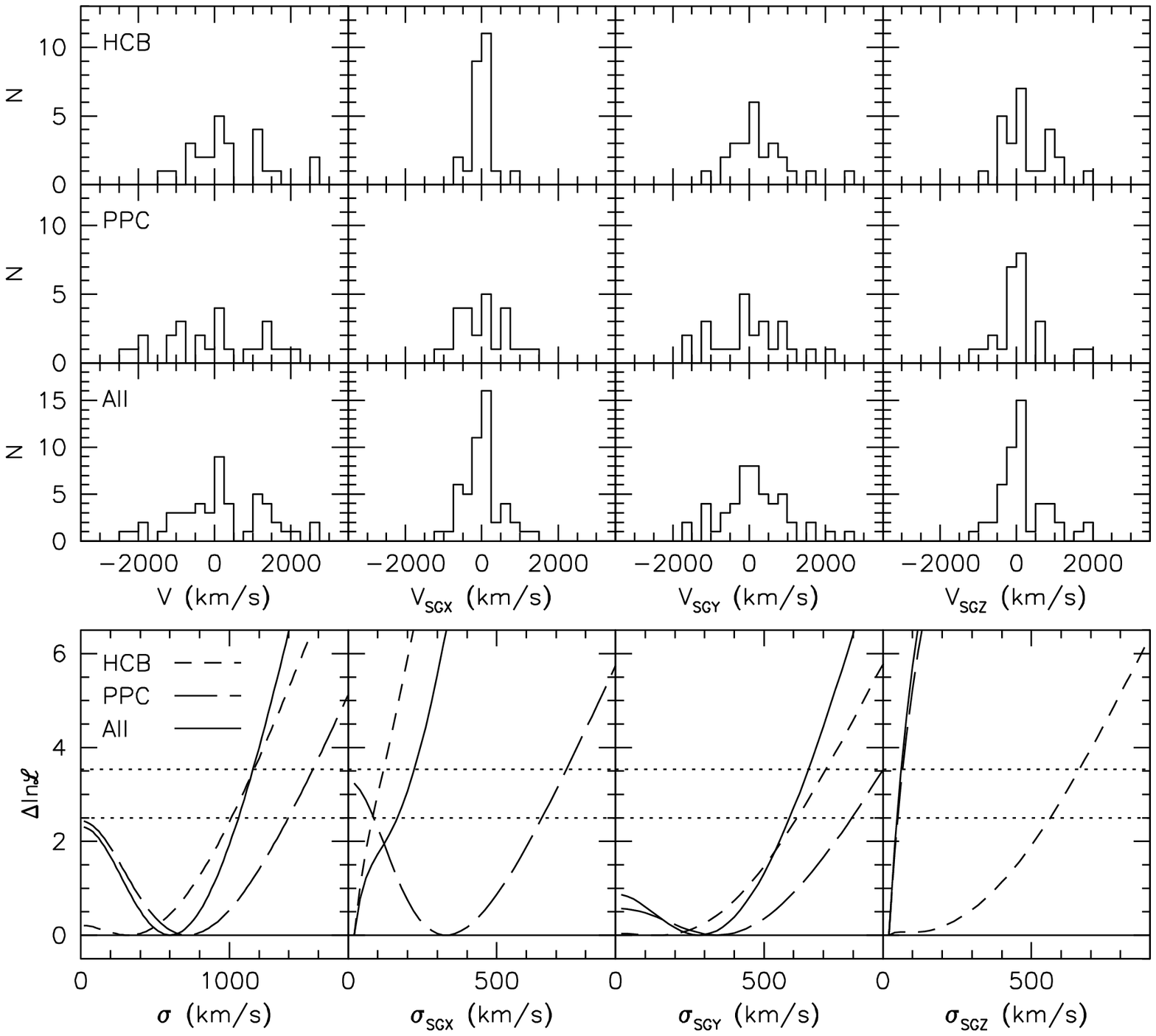}
\caption{Upper panels: The histograms of the peculiar velocities for the
HCB and PPC regions and the whole sample, both radially and projected in
Supergalactic (X,Y,Z) coordinates. Lower panels: The relative
likelihood, $\Delta\ln{\cal L}=\ln{\cal L}_{\rm max}-\ln{\cal L}$, as a
function of the assumed intrinsic dispersion, both overall and in each
Supergalactic coordinate. The solid curve is for the whole sample; the
dashed and long-dashed curves are for the HCB and PPC regions
respectively. The upper dotted line is the 1$\sigma$ confidence level
for the whole sample, while the lower dotted line is the 1$\sigma$
confidence level for both individual regions.}
\label{fig:rmslike}
\end{figure}

We can estimate the intrinsic dispersion of the peculiar velocity field
using the maximum likelihood approach described in Paper~VI (see
Section~2.1 and Appendix~A; cf.\ Watkins 1997). The upper panels of
Figure~\ref{fig:rmslike} show the distributions of peculiar velocities,
both radially and in Supergalactic coordinates, for the HCB and PPC
regions separately and for the whole sample. The peculiar velocities in
all cases have means close to zero, and the question is how large an
intrinsic dispersion is required, combined with the observational
uncertainties, to reproduce the observed scatter in the peculiar
velocities. The lower panels of Figure~\ref{fig:rmslike} show the
relative likelihood, $\Delta\ln{\cal L}=\ln{\cal L}_{\rm max}-\ln{\cal
L}$, as a function of the assumed intrinsic dispersion. The most likely
estimate of the three-dimensional velocity dispersion for the whole
sample is about 600\kms, but the 1$\sigma$ range is 0--1200\kms. The
most likely dispersions for the HCB and PPC regions separately are about
300\kms\ and 700\kms\ respectively. Hence the intrinsic dispersion of
the clusters' peculiar velocities is not well-determined by this data,
due to the large uncertainties in the observed peculiar velocities.

\subsection{Comparisons with other results}

\begin{figure}
\plotone{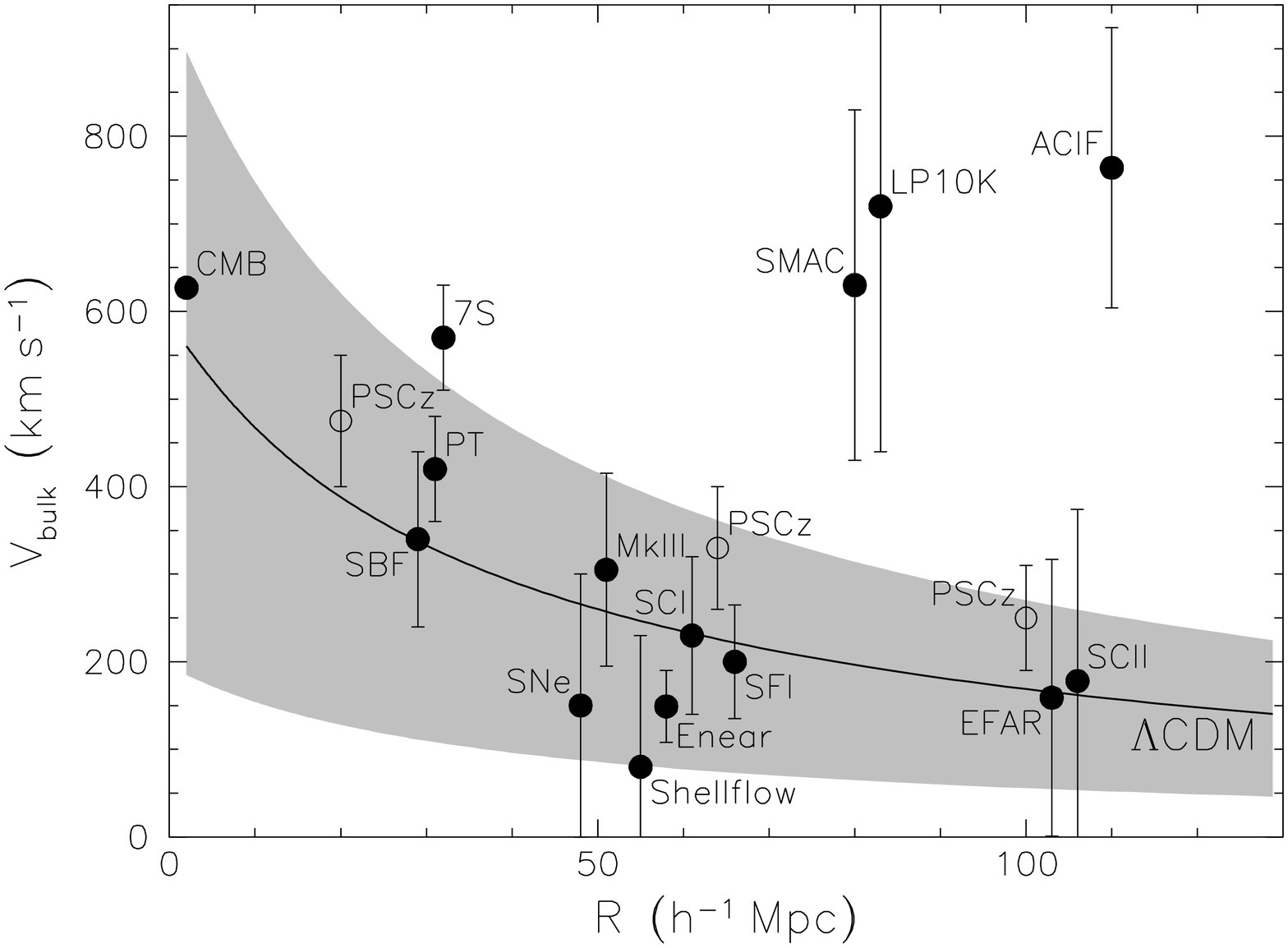}		
\caption{Bulk motion amplitude as a function of scale. The theoretical
curve is the expectation for the bulk motion within a spherical volume
of radius $R$ in a $\Lambda$CDM model ($\Gamma$=0.25, $\sigma_8$=1.0,
$h$=0.7); the grey region shows the 90\% range of cosmic scatter. The
bulk motions determined in various studies are shown at the `effective
scale' of each sample (which is generally only approximate). The bulk
motions shown are for the Local Group w.r.t.\ the CMB (Kogut \etal
1993), 7S (Lynden-Bell \etal\ 1988), ACIF (Lauer \& Postman 1994; Colless
1995), SFI (Giovanelli \etal\ 1998a), SCI (Giovanelli \etal\ 1998b),
SCII (Dale \etal\ 1999a), MkIII (Dekel \etal\ 1999), SMAC (Hudson et al.\
1999), LP10K (Willick 1999), Shellflow (Courteau \etal\ 2000), SNe
(Riess 2000), ENEAR (da Costa \etal\ 2000), SBF (Tonry \etal\ 2000), PT
(Pierce \& Tully 2000), and EFAR (this work). Also shown are the
predicted bulk motions derived from the PSCz redshift survey (Saunders
\etal\ 2000; Dekel 2000).}
\label{fig:pvscale}
\end{figure}

A comparison of the EFAR bulk motion to other measurements of bulk
motions on various scales, and to theoretical predictions, is given in
Figure~\ref{fig:pvscale}. The figure shows the reported bulk motions
from a number of other observational studies as a function of the
effective scale of the sample. Also shown is the theoretical prediction
for the bulk motion measured with a top-hat window function of radius
$R$ (in h$^{-1}$\,Mpc) for a fairly `standard' flat $\Lambda$CDM
cosmology having a power spectrum with shape parameter $\Gamma$=0.25,
normalisation $\sigma_8$=1.0 and Hubble constant $h$=0.7 (corresponding
to $\Omega_0$=0.36 and $\Omega_\Lambda$=0.64; see, e.g., Coles \&
Lucchin 1995, p.399).

This comparison is limited by a number of factors: (i)~The finite,
sparse and non-uniform observed samples do not have top-hat window
functions, and their effective scales $R$ are not well-defined (compare
this figure with the similar figure in Dekel (2000)); this uncertainty
is ameliorated by the slow decrease in the expected bulk motion with
scale. (ii)~Only the amplitudes of the bulk motions are compared, and
not the directions; however, the observed bulk motions that are
significantly different from zero have a common direction to within
about 30\degree, close to the direction of the CMB dipole. (iii)~The
uncertainties in the measured bulk motions are only crudely estimated in
some studies, and ignore or under-estimate the systematic biases.
Despite these limitations, the figure does show that, allowing for both
observational uncertainties and cosmic variance, the measured bulk
motions are in most cases quite consistent with the theoretical
predictions (which vary relatively little for any model that is
consistent with the currently-accepted ranges of the cosmological
parameters). In this section and the next we determine the extent to
which the EFAR results are consistent with the models and with the
possibly-discrepant results of Lauer \& Postman (1994; ACIF) and Hudson
\etal\ (1999; SMAC). The bulk flow obtained by Willick (1999; LP10K) is
similar to the SMAC result, and is not considered explicitly.

\begin{figure}
\plotone{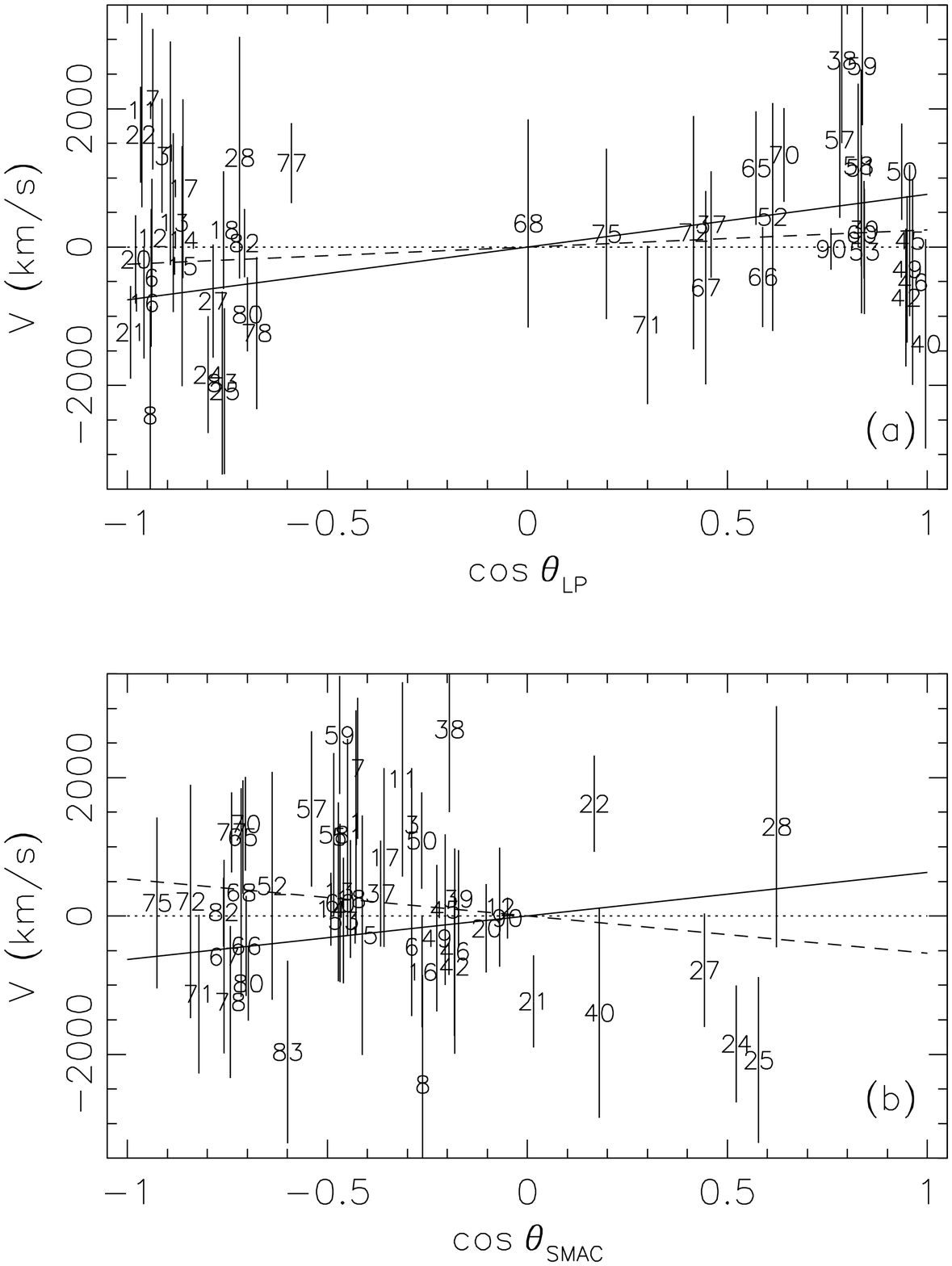}
\caption{The peculiar velocities of the EFAR clusters versus the cosine
of their angle with respect to the direction of (a)~the Lauer \& Postman
dipole, ($l$,$b$)=(341\degree,49\degree), and (b)~the SMAC dipole,
($l$,$b$)=(260\degree,$-$1\degree). Each cluster is indicated by its
CAN. The solid line shows the claimed relation; the dotted line is the
best fit to the EFAR data (see text for details).}
\label{fig:complpsmac}
\end{figure}

We can test whether the observed EFAR peculiar velocity field is
consistent with the bulk motions claimed by other authors. The bulk
motion of the Lauer \& Postman (1994) cluster sample in the CMB frame,
based on brightest cluster galaxy distances as re-analysed by Colless
(1995), is 764\kms\ in the direction ($l$,$b$)=(341\degree,49\degree).
This direction is only 39\degree\ from the median direction of the HCB
clusters in the EFAR sample, and its antipole is just 15\degree\ from
the median direction of the PPC clusters. Consequently the EFAR sample
is able to provide a strong test of the existence of the Lauer \&
Postman bulk motion. Figure~\ref{fig:complpsmac}a shows the peculiar
velocities of the EFAR sample as a function of the cosine of their angle
with respect to the direction of the Lauer \& Postman dipole. The
best-fit bulk flow in the Lauer \& Postman direction has
$V$=250$\pm$209\kms, and is consistent with zero at the 1.2$\sigma$
level. A $\chi^2$~test finds that a pure Lauer \& Postman bulk motion of
764\kms\ in this direction is consistent with the data at only the 0.2\%
level.

The bulk motion of the SMAC sample, for which peculiar velocities are
derived from FP distances by Hudson \etal\ (1999), is 630\kms\ in the
direction ($l$,$b$)=(260\degree,$-$1\degree). The median direction of
the HCB clusters is 57\degree\ from the antipole of this motion, and the
median direction of the PPC clusters is 76\degree\ from the antipole.
Hence the EFAR sample is less well-suited to testing for bulk motions in
this direction. Nonetheless, the formal rejection of the SMAC motion is
even stronger than for the Lauer \& Postman motion.
Figure~\ref{fig:complpsmac}b shows the peculiar velocities of the EFAR
sample as a function of the cosine of their angle with respect to the
SMAC dipole. The best-fit bulk flow along the SMAC direction has
$V$=$-$536$\pm$330\kms\ (i.e.\ in the opposite direction), and is
consistent with zero bulk motion at the 1.6$\sigma$ level. A
$\chi^2$~test finds that a pure SMAC bulk motion of 630\kms\ in this
direction is consistent with the data at only the 0.04\% level.

It is worth noting that an observed bulk flow amplitude of zero would be
consistent with the Lauer \& Postman flow at less than the 0.2\% level,
but consistent with the SMAC flow at the 3.2\% level---if the real bulk
flow is small, therefore, the apparently high significance of the
rejection of the SMAC flow may be the result of the large uncertainty in
the observed amplitude of the flow.

\begin{figure}
\plotone{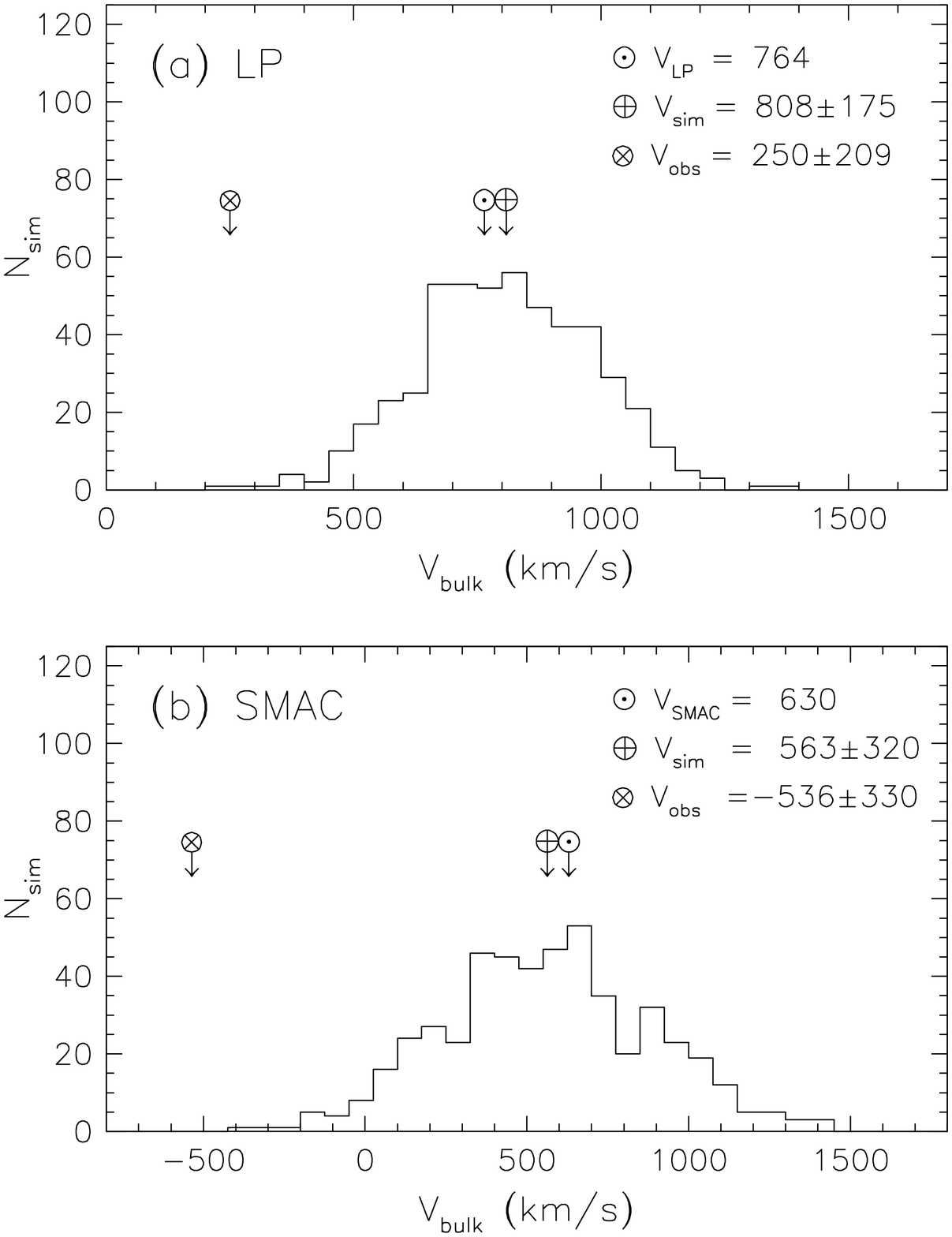}
\caption{Simulations of the recovery from the EFAR dataset of (a)~the
Lauer \& Postman (1994; LP) bulk flow and (b)~the Hudson \etal\ (1999;
SMAC) bulk flow. The histograms are the distributions of the recovered
bulk flow amplitude in the directions of the LP and SMAC dipoles. The
labelled arrows show the true amplitude ($V_{\rm LP}$ or $V_{\rm
SMAC}$), the mean of the recovered amplitudes ($V_{\rm sim}$) and its
rms scatter, and the observed amplitude ($V_{\rm obs}$) and its
uncertainty.}
\label{fig:simflows}
\end{figure}

These $\chi^2$~tests do not take into account the correlated errors in
the peculiar velocity estimates. We therefore carry out Monte Carlo
simulations of the EFAR dataset, including the effects of the correlated
errors, in order to check the consistency of the observed peculiar
velocities with the claimed bulk flows of Lauer \& Postman (LP) and
Hudson \etal\ (SMAC). Figure~\ref{fig:simflows} shows the distributions
of the bulk flow amplitudes recovered from 500 simulations of the LP and
SMAC bulk motions. The mean values of the recovered bulk flow amplitude
($V_{\rm sim}$) are very close to the true values ($V_{\rm LP}$ or
$V_{\rm SMAC}$), although in each case there is a small but
statistically significant bias. However the value of the bulk flow
amplitude derived from the actual EFAR dataset ($V_{\rm obs}$) is in
both cases far out on the wing of the distribution: only one of the 500
simulations of the Lauer \& Postman flow, and none of the 500
simulations of the SMAC flow, yields a bulk flow amplitude less than the
observed value. Hence the observations are consistent with a pure Lauer
\& Postman bulk flow only at the 0.2\% level, and with a pure SMAC bulk
flow at less than the 0.2\% level. The correlated errors in the peculiar
velocities do not significantly alter the results obtained from the
$\chi^2$~tests.

\subsection{Comparisons with theoretical models}

The above comparisons assume pure bulk flows and ignore the greater
complexity of the real velocity field. We can make more realistic
comparisons if we adopt a more detailed model for the velocity field. In
principle this approach also allows us to use the observed peculiar
velocities to discriminate between different cosmological models. The
velocity field models are characterised by a mass power spectrum, which
determines the velocity field on large scales where the dynamics are
linear, and a small-scale rms `thermal' motion, $\sigma_*$, which
approximates the effects of non-linear dynamics on small scales. Given
such a model, the method for computing the expected bulk flow in a
particular sample, and for estimating the probability of an observed
bulk flow, has been developed by Kaiser (1988) and Feldman \& Watkins
(1994, 1998).

As shown by Feldman \& Watkins (1994), the covariance matrix for the
maximum likelihood estimator of the bulk flow in a sample is given by
the sum of a `noise' term, which depends on the spatial distribution of
the clusters, the errors in their peculiar velocities and the thermal
rms motions, and a `velocity' term, which also depends on the power
spectrum of the assumed cosmological model. We adopt a thermal rms
motion of $\sigma_*$=250\kms. Although this value is not well-determined
it has little effect on the results (as we show below), since it enters
in quadrature sum with the uncertainties on the cluster peculiar
velocities, which are generally much larger (see
Table~\ref{tab:efarpv}). Our adopted cosmological model has a CDM-like
power spectrum with $\Gamma$=0.25 and $\sigma_8$=1.0, consistent with
the power spectrum measured from the APM galaxy survey (Baugh \&
Efstathiou 1993) and the PSCz redshift survey (Sutherland \etal\ 1999).
This corresponds to the currently-favoured flat $\Lambda$CDM cosmology
with $H_0$$\approx$70\,km\,s$^{-1}$\,Mpc$^{-1}$, $\Omega_0$$\approx$0.35
and $\Omega_\Lambda$$\approx$0.65.

\begin{figure}
\plotone{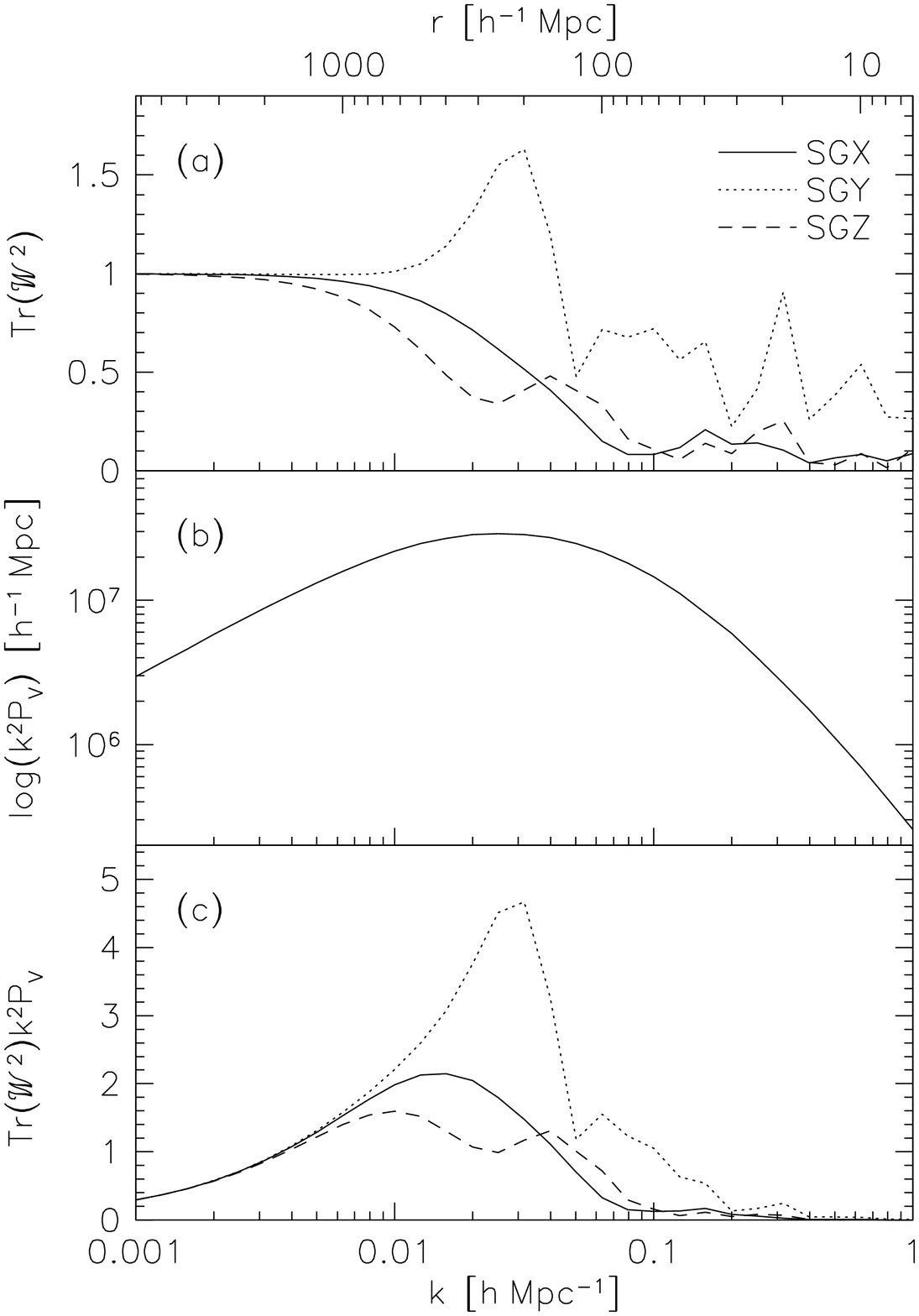}
\caption{(a)~The trace of the squared tensor window function for the
EFAR sample along the Supergalactic X, Y and Z axes; (b)~the power
spectrum for a CDM-like model with $\Gamma$=0.25 and $\sigma_8$=1.0; and
(c)~the contributions of different scales to the covariance in the
measured bulk velocity, given by the product of the power spectrum and
the squared tensor window function.}
\label{fig:pkwk}
\end{figure}

The survey's sensitivity to the power spectrum is determined by its
window function. Figure~\ref{fig:pkwk}a shows the window function for
the EFAR sample along the Supergalactic X, Y and Z axes; the Y axis in
particular shows the effect of correlated errors resulting from not
having a full-sky sample. The model power spectrum is shown in
Figure~\ref{fig:pkwk}b. The product of the power spectrum and the window
function, shown in Figure~\ref{fig:pkwk}c, gives the relative
contributions of different scales to the covariance in the measured bulk
velocity. The bulk velocity depends on a broad range of scales, with the
largest contributions coming from scales of a few hundred h$^{-1}$\,Mpc.

For the EFAR survey the `noise' component of the covariance matrix (in
Supergalactic coordinates) is
\begin{equation}
\label{eqn:errcov}
R^{\epsilon}_{ij} = \left[
\begin{array}{rrr}
+101655 &  +47914 &  -24001 \\
 +47914 &  +65373 &  -39617 \\
 -24001 &  -39617 &  +87567
\end{array} \right]
\end{equation}
while the `velocity' component is
\begin{equation}
\label{eqn:velcov}
R^{v}_{ij} = \left[
\begin{array}{rrr}
 +37169 &  +17211 &    -344 \\
 +17211 &  +23165 &   -6084 \\
   -344 &   -6084 &  +20980
\end{array} \right]~.
\end{equation}
Thus the overall covariance matrix {\boldmath $R$} is
\begin{equation}
\label{eqn:totcov}
R_{ij} = R^{\epsilon}_{ij} + R^{v}_{ij} = \left[
\begin{array}{rrr}
+138824 &  +65125 &  -24345 \\
 +65125 &  +88538 &  -45701 \\
 -24345 &  -45701 & +108547 \\
\end{array} \right]~.
\end{equation}
It is immediately apparent that (for the model considered here) the
covariance matrix is dominated by the `noise' term.

The maximum likelihood estimate, {\boldmath $U$}, for the bulk flow
of the sample clusters is given by
\begin{equation}
\label{eqn:mlbulk}
U_i = R^{\epsilon}_{ij}\sum_n\frac{\hat{r}_{n,j}v_n}{\sigma_n^2+\sigma_*^2}
\end{equation}
where $U_i$ is the $i$th component of the bulk flow, $R^{\epsilon}_{ij}$
is the `noise' covariance matrix, $\hat{r}_{n,j}$ is the $j$th component
of the unit vector of the $n$th cluster, $v_n$ and $\sigma_n$ are the
cluster's peculiar velocity and its uncertainty, and $\sigma_*$ is the
assumed rms thermal motion of the model. For the EFAR sample, the
maximum likelihood bulk flow vector in Supergalactic coordinates is
($-$24, $-$6, $+$717)\kms, almost entirely in the SGZ axis. In Galactic
coordinates this is 718\kms\ in the direction
($l$,$b$)=(45.4\degree,$+$5.9\degree). 

However this formal result is rather ill-determined, since it is far
from the main axis of the EFAR sample (cf.\ Figure~\ref{fig:pvsky}). An
indication of the uncertainty can be obtained by ignoring the
cross-correlations in the covariance matrix and estimating the rms error
as $({\rm Trace}(R^{\epsilon}))^{1/2}$=505\kms. In the context of the
assumed cosmological model, the probability of measuring a bulk flow
vector {\boldmath $U$} can be obtained by computing the $\chi^2$ statistic
from the covariance matrix as
\begin{equation}
\label{eqn:chi2vel}
\chi^2=U_i R^{-1}_{ij} U_j ~.
\end{equation}
The probability (given the cosmological model and the properties of the
sample) of observing a bulk flow with a value of $\chi^2$ greater than
this is given by the appropriate integral over the $\chi^2$ distribution
with 3 degrees of freedom (the 3 components of {\boldmath $U$}). For the
EFAR sample this procedure yields $\chi^2$=6.1 with 3 degrees of
freedom, and hence the observed bulk flow is consistent with the model
at the 11\% confidence level. If the rms thermal motion $\sigma_*$ is
set to be zero rather than 250\kms, the observations are still
consistent with the model at the 9\% confidence level.

The expectation value for the bulk motion (given the cosmological model
and the properties of the sample) can be obtained as
\begin{equation}
\label{eqn:expvel}
V = \frac{(\sigma_1\sigma_2\sigma_3)^{-1}}{(2\pi)^{3/2}}
    \int |V|\exp\left(-\sum_i\frac{V_i^2}{2\sigma_i^2}\right)\,d^3V ~,
\end{equation}
where $\sigma_1$, $\sigma_2$ and $\sigma_3$ are the lengths of the axes
of the covariance ellipsoid obtained from the eigenvalues of the
covariance matrix. The directions of these axes are given by the
(orthogonal) eigenvectors of the covariance matrix. For the EFAR sample
and our adopted cosmological model, these eigenvalues and eigenvectors
(in Supergalactic coordinates) are
\begin{equation}
\label{eqn:evalevec}
\begin{array}{rrr}
\sigma_1=454\kms\ & \mbox{\boldmath $e_1$}=(+0.7026,+0.5604,-0.4385) ~, \\
\sigma_2=309\kms\ & \mbox{\boldmath $e_2$}=(+0.5679,-0.0703,+0.8201) ~, \\
\sigma_3=185\kms\ & \mbox{\boldmath $e_3$}=(-0.4287,+0.8253,+0.3676) ~. 
\end{array}
\end{equation}
The corresponding directions in Galactic coordinates are 
{\boldmath $e_1$}=(172.6\degree,$+$30.6\degree), 
{\boldmath $e_2$}=(82.0\degree,$+$1.2\degree), and 
{\boldmath $e_3$}=(350.0\degree,$+$59.4\degree). 
We therefore find an expectation value for the amplitude of the bulk
flow of 619\kms, so that the observed value is not much larger than that
expected from our model, as the $\chi^2$ statistic indicates. It is
worth noting that the expected bulk flow amplitude is strongly dominated
by the `noise' term in the covariance matrix. For our adopted
cosmological model in the absence of noise, we would expect to measure a
bulk flow amplitude from the EFAR sample of only 355\kms, whereas in the
absence of any cosmological velocities, the noise in our measurement
would still lead us to expect a bulk flow amplitude of 553\kms.

We obtain a smaller upper limit on the bulk motion we if consider only
the component of the bulk flow along the minimum-variance axis of the
covariance ellipsoid. Unsurprisingly, this axis, {\boldmath
$e_3$}=(350.0\degree,$+$59.4\degree), is just 20\degree\ away from the
median axis of the 50 clusters in the peculiar velocity sample,
$\langle(l,b)\rangle$=(7\degree,$+$42\degree). The expected bulk flow
amplitude along this axis is 147\kms\ (124\kms\ from noise alone,
76\kms\ from model alone), while the maximum likelihood estimate of the
observed bulk motion is 269\kms. Since $\sigma_3$=185\kms, this gives
$\chi^2$=2.11 with 1 degree of freedom, implying that the observed bulk
motion in this direction is consistent with the model at the 15\%
confidence level.

Thus there is no evidence that the bulk motion of the EFAR sample is
inconsistent with a cosmological model having a CDM-like power spectrum
with $\Gamma$=0.25 and $\sigma_8$=1.0, consistent with the best current
determinations. In fact, repeating this analysis, we find that the
observations are consistent with a wide range of cosmological models,
including both standard CDM and open, low-density CDM models.

We can also ask to what extent the EFAR sample is capable of testing
whether the bulk motions measured by Lauer \& Postman (1994), SMAC
(Hudson \etal\ 1999) and LP10K (Willick 1999) are consistent with the
velocity field model. To do so we use the $\chi^2$ statistic computed
according to equation~\ref{eqn:chi2vel}, inserting the EFAR covariance
matrix for {\boldmath $R$} and the observed Lauer \& Postman, SMAC or
LP10K bulk motions for {\boldmath $U$}. If the EFAR bulk motion had been
found to be identical to the SMAC result, it would have been consistent
with the velocity field model at the 25\% level; if it had been found to
be identical to the LP10K result it would have been consistent with the
model at the 9\% level. However a bulk motion identical to the Lauer \&
Postman result would have been rejected at the 0.09\% level. Hence, as
expected, the directionality of the EFAR sample means that while it
would have provided a strong indication of an inconsistency with the
model if the Lauer \& Postman result had been recovered, recovery of the
SMAC or LP10K results would not have implied a problem with the model.

\begin{figure}
\plotone{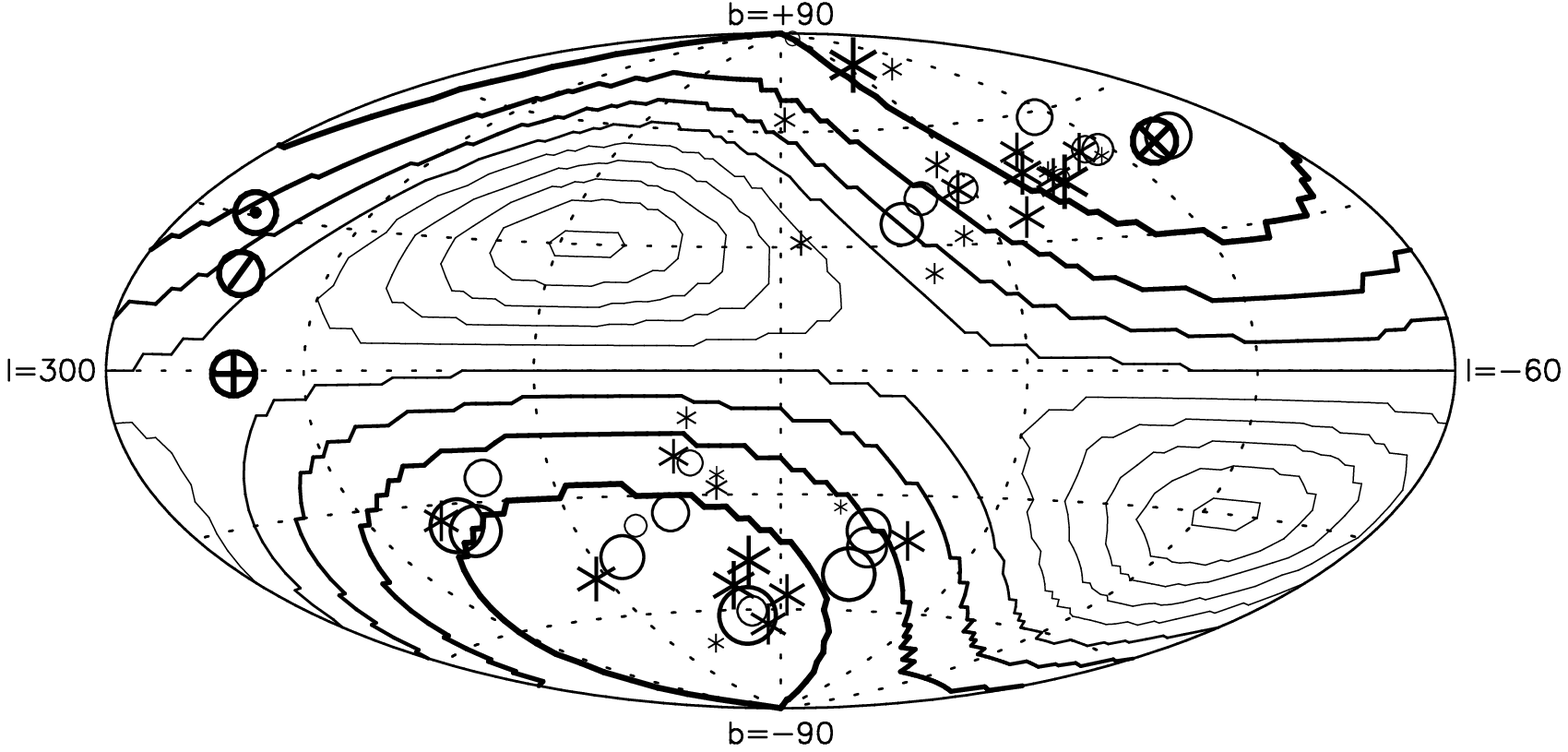}
\caption{Contour plots of the bulk motion amplitude, in each direction
on the sky, that would be rejected at the 1\% level or better by the
EFAR sample. The assumed power spectrum is CDM-like, with $\Gamma=0.25$
and $\sigma_8$=1.0, and the rms thermal motions of the clusters is
assumed to be $\sigma_*$=250\kms. The EFAR clusters with positive
(negative) peculiar velocities are indicated by asterisks (circles).
Other symbols show the directions with respect to the CMB frame of the
Local Group dipole ($\odot$), the Lauer \& Postman (1994) dipole
($\otimes$), the SMAC (Hudson \etal\ 1999) dipole ($\oplus$), and the
LP10K (Willick 1999) dipole ($\oslash$). The contours run in steps of
100\kms\ from 700\kms\ to 1500\kms, with the lowest contour being the
thickest.}
\label{fig:mapchiv}
\end{figure}

We can generalise this analysis to illustrate how the directionality of
the EFAR sample affects the constraints it could place on observed bulk
motions in different directions. Figure~\ref{fig:mapchiv} shows, in each
direction on the sky, the amplitude of the observed bulk motion that
would be rejected as inconsistent with the velocity field model at the
1\% confidence level using equation~\ref{eqn:chi2vel}.

It is important to emphasise that although it would not have been
surprising, under this model, to have recovered the SMAC motion from the
EFAR sample, in fact the tests of the previous section indicated that
the actual motions recovered from the EFAR sample are highly
inconsistent with a pure SMAC bulk flow. As already noted, however,
because those tests do not use a full velocity field model and do not
account for the window function of the sample, they will tend to
over-estimate the degree of inconsistency. The best test is a
simultaneous consistency check between both datasets and the model
(Watkins \& Feldman 1995), determining the joint probability of deriving
both the observed EFAR bulk motion from the EFAR sample and the observed
SMAC motion from the SMAC sample under the assumptions of the velocity
field model. This type of test has already been carried out for the SMAC
sample with respect to various other samples by Hudson \etal\ (2000),
who find consistency with all the other peculiar velocity surveys with
the possible exception of Lauer \& Postman, and a marginal conflict with
a flat $\Lambda$CDM model similar to that used here. Once the SMAC
peculiar velocities have been published, a similar test can be carried
out to check the consistency of the EFAR and SMAC survey results.

\section{CONCLUSIONS}
\label{conclusions}

We have measured peculiar velocities for 84 clusters of galaxies in two
large, almost diametrically opposed, regions at distances between 6000
and 15000\kms. These velocities are based on Fundamental Plane (FP)
distance estimates for early-type galaxies in each cluster. We fit the
FP to the best-studied 29 clusters using a maximum likelihood algorithm
which takes account of both selection effects and measurement errors and
yields FP parameters with smaller bias and variance than other fitting
procedures. We obtain a best-fit FP with coefficients consistent with
the best existing determinations. Apparent differences in the FPs
obtained in previous studies can be reconciled by allowing for the
biases imposed by the various fitting methods. We then fix the FP
parameters at their best-fit values and derive distances for the whole
cluster sample. The resulting peculiar velocities show no evidence for
residual systematic errors, and, for the small numbers of clusters in
common, are consistent with those measured by other authors.

We have examined the bulk motion of the sample regions using the 50
clusters with the best-determined peculiar velocities. We find the bulk
motions in both regions are small, and consistent with zero at about the
5\% level. We use both direct $\chi^2$ comparison and the more
sophisticated window function covariance analysis developed by Kaiser
(1988) and Feldman \& Watkins (1994, 1998) to compare our result with
the predictions of standard cosmological models and the results of other
studies. We find that the bulk motion of our sample is consistent (at
about the 10\% level) with the prediction of a $\Lambda$CDM model with
parameters $\Gamma$=0.25, $\sigma_8$=1.0 and $h$=0.7; indeed the motion
is consistent with most cosmological models having parameters that are
broadly consistent with the observed shape and normalisation of the
galaxy power spectrum.

We examine whether our results can be reconciled with the
large-amplitude bulk motions on similar scales found in some other
studies. Our sample lies close to the direction of the large-amplitude
dipole motion claimed by Lauer \& Postman (1994), so that we are able to
make an effective test of the bulk motion in this direction. We find
that a pure Lauer \& Postman bulk motion is inconsistent with our data
at the 0.2\% confidence level. This strong rejection of the Lauer \&
Postman result is supported by the window function covariance analysis.
We find an even stronger inconsistency between the EFAR peculiar
velocities and the result of the SMAC survey (Hudson \etal\ 1999), with
a pure SMAC bulk motion ruled out at the 0.04\% confidence level. This
is a surprisingly strong result, given that the main axis of the EFAR
sample lies at a large angle to the direction of the SMAC dipole. It
will be important to carry out a simultaneous consistency check of both
datasets with a full velocity field model using the generalised
covariance analysis described by Watkins \& Feldman (1995) and Hudson
\etal\ (2000).

To summarise current observations of bulk motions on scales larger than
6000\kms: (i)~The EFAR and SCII (Dale \etal\ 1999a) surveys find small
bulk motions, close to the predictions of cosmological models that are
constrained to be consistent with other large-scale structure
observations. (ii)~The SMAC survey (Hudson \etal\ 1999) finds a bulk
motion with a much larger amplitude. However a full accounting for the
uncertainties and window function of the survey shows that it is in fact
only marginally inconsistent with the models (at about the 2$\sigma$
level; Hudson \etal\ 2000). (iii)~The LP10K survey finds a bulk motion
very similar to the SMAC dipole, but the smaller sample size means that
the uncertainties are larger and consequently the result is not
inconsistent. (iv)~The Lauer \& Postman (1994) result {\em is}
inconsistent with such models at the 3--5\% level (Feldman \& Watkins
1994). However it is also inconsistent with the EFAR results (at the
0.2\% confidence level) and with the other surveys combined (at the
0.6\% level; Hudson \etal\ 2000), and therefore should be treated with
reserve. We conclude that existing measurements of large-scale bulk
motions provide no significant evidence against standard models for the
formation of structure.

\section*{Acknowledgements}

This work was partially supported by NSF Grant AST90-16930 to DB,
AST90-17048 and AST93-47714 to GW, and AST90-20864 to RKM. RPS was
supported by DFG grants SFB 318 and 375. The collaboration benefitted
from NATO Collaborative Research Grant 900159 and from the hospitality
and financial support of Dartmouth College, Oxford University, the
University of Durham and Arizona State University. Support was also
received from PPARC visitors grants to Oxford and Durham Universities
and a PPARC rolling grant `Extragalactic Astronomy and Cosmology in
Durham'.

\end{document}